\documentclass[fleqn,usenatbib]{mnras}

\usepackage{newtxtext,newtxmath}

\usepackage[T1]{fontenc}

\DeclareRobustCommand{\VAN}[3]{#2}
\let\VANthebibliography\thebibliography
\def\thebibliography{\DeclareRobustCommand{\VAN}[3]{##3}\VANthebibliography}

\usepackage{graphicx}	
\usepackage{amsmath}	
\usepackage{bm}

\newcommand{\Gaia}{\textit{Gaia} }
\newcommand{\rev}[1]{#1}

\title[A single cosmic clock for GSE and the bar]{Impact of Gaia-Sausage-Enceladus merger revealed by a single cosmic clock: thick-to-thin disc transition, gas disc shrinking and bar formation}

\author[N. Funakoshi et al.]{Natsuki Funakoshi,$^{1,2}$\thanks{E-mail: \href{mailto:n.funakoshi@ucl.ac.uk}{n.funakoshi@ucl.ac.uk}}
Daisuke Kawata,$^{1,3}$
Jason L. Sanders,$^{4}$
Ioana Ciuc{\u{a}},$^{5}$
Robert J.~J. Grand,$^{6}$
\newauthor
Noriyuki Matsunaga,$^{7}$
Brian Thorsbro,$^{2,8}$
HanYuan Zhāng$^{9}$
\\
$^{1}$Space and Climate Physics, Mullard Space Science Laboratory, University College London Holmbury St. Mary, Dorking, Surrey, RH5 6NT, UK\\
$^{2}$Division of Astrophysics, Department of Physics, Lund University, Box 118, 221 00 Lund, Sweden\\
$^{3}$National Astronomical Observatory of Japan, 2-21-1 Osawa, Mitaka, Tokyo 181-8588, Japan \\
$^{4}$Department of Physics and Astronomy, University College London, London, WC1E 6BT, UK\\
$^{5}$Kavli Institute for Particle Astrophysics and Cosmology, Stanford University, 452 Lomita Mall, Stanford, CA 94305, UK\\
$^{6}$Astrophysics Research Institute, Liverpool John Moores University, 146 Brownlow Hill, Liverpool L3 5RF, UK\\
$^{7}$Department of Astronomy, The University of Tokyo, 7-3-1, Hongo, Bunkyo-ku, 113-0033, Tokyo, Japan\\
$^{8}$Université Côte d’Azur, Observatoire de la Côte d’Azur, CNRS, Laboratoire Lagrange, 06000 Nice, France\\
$^{9}$Institute of Astronomy, University of Cambridge, Madingley Road, Cambridge CB3 0HA, UK
}

\date{Accepted XXX. Received YYY; in original form ZZZ}

\pubyear{\the\year{}}

\begin{document}
\label{firstpage}
\pagerange{\pageref{firstpage}--\pageref{lastpage}}
\maketitle

\begin{abstract}
Whether the formation of the Galactic bar was connected to the Gaia-Sausage-Enceladus (GSE) merger, the last significant merger experienced by the Milky Way, remains an open question. \rev{The} bar formation epoch has been inferred from the star formation history of the nuclear stellar disc \rev{(NSD)} traced by Mira variables, and the GSE merger epoch from the age-metallicity relation of red giants. However, these epochs rest on different stellar tracers with different age scales and cannot be compared directly. 
\rev{We} recalibrate the Mira period-age relation onto the age scale of APOGEE red giants, with ages from a Bayesian neural network trained on \textit{Kepler} asteroseismic ages. We \rev{fit the Gaia DR3 proper motions of 2\,394 O-rich disc Mira variables with an action-based quasi-isothermal distribution function whose parameters vary with period, and derive the period–age relation by matching} their vertical velocity dispersion as a function of period to that \rev{of the} APOGEE red giants as a function of age in \citet{2025MNRAS.543.2275F}.
This places \rev{the NSD star-formation rise and peak} at $8.5-10.5$ Gyr ago on the red giant age scale, overlapping with the final stage of the GSE-induced starburst, thick-to-thin disc transition and the contraction of the star-forming gas disc. \rev{If NSD star formation traces bar formation,} these coincidences suggest a possible scenario in which the GSE seeded the bar formation, \rev{while} fresh gas supplied by the GSE \rev{and} outer disc gas driven inward by its tidal torques, increased the baryonic dominance of the inner Galaxy. 
\end{abstract}



\section{Introduction}
The Galactic bar is one of the dominant non-axisymmetric structures in the inner Milky Way \citep[e.g.][]{1991MNRAS.252..210B,1991ApJ...379..631B,1995ApJ...445..716D} and plays a central role in shaping the dynamical and secular evolution of the Galactic disc \citep[e.g.][]{2004ARA&A..42..603K,2013MNRAS.429.1949A,2014RvMP...86....1S}. Its present-day properties, such as its size and orientation \citep[e.g.][]{2013MNRAS.435.1874W,2015MNRAS.450.4050W} and its pattern speed (\citealp[e.g.][]{2017MNRAS.465.1621P,2019MNRAS.488.4552S,2019MNRAS.490.4740B,2022MNRAS.512.2171C};
see Fig.~10 of \citealp{2025NewAR.10001721H} for a recent compilation) have been increasingly constrained by photometric, spectroscopic and astrometric surveys.
However, the formation history of the Galactic bar remains relatively poorly constrained. 

Our understanding of how galactic bars form has been shaped by dynamical theory and numerical simulations, combined with observations of the demographics and properties of barred galaxies \citep[e.g.][]{2013MNRAS.429.1949A,2014RvMP...86....1S,2025MNRAS.538.1587F}. Together, these studies suggest that bars can form either through internal disc instabilities and/or tidal perturbations associated with interactions or mergers. In isolated disc galaxies, a stellar bar may arise spontaneously from global gravitational instabilities when the disc becomes sufficiently self-gravitating, and dynamically cold \citep[e.g.][]{1971ApJ...168..343H,1973ApJ...186..467O,1982MNRAS.199.1069E, 2018MNRAS.473.2608Z, 2019MNRAS.482.1983F}. In this secular picture, the bar formation time depends on the internal evolution of the galaxy, including the growth of the stellar disc, its dynamical temperature, i.e. the velocity dispersion of stars, gas content and the stabilising influence of the bulge and dark matter halo \citep[e.g.][]{2002ApJ...569L..83A,2012ApJ...757...60K,2025MNRAS.538.1587F,2025ApJ...978...37A}.  
By contrast, interactions and mergers can tidally disturb the stellar disc, driving non-axisymmetric structures and triggering bar formation \citep[e.g.][]{1987MNRAS.228..635N,1990A&A...230...37G, 1998ApJ...499..149M, 2016ApJ...826..227L, 2019MNRAS.483.2721P}. In such simulations, tidally induced bars often form during or shortly after a close passage of the perturber \citep{2024MNRAS.531.1520M,2025A&A...702A...7L}. How strongly an encounter drives bar formation depends on the strength of the tidal perturbation it produces, the orbit of the perturber and the dynamical state of the host disc \citep[e.g.][]{1972ApJ...178..623T,2018ApJ...857....6L}.

The Milky Way is now known to have experienced the last significant merger, the Gaia-Sausage-Enceladus (GSE) merger \citep[e.g.][]{2003ApJ...585L.125B,2010A&A...511L..10N,2018MNRAS.478..611B,2018ApJ...863..113H,2018Natur.563...85H} roughly $8-11~\mathrm{Gyr}$ ago \citep{2019NatAs...3..932G,2020NatAs...4..382C,2020ApJ...897L..18B}. The GSE merger is thought to have affected the dynamical and chemical evolution of the Galactic disc \citep[e.g.][]{2018MNRAS.474.3629G,2020MNRAS.494.3880B,2020A&A...636A.115D,2024MNRAS.528L.122C,2025MNRAS.543.2275F,2026MNRAS.548f2154O}. Given that external perturbations can trigger bar formation, the GSE merger is a natural candidate for an early perturbation that may have triggered the Galactic bar formation.

In a Milky Way-like cosmological simulation \citep[\texttt{Auriga};][]{2024MNRAS.532.1814G}, \citet{2024MNRAS.531.1520M} found that a GSE-like merger induced tidal perturbations on the disc, promoted the initial seed of the bar and inward motion of the gas and triggered a starburst, followed by the formation of a stable stellar bar at $\sim 1~\mathrm{Gyr}$ after the first pericentric passage of the GSE-like progenitor. 
In a controlled rerun without the GSE-like merger, the bar formed $\sim 2~\mathrm{Gyr}$ later than in the original simulation, suggesting that the GSE-like merger triggered bar formation by shifting it to earlier time.
This result motivates a direct comparison between the epoch of the GSE merger and that of bar formation in the Milky Way.

The formation epoch of the Milky Way bar can be indirectly constrained using the stellar populations of the nuclear stellar disc (NSD), a compact stellar disc structure in the central few hundred parsecs of the Galaxy \citep[e.g.][for review]{2002A&A...384..112L,2021A&A...650A.191S, 2025A&ARv..33....7S}. The bar formation is considered to drive gas into the central regions of the Galaxy, where it settles into a star-forming nuclear disc \citep[e.g.][]{2020MNRAS.492.4500B}. In this picture, the formation of the NSD is closely linked to the formation of the bar, and the initial build-up of the NSD can provide an indirect but powerful constraint on the bar formation epoch. To apply this idea, \citet{2024MNRAS.530.2972S} used Mira variables in the Galactic centre as age and kinematic tracers, adopting the \rev{Mira period–age relation calibrated by \citet{2023MNRAS.521.1462Z} using stellar kinematics.} \rev{In this calibration, Miras were grouped by pulsation period, and the velocity dispersion of each period group was inferred from \Gaia proper motions using a dynamical model. These velocity dispersions were then converted into ages using an age–velocity-dispersion relation established from red giant and sub-giant stars with isochrone-based ages \citep{2018MNRAS.475.1093Y}, thereby providing an indirect calibration between Mira period and age.}
They found that Miras whose kinematics are consistent with the NSD show a rapid increase in number at ages of around $8~\mathrm{Gyr}$, suggesting a marked rise in NSD star formation at that epoch \rev{\citep[see also][]{2020NatAs...4..377N}}. If this early NSD build-up was driven by bar-induced gas inflow, this points to a bar formation epoch roughly $8~\mathrm{Gyr}$ ago \citep{2024MNRAS.530.2972S}.
\rev{Other, more indirect inferences favour a younger bar. For examples, \citet{2022A&A...659A..80W} suggested a bar age of $\sim6-8$~Gyr from the ages of inner-ring stars, which are thought to have formed in a ring created by the bar.
\citet{2024A&A...681L...8N} found a sharp decline in the [Fe/H] of the most metal-rich stars in the solar neighbourhood at ages younger than $\sim3$~Gyr. They interpreted this as the suppression of metal-rich star formation in the inner disc following bar formation, which they placed at $3-4$~Gyr ago. \citet{2026A&A...705A..92R} also attributed an enhanced star formation rate at $\sim4$~Gyr ago to bar formation. These estimates rely on different tracers and assumptions, and the bar formation epoch remains debated. In this paper, we adopt the onset of NSD star formation as our indicator of the bar formation epoch, because the NSD is expected to form from bar-driven gas inflow and therefore traces bar formation more directly than populations in the wider disc. Our conclusions are conditional on this assumption.}

An independent constraint on the GSE merger epoch has been obtained from the age-metallicity relation of Galactic disc stars. \citet{2024MNRAS.528L.122C} analysed APOGEE (The Apache Point Observatory Galactic Evolution Experiment) red giant stars with ages inferred using a Bayesian machine-learning framework, called BINGO \citep[Bayesian INference for Galactic archaeOlogy][]{2021MNRAS.503.2814C}, calibrated on \textit{Kepler} asteroseismic ages \citep{2021A&A...645A..85M}. They identified the Great Galactic Starburst (GGS) phase, which shows rapid chemical enrichment and indicates the epoch of the gas-rich GSE merger around $10-12$~Gyr ago. In \citet{2025MNRAS.543.2275F}, we used the same red giant sample and found a hint that the star-forming gas disc shrank at 10~Gyr ago, shortly after the GSE merger \citep[see also][]{2025arXiv251209987Z}, which is qualitatively similar to the inward gas motion seen in the GSE-like merger simulation by \citet{2024MNRAS.531.1520M}. If the Galactic bar formed during or shortly after the GSE merger, it would match the scenario proposed by \citet{2024MNRAS.531.1520M}: a GSE-like merger triggered the bar formation onset. 

Still, a direct comparison between these epochs is complicated by the use of different stellar age calibrations. The bar formation epoch inferred from NSD Miras in \citet{2024MNRAS.530.2972S} is \rev{based on a Mira period–age relation that is kinematically calibrated using an age–velocity-dispersion relation established from red giant and sub-giant stars with isochrone-based ages} \citep{2012MNRAS.427..127B,2018MNRAS.475.1093Y,2023MNRAS.521.1462Z}, whereas the GSE-merger epoch inferred from APOGEE red giants relies on ages calibrated against \textit{Kepler} asteroseismic measurements \citep{2021A&A...645A..85M,2021MNRAS.503.2814C,2024MNRAS.528L.122C}. To assess whether or not the GSE merger and bar formation occurred at similar epochs, the Mira period--age scale and red giant age scales must first be brought onto a consistent age scale. 

In this study, we recalibrate the Mira period-age relation onto the APOGEE/BINGO age scale and use it to examine whether the bar formation epoch is comparable to the GSE merger epoch. We compare the kinematics of the two stellar tracers in the Galactic disc: Miras and APOGEE red giants. 
In Section~\ref{sec:data}, we describe the APOGEE red giant and Mira sample used in this work. In Section~\ref{sec:modelling}, we present our age-scale calibration method: we model the period-dependent kinematics and disc structure of Mira variables as described in Section~\ref{sec:DF}, and compare them with the age-dependent velocity dispersions of APOGEE red giants derived in \citet{2025MNRAS.543.2275F} as described in Section~\ref{sec:pa}. This comparison allows us to calibrate the Mira period-age relation onto the APOGEE red giant age scale. In Section~\ref{sec:result}, \rev{we present the calibrated period-age relation of Mira variables.}
\rev{In Section~\ref{sec:discussion}, we place the bar formation epoch on the APOGEE/BINGO age scale and compare it}
with the GSE merger epoch identified from APOGEE red giants, including the epoch of the star-forming gas disc contraction found in \citet{2025MNRAS.543.2275F}. 
We discuss the implications of our results for the formation epoch of the Milky Way bar and its possible connection to the GSE merger. We show that the inferred formation epoch of the Milky Way bar is broadly consistent with the timeline seen in the \texttt{Auriga} simulation of \citet{2024MNRAS.531.1520M}, in which a GSE-like merger triggers bar formation. Finally, in Section~\ref{sec:summary}, we summarise our conclusions.

\section{Data}\label{sec:data}
\subsection{APOGEE red giants}\label{sec:data:apogee}
The APOGEE red giant sample used in this paper was identical to that used in \citet{2025MNRAS.543.2275F}. We therefore only summarise the main properties of the sample here.

The sample was based on the red giant catalogue of \citet{2024MNRAS.528L.122C}, who applied the BINGO Bayesian-neural-network-based age model developed by \citet{2021MNRAS.503.2814C} to APOGEE-2 DR17 stars \citep{2015AJ....150..173N, 2017AJ....154...94M,2019PASP..131e5001W}. BINGO was trained with APOKASC-2 stars with asteroseismic age \citep{2021A&A...645A..85M}, and infers stellar ages and their statistical uncertainties from APOGEE stellar parameters and chemical abundances. The training set was restricted to red giant branch stars and high-mass red clump stars, for which the asteroseismic ages are considered reliable. \citet{2024MNRAS.528L.122C} selected disc red giants similar to those in the training sample from the APOGEE-2 DR17 catalogue, and applied BINGO only to the selected giants.

We cross-matched this APOGEE red giant sample with BINGO ages against \Gaia DR3 \citep{2023A&A...674A..37G} to obtain proper motions and parallaxes, and adopted $V_{\rm{helio}}$ from the APOGEE catalogue as the line-of-sight velocity. We applied the same quality cuts as in \citet{2025MNRAS.543.2275F}: $\texttt{parallax}/\texttt{parallax\_error}>5$, renormalised unit weight error \texttt{RUWE}$~<1.4$, an age uncertainty smaller than $0.02~\mathrm{dex}$ in the base-10 logarithmic age and a probability higher than 95\% of being either red giant branch stars or high-mass red clump stars \citep{2024MNRAS.528L.122C}. We also removed stars with vertical distances larger than five times the dispersion of the vertical distribution of stars with similar ages. After these selections, the final APOGEE red giant sample contains 16\,617 stars.

The sample consisted mainly of Galactic disc stars with $\mathrm{[Fe/H]}>-1.0$. This was a consequence of the BINGO age-uncertainty selection because low-metallicity stars were poorly represented in the asteroseismic training set and therefore rarely satisfied the age-uncertainty cut. This selection was appropriate for the present analysis, since the aim was to compare the Mira variables with an APOGEE reference sample representing the Galactic disc. As discussed in \citet{2021MNRAS.503.2814C}, \citet{2024MNRAS.528L.122C} and \citet{2025MNRAS.543.2275F}, the BINGO ages are on their own relative age scale, which may not correspond exactly to an absolute age scale of the Universe.

\subsection{Mira variables}

We based our sample on the O-rich Mira variable catalogue constructed by \citet{2023MNRAS.521.1462Z}, who collected long-period variable (LPV) candidates from \Gaia DR3 \citep{2016A&A...595A...1G, 2023A&A...674A...1G,2023A&A...674A..15L,2024MNRAS.530.2972S}. Starting from the full LPV catalogue cross-matched with 2MASS \citep[Two Micron All Sky Survey;][]{2006AJ....131.1163S}, \citet{2023MNRAS.521.1462Z} applied cuts on pulsation period ($80<P<1\,000~\mathrm{days}$) following \citet{2009MNRAS.399.1709M}, and in the G-band semi-amplitude computed from a Fourier fit ($>0.5$ mag, similar to the cut of 0.43 mag used by \citealt{2019MNRAS.483.3022G}). \citet{2023MNRAS.521.1462Z} applied further various selection cuts to extract O-rich Miras and minimise the contaminations. See \citet{2023MNRAS.521.1462Z} for more details.

Following \citet{2023MNRAS.521.1462Z}, distance moduli, $m=5\log_{10}(d/10~\mathrm{pc})$, where $d$ is the heliocentric distance of the star, were estimated from the period-luminosity relation (PLR) of O-rich Mira variables using the Wesenheit magnitude $W_{KJK} = K_s-0.473(J - K_s)$, with absolute magnitudes calibrated using Large Magellanic Cloud Mira variables:
\begin{equation*}
    M_{K J K}= \begin{cases}-7.53-4.05\left(\log _{10} P-2.3\right), & \log _{10} P<2.6, \\ -8.75-6.99\left(\log _{10} P-2.6\right), & \log _{10} P \geq 2.6,\end{cases}
\end{equation*}
where $P$ is pulsation period in days. 
The extinction coefficient was taken from \citet{2019ApJ...877..116W}. 
The distance modulus uncertainty $\sigma_m$ was computed by combining in quadrature the intrinsic scatter of the PLR with uncertainties propagated from the 2MASS photometric and period measurements. The dominant contribution came from the single-epoch 2MASS photometry, with a typical scatter of $>0.22$~mag, while period uncertainties contributed a median of~0.06 mag. We retained only stars with $\sigma_m<0.6$~mag, following the quality criterion of \citet{2023MNRAS.521.1462Z}, to ensure reliable distance estimates.

We cross-matched the sample with \Gaia DR3 to obtain astrometric data, including proper motions and the \texttt{RUWE} as an astrometric quality indicator. Stars with \texttt{RUWE}$~>1.4$ were removed following \citet{2021A&A...649A...2L}.
Mira variables present particular challenges for \textit{Gaia}'s astrometric pipeline for several reasons: they are extremely red sources for which \textit{Gaia}'s image parameter 
determination is less well characterised \citep{2021A&A...649A..11R}; they vary in both magnitude and colour, whereas the current \Gaia astrometric pipeline assumes a fixed colour for each star, potentially introducing systematic errors \citep{2003A&A...399.1167P}; and they have physical radii of order 1~AU or larger with optical photocentre wobbles on scales of up to 10\% of the stellar radius \citep{2011A&A...528A.120C,2018A&A...617L...1C,2017A&A...600A.137F,2024A&A...690A.125B}. The net effect is that the formal \Gaia astrometric uncertainties underestimate the true uncertainties. To account for this effect, we inflated the reported proper motion uncertainties by a factor of approximately 1.6 and by an additional factor of $1 + \exp[-(m - 8.5)/0.8]$, where $m$ is distance modulus, as recommended by \citet{2023MNRAS.523.2369S,2023MNRAS.521.1462Z}.

Although we did not make any spatial cuts to the APOGEE sample described in Section~\ref{sec:data:apogee}, we \rev{restricted the} Mira sample \rev{to} the Galactocentric radii $R>6$~kpc and the heliocentric distances $d<4$~kpc.
\rev{The APOGEE sample is strongly concentrated around the solar neighbourhood, with limited coverage at $d=4$~kpc and $R<6$~kpc, whereas the Mira sample is more broadly distributed across the Galactic disc and is less strongly concentrated around the solar neighbourhood.}
\rev{Restricting the Mira sample to this volume improves the comparability of the two tracers by limiting the analysis to a region where the APOGEE red giant sample provides substantial spatial coverage.
These cuts therefore restrict the Mira sample to a volume where the APOGEE red giant sample provides substantial spatial coverage, improving a more consistent comparison between the two tracers. }
\rev{The cut at $R > 6$~kpc also excludes the region} dominated by the Galactic bar, whose non-axisymmetric kinematics cannot be captured by our axisymmetric dynamical model \rev{and could introduce systematic biases into our fits.} \rev{In addition, this cut avoids regions of high extinction} towards the Galactic centre, where dust extinction corrections based on Wesenheit magnitudes are more uncertain \citep{2023MNRAS.523.2369S}. 
Finally, we excluded stars at vertical distances exceeding five times the vertical dispersion of Mira stars with similar periods in our sample, following the same outlier rejection procedure applied to the APOGEE sample in \citet{2025MNRAS.543.2275F}.

\begin{figure}	\includegraphics[width=\columnwidth]{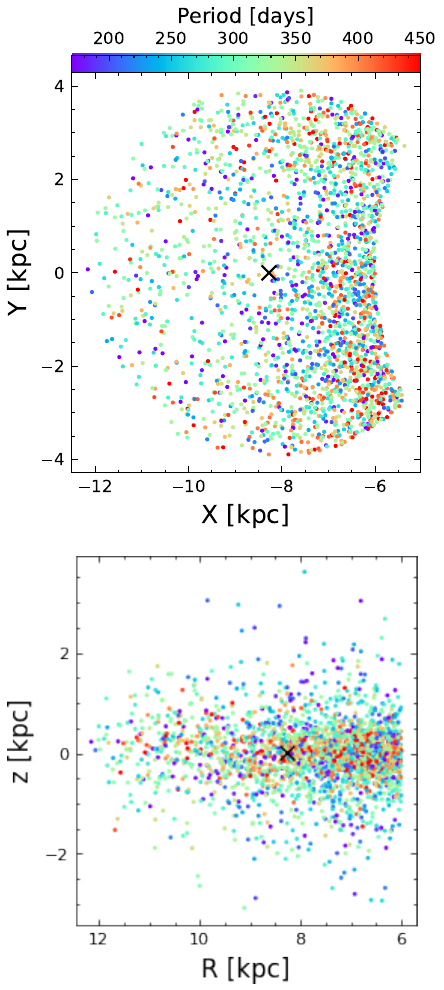}
    \caption{Spatial distribution of the Mira variable stars used in our study. The upper panel shows the face-on view of the stellar density distribution. The lower panel presents the distribution in Galactocentric radius and vertical height. In both panels, colour denotes stellar pulsation period. The position of the Sun is indicated by a black cross in both panels, located at ($X,Y$)=($-8.275$,0)~kpc in the upper panel and ($R,z$)= (8.275,0.0208)~kpc in the lower panel.}
    \label{fig:xyrz}
\end{figure}

After applying all the selection criteria described above, 
the final Mira variable sample contained 2\,394 stars with measurements of ($\ell,b,m,\bm{\mu},P$), where $\ell$ and $b$ are the Galactic longitude and latitude, $m$ is the distance modulus, $\bm{\mu}$ is the proper-motion vector and $P$ is the stellar pulsation period. The spatial distribution of our selected Mira sample is shown in Fig.~\ref{fig:xyrz}. 

For Mira variables, the available line-of-sight velocity measurements are not always reliable tracers of the centre-of-mass line-of-sight velocities. \Gaia RVS measurements are limited for these very red and strongly variable stars, which can be faint in the optical during parts of their pulsation cycles. In addition, the observed spectral-line velocities may be affected by the expansion and contraction of the pulsating atmosphere, producing phase-dependent offsets from the true centre-of-mass line-of-sight velocity. 
We therefore chose not to use line-of-sight velocities of Mira variables in the distribution-function fitting for the Mira sample.

\section{Modelling}\label{sec:modelling}
The goal of this study is to derive a period--age relation for Mira variables by linking their pulsation periods to the ages of APOGEE giant stars inferred with BINGO \citep{2024MNRAS.528L.122C}. To achieve this, we adopted a two-step strategy.

First, we quantified how the structure and kinematics of the Galactic disc traced by the Mira variables depended on their period. 
This allowed us to measure the period dependence of the disc distribution function, $\mathrm{DF}(P)$, for Miras.
Second, we compared the period-dependent distribution functions of Miras with the age-dependent distribution functions of APOGEE red giants. For the APOGEE red giants, whose selection is described in Section~\ref{sec:data:apogee}, we used the best-fit distribution functions as a function of age, $\mathrm{DF}(\mathrm{age})$, obtained by \citet{2025MNRAS.543.2275F}. Under the assumption that stars of the same age share the same distribution function, we derived the period--age relation for Miras by identifying the age at which the APOGEE red giant distribution function matched the Mira distribution function at a given period.
In practice, however, we did not use the full distribution function for this comparison. We consider that the most reliable kinematic tracer for Miras is the vertical velocity dispersion $\sigma_{z,0}$ and therefore derived the period--age relation by matching the age-$\sigma_{z,0}$ relation of the APOGEE red giants to the period-$\sigma_{z,0}$ relation of the Miras. The validity of this choice is assessed in Appendix~\ref{sec:validation} using mock data, which show that, while the Mira data alone do not robustly constrain all distribution-function parameters, they are sufficient to recover the period--$\sigma_{z,0}$ relation reliably.

\subsection{Distribution function modelling for Mira variables}\label{sec:DF}
We aimed to fit the kinematic data of Mira with a probability distribution function for the proper motions of stars, $p(\bm{\mu} \mid \ell, b, m, P)$.
Since stellar-density measurements are highly sensitive to observational selection functions and are difficult to obtain accurately, and since line-of-sight velocities for Miras are not sufficiently reliable, we focused solely on fitting the action-based distribution functions of proper motions, following \citet{2023MNRAS.521.1462Z} and \citet{2024MNRAS.530.2972S}.

We adopted the same solar position and motion as those assumed in \citet{2025MNRAS.543.2275F}. The Galactocentric radius of the Sun was set to $R_0=8.275$~kpc \citep{2021A&A...647A..59G}, and its height above the Galactic mid-plane was taken to be $z_0=20.8$~pc \citep{2019MNRAS.482.1417B}. For the solar peculiar motion with respect to the Local Standard of Rest, we used $(u_0, v_0, w_0)=(9.65, 14.00, 8.59)~\mathrm{km~s^{-1}}$ \citep{2024MNRAS.529.1035A}. Here, $u_0$, $v_0$ and $w_0$ denote the velocity components toward the Galactic centre, in the direction of Galactic rotation and toward the north Galactic pole, respectively. The circular velocity at the solar radius was fixed to $V_\mathrm{c}(R_0)=231.21~\mathrm{km~s^{-1}}$, as obtained from the axisymmetric Milky Way gravitational potential described by \citet{2017JOSS....2..388P}.

The probability distribution function for Miras is described as follows: 
\begin{equation}\label{eq:probabiliry}
p(\bm{\mu} \mid \ell, b, m, P)=\frac{p(\ell, b, m, \bm{\mu},  P)}{p(\ell, b, m, P)}
 =\frac{\int \mathrm{d} v_{\|}~p\left(\ell, b, m, \bm{\mu}, v_{\|},P\right)}{\int \mathrm{d}^3 \bm{v} ~p\left(\ell, b, m, \bm{v},P \right)}.
\end{equation}
This expression gives the probability distribution of proper motions conditioned on the position and period of the stars. By conditioning on these observables, the resulting distribution is expected to be less sensitive to the selection function. Since line-of-sight velocities are not used for Miras, we marginalised the distribution function over the line-of-sight velocity.
We accounted for the observational uncertainties by convolving the model probability density with the measurement errors in proper motion, distance modulus and pulsation period. The integrand in the numerator becomes
\begin{equation}
\begin{aligned}
&p\left(\ell, b, m, \bm{\mu}, v_{\|}, P\right) \\
&\quad = \int \mathrm{d}^2 \bm{\mu}^{\prime} 
\mathrm{d} m^{\prime} \mathrm{d} P^{\prime} \mathcal{N}\left(\bm{\mu} \mid \bm{\mu}^{\prime}, \mathbf{\Sigma}_\mu\right) \mathcal{N}\left(m \mid m^{\prime}, \sigma_m^2\right) \mathcal{N}\left( P \mid P^{\prime}, \sigma_{P}^2\right)\\ &\hspace{5.5cm} p\left(\ell, b, m^{\prime}, \bm{\mu}^{\prime}, v_{\|},P^\prime\right),
\end{aligned}
\end{equation}
where $\mathcal{N}\left(x\mid\mu,\sigma^2\right)$ denotes a Gaussian distribution with mean $\mu$ and variance $\sigma^2$. The sky position of stars, $(\ell,b)$, was assumed to be accurately measured and were not included in the marginalisation.
Equivalently, the integrand in the denominator can be written as
\begin{equation}
\begin{aligned}
&p\left(\ell, b, m, \bm{v}, P\right) \\
&\quad = \int \mathrm{d} m^{\prime}
\mathrm{d} P^{\prime}
\mathcal{N}\left(m \mid m^{\prime}, \sigma_m^2\right)
\mathcal{N}\left(P \mid P^{\prime}, \sigma_P^2\right)
p\left(\ell, b, m^{\prime}, \bm{v}, P^{\prime}\right).
\end{aligned}
\end{equation}
\rev{Here, $\bm{v}$ denotes the intrinsic three-dimensional velocity in the model distribution function, rather than an observed velocity.}

Then, we related the distribution function in observable coordinates to the distribution function in action space as follows:
\begin{align}
p\left(\ell, b, m^{\prime}, \bm{\mu}^{\prime}, v_{\|}^{\prime},P\right)&=\left|\frac{\partial(\bm{J}, \bm{\theta})}{\partial\left(\ell, b, m, \bm{\mu}, v_{\|}\right)}\right| f_P(\bm{J}) \propto s^5 \cos b \text { }f_P(\bm{J}) \text {, }\label{eq:propto_s5}\\
p\left(\ell, b, m^{\prime}, \bm{v}^{\prime},P\right)&=\left|\frac{\partial(\bm{J}, \bm{\theta})}{\partial\left(\ell, b, m, \bm{v}\right)}\right| f_P(\bm{J}) \propto s^3 \cos b \text { }f_P(\bm{J}) \text {, }\label{eq:propto_s3}
\end{align}
where $\bm{J}=(J_R,J_\phi,J_z)$ is the vector of actions corresponding to the six-dimensional observable coordinates, with corresponding angle coordinates $\bm{\theta}$, and $s$ is the distance corresponding to the distance modulus. The factor $s^5\cos{b}$ in equation~\eqref{eq:propto_s5} and $s^3\cos b$ in equation~\eqref{eq:propto_s3} are derived from the Jacobian between the Galactic coordinates and Cartesian coordinate, $(\bm{x},\bm{v})$. The factor $s^3\cos{b}$ appears in the transformation of the volume element in three-dimensional position space in both equations~\eqref{eq:propto_s5} and \eqref{eq:propto_s3}. The additional factor of $s^2$ in equation~\eqref{eq:propto_s5} comes from the transformation of the area element in proper motion velocity space. On the other hand, in equation~\eqref{eq:propto_s3}, because the three-dimensional velocity, $\bm{v}$, is sampled in Cartesian coordinates, as explained in Appendix~\ref{sec:IS}, only the spatial volume Jacobian, $s^3\cos b$, appears in this expression. As noted by \citet{2023MNRAS.521.1462Z}, the Jacobian between $(\bm{x},\bm{v})$ and $(\bm{J},\bm{\theta})$ is unity because the transformation is canonical.

To model the Galactic disc, we adopted the quasi-isothermal distribution function $f(\bm{J})$ of \citet{2010MNRAS.401.2318B}, as implemented in the \texttt{AGAMA} software package \citep{2018arXiv180208255V}, which uses the following functional form:
\begin{equation}\label{eq:DF}
\begin{gathered}
f_P(\bm{J})=\frac{\tilde{\Sigma} \Omega}{2 \pi^2 \kappa^2} \times \frac{\kappa}{\tilde{\sigma}_r^2} \exp \left(-\frac{\kappa J_r}{\tilde{\sigma}_r^2}\right) \times \frac{\nu}{\tilde{\sigma}_z^2} \exp \left(-\frac{\nu J_z}{\tilde{\sigma}_z^2}\right) \times B\left(J_\phi\right), \\
B\left(J_\phi\right)= \begin{cases}1 & \text { if } J_\phi \geq 0, \\
\exp \left(\frac{2 \Omega J_\phi}{\tilde{\sigma}_r^2}\right) & \text { if } J_\phi<0,\end{cases} \\
\tilde{\Sigma}\left(R_{\mathrm{c}}\right) \equiv \Sigma_0 \exp \left(-R_{\mathrm{c}} / R_{\text {disc }}(P)\right), \\
\tilde{\sigma}_r^2\left(R_{\mathrm{c}}\right) \equiv \sigma_{r, 0}^2(P) \exp \left(-2\left(R_{\mathrm{c}}-R_0\right) / R_{\sigma, R}(P)\right), \\
\tilde{\sigma}_z^2\left(R_{\mathrm{c}}\right) \equiv \sigma_{z, 0}^2(P) \exp \left(-2\left(R_{\mathrm{c}}-R_0\right) / R_{\sigma, z}(P)\right),
\end{gathered}
\end{equation}
where $R_{\mathrm{c}}$ is the radius of the circular orbit corresponding to the star’s angular momentum, defined by $J_{\phi} \equiv L_\mathrm{z} = R_{\mathrm{c}} V_{\mathrm{c}}(R_{\mathrm{c}})$. In an axisymmetric potential, $L_\mathrm{z}$ is conserved and may equivalently be written as $L_\mathrm{z} = R V_\phi$, using the star’s present Galactocentric radius and azimuthal velocity. The symbols $\kappa$, $\Omega$ and $\nu$ represent the epicyclic, angular and vertical orbital frequencies, respectively. The model involves five key free parameters: the disc scale length, $R_{\textrm{disc}}$; the radial and vertical velocity dispersions at the solar radius, $\sigma_{R,0}$ and $\sigma_{z,0}$; and the scale lengths, $R_{\sigma,R}$ and $R_{\sigma,z}$\rev{, which characterize the radial profile of the radial and vertical velocity dispersions, respectively. Larger values of these scale lengths correspond to more gradual declines in velocity dispersion with increasing circular-orbit radius}. These parameters describe the shape of the distribution function and approximate the behaviour of the corresponding physical quantities, but they should not be interpreted as exact measurements of those quantities \citep[see also,][]{2011MNRAS.413.1889B,2013ApJ...779..115B}. We adopted a fixed axisymmetric gravitational potential of the Milky Way from \citet{2017JOSS....2..388P} and computed the actions, $\bm{J}=\bm{J}(\bm{x},\bm{v})$, of each star using the St\"{a}ckel fudge approximation as implemented in \texttt{AGAMA}.

\rev{Following \citet{2023MNRAS.521.1462Z} and \citet{2025MNRAS.543.2275F}, we evaluated the integrals in equation~\eqref{eq:probabiliry} by importance sampling (details in Appendix~\ref{sec:IS}). The key difference from those works is that the reference distribution function is the age-dependent best fit of \citet{2025MNRAS.543.2275F}, converted to period by iterating the period–age relation to convergence.}

\subsubsection{Likelihood, priors and smoothing}
We maximised the following log-likelihood using MCMC:
\begin{equation}
\ln L=\sum_k^{\text {stars }} \ln p\left(\bm{\mu}_k\mid \ell_k, b_k, m_k, P_k\right),
\end{equation}
for the five parameters: $R_{\textrm{disc}}$, $\sigma_{R,0}$, $\sigma_{z,0}$, $R_{\sigma,R}$ and $R_{\sigma,z}$. In the real Galactic disc, these parameters are expected to vary with stellar age. Since the pulsation period of a Mira variable is assumed to be correlated with its age, we modelled the period dependence of the five distribution-function parameters, i.e. $R_\mathrm{disc}(P),\sigma_{R,0}(P),\sigma_{z,0}(P),R_{\sigma,R}(P)$ and $R_{\sigma,z}(P)$. Following the approach used by \citet{2024MNRAS.530.2972S} and \citet{2025MNRAS.543.2275F}, we represented the period dependence of each parameter using a cubic spline function.

We focused our study on the Miras with periods in the range $170<P<450~\mathrm{days}$. This range was defined by excluding approximately 5\% of the Mira sample at each end of the period distribution. The Mira sample was concentrated around $P\simeq 330~\mathrm{days}$, while the number of stars decreased substantially toward both shorter and longer periods. Therefore, the disc parameters inferred outside $170<P<450~\mathrm{days}$ were not considered reliable.

For the spline model, we placed six knots uniformly over the range $170<P<450~\mathrm{days}$. In addition, we placed two knots at shorter periods and three knots at longer periods, giving a total of 11 knots for each spline. These outer five knots were included only to stabilise the spline fit near the edges of the period range used in this study. 
This treatment was introduced to avoid applying a hard cut to the data at $170<P<450~\mathrm{days}$. Stars outside this period range were still included in the likelihood calculation. In particular, because the observational uncertainty in $P$ was included through Monte Carlo sampling, stars with observed periods outside $170<P<450~\mathrm{days}$ could still contribute to the likelihood within this range if some of their sampled periods fell inside it. Therefore, although we discussed the final spline curves only within $170<P<450~\mathrm{days}$, the information from stars outside of the boundaries was not completely discarded.

Since the period dependence of each of the five distribution-function parameters was represented by a cubic spline with 11 knots, the model contained $5\times 11=55$ free parameters. These 55 parameters were fitted simultaneously in the MCMC.
The following uniform prior distributions are assigned to all the knots,
\begin{equation}\label{eq:prior}
\begin{aligned}
&\ln R_{\mathrm{disc}}\mathrm{/kpc} \sim \mathcal{U}\left(\ln (0.01), \ln(10)\right), \\
&\ln \sigma_{R,0} \mathrm{/km~s^{-1}}\sim \mathcal{U}\left(\ln (0.01), \ln(100)\right), \\
&\ln \sigma_{z, 0} \mathrm{/km~s^{-1}}\sim \mathcal{U}\left(\ln (0.01), \ln(100)\right), \\
&\ln R_{\sigma, R} \mathrm{/kpc}\sim \mathcal{U}\left(\ln (0.01), \ln(60)\right),\\
&\ln R_{\sigma, z} \mathrm{/kpc}\sim \mathcal{U}\left(\ln (0.01), \ln(60)\right).
\end{aligned}
\end{equation}
To suppress oscillations between spline knots, we introduced a smoothing prior $P_{\mathrm{smooth}}$,
\begin{equation}\label{eq:smoothing}
    \ln P_{\mathrm{smooth}}=-\sum_{\theta} \sum_{n=0}^{N_{\mathrm{knots}}-2}\frac{(\ln \theta_{n+1}-\ln \theta_{n})^2}{2(\sigma_{\theta})^2},
\end{equation}
where $\sigma_{\theta}$ is the scaling parameter, set to 0.3 for all parameters, $\theta=(R_\mathrm{disc}$, $ \sigma_{R,0}$, $\sigma_{z,0}$, $R_{\sigma,R}$, $R_{\sigma,z})$, with $\theta_{n+1}$ and $\theta_{n}$ indicating the parameter values at the $(n+1)$-th and $n$-th knots, respectively. $N_\mathrm{knots}$ is the number of knots, which was set to 11 including knots outside of $170<P<450~\mathrm{days}$.
Employing the smoothing prior helps reduce the risk of overfitting. 

For the MCMC fitting, we used the code developed by \citet{2024MNRAS.530.2972S}
and modified it for this application. This code uses Hamiltonian Monte Carlo with the NUTS sampler \citep{2011arXiv1111.4246H}, implemented in \texttt{NumPyro} \citep{2019arXiv191211554P}, which utilises \texttt{JAX} \citep{jax2018github} to accelerate the computation.

\subsection{Period--age relation modelling}\label{sec:pa}

Once the Mira kinematics were fitted with the distribution function as a function of the period, $f_P$, we compared it with the age-dependent distribution function, $f_\tau$, obtained from the APOGEE red giants in \citet{2025MNRAS.543.2275F}, and derived the period--age relation for Miras by requiring $f_\tau$ and $f_P$ to be consistent. Following \citet{2023MNRAS.521.1462Z}, we modelled the age of Miras as a function of pulsation period using the following functional form;
\begin{equation}
\tau(P)=
\frac{\tau_0}{2}
\left[
1+\tanh\left(\frac{330-P}{P_s}\right)
\right].
\label{eq:period_age_model}
\end{equation}
Unlike \citet{2023MNRAS.521.1462Z}, who additionally fitted the intrinsic scatter of the period--age relation, we did not include an explicit intrinsic-scatter parameter in equation~\eqref{eq:period_age_model}.
In reality, Miras with the same pulsation period may span a range of ages, for example because of differences in metallicity or other stellar properties. In our modelling, however, all Miras at a given period were assumed to follow the same period-dependent distribution function and were therefore assumed to be the same age. \rev{An intrinsic age dispersion at a fixed pulsation period may introduce a systematic offset in the inferred Mira ages from kinematics. A population spanning a range of ages is not, in general, equivalent to a single-age population described by a quasi-isothermal distribution function with parameters evaluated at the mean age. However, the intrinsic age dispersion could depend on metallicity in a complex way, and quantitative assessment of the impact of the intrinsic scatter and/or systematic dependence to the stellar properties requires careful calibrations.  Therefore, we postpone such assessments to a future study, and in this paper, we present the results under the simple assumption of the period-age relation.}

We now describe how we constructed the likelihood for fitting the period--age relation. This fit used the vertical velocity dispersion to calibrate the period-age relation among the distribution-function parameters
because we used only the proper motions of Miras, and the proper motions provide more direct information on the vertical motion, especially for stars at low Galactic latitudes, where our disc Mira sample is concentrated.
We therefore compared only the vertical velocity dispersion as a function of period for Miras with that as a function of age for APOGEE red giants.
We obtained the period-dependent vertical velocity dispersion, $\ln\sigma_{z,0}^{\rm Mira}(P)$, from the Mira distribution-function fit, while we obtained the age-dependent vertical velocity dispersion for APOGEE red giants, $\ln\sigma_{z,0}^{\rm RG}(\tau)$, from \citet{2025MNRAS.543.2275F}. For a given set of period--age parameters, $\theta=(\tau_0,P_s)$ in equation~\eqref{eq:period_age_model}, each Mira period knot, $P_n$, was mapped to an age $\tau=\tau(P;\tau_0,P_s)$. The likelihood to measure the consistency between $\ln\sigma_{z,0}^{\rm Mira}(\tau(P;\tau_0,P_s))$ and $\ln\sigma_{z,0}^{\rm RG}(\tau(P))$ was evaluated at the six Mira period knots, $P_n$, within $170<P<450$ days, using the period-dependent distribution-function parameters obtained from the distribution-function fit described in Section~\ref{sec:DF}. We used a \rev{multivariate} Gaussian likelihood for the residual vector, $\Delta$,
\begin{equation}
\ln L(\theta)=-\frac{1}{2}
\left[\Delta^{\mathsf{T}} C_{\rm tot}^{-1} \Delta+\ln |C_{\rm tot}|\right],
\label{eq:period_age_likelihood}
\end{equation}
where the residual vector represented the difference between the Mira and red giant mean vertical velocity dispersions at the period knots as follows,
\begin{equation}
\Delta_n(\theta)=\mu^{\rm Mira}_{\ln \sigma_z}(P_n)-\mu^{\rm RG}_{\ln \sigma_z}(\tau(P_n))
\end{equation}
and the total covariance matrix was given by
\begin{equation}
C_{\rm tot}(\theta)=
C_{\rm Mira}(P_n,P_m)
+
C_{\rm RG}(\tau(P_n),\tau(P_m)).
\label{eq:period_age_cov_total}
\end{equation}
\rev{Before fitting the period--age relation}, we computed the posterior
mean,
$
\mu_{\ln\sigma_z}^{\rm Mira}(P_n)
=
\left\langle
\ln\sigma_{z,0}^{\rm Mira}(P_n)
\right\rangle,
$
and the covariance matrix,
$
C_{\rm Mira}(P_n,P_m)
=
{\rm Cov}
\left[
\ln\sigma_{z,0}^{\rm Mira}(P_n),
\ln\sigma_{z,0}^{\rm Mira}(P_m)
\right],
$
from the posterior samples of the Mira distribution-function fit. The covariance
matrix describes both the uncertainty in the Mira period--$\ln\sigma_{z,0}$
relation and the correlations between different period knots. \rev{We precomputed these quantities from the Mira posterior samples} and kept fixed during the \rev{subsequent} MCMC fit of the period--age
relation.

The APOGEE red giants age--$\ln\sigma_{z,0}$
reference relation, $\ln\sigma_{z,0}^{\rm RG}(\tau)$ was obtained from \citet{2025MNRAS.543.2275F}. From this relation, we precomputed the mean and covariance of $\ln\sigma_{z,0}^{\rm RG}$ on a fine age grid before running the MCMC. The mean was defined as
$
\mu_{\ln\sigma_z}^{\rm RG}(\tau)
=
\left\langle
\ln\sigma_{z,0}^{\rm RG}(\tau)
\right\rangle,
$
and the covariance as
$
C_{\rm RG}(\tau,\tau^\prime)
=
{\rm Cov}
\left[
\ln\sigma_{z,0}^{\rm RG}(\tau),
\ln\sigma_{z,0}^{\rm RG}(\tau^\prime)
\right].
$
This covariance describes both the uncertainty in the red giant
age--$\ln\sigma_{z,0}$ relation and the correlations between different ages.
The precomputed mean and covariance were stored as lookup tables.
For each trial period--age relation during the MCMC, these lookup tables were used to evaluate $\mu_{\ln\sigma_z}^{\rm RG}(\tau(P_n))$ and
$C_{\rm RG}(\tau(P_n),\tau(P_m))$ by interpolation at the ages corresponding to the Mira period knots.

We sampled the posterior distribution of the two parameters $\tau_0$ and $P_s$ with MCMC using \texttt{emcee} \citep{2013PASP..125..306F}. We adopted the following uniform priors:
\begin{equation}\label{eq:prior_p-a}
\begin{aligned}
&\tau_0\mathrm{/Gyr} \sim \mathcal{U}\left(10, 30\right), \\
&\ln P_s \mathrm{/days}\sim \mathcal{U}\left(\ln (100), \ln(1600)\right).
\end{aligned}
\end{equation}

\section{Results}\label{sec:result}
\subsection{Mira variables' kinematics as a function of period}\label{sec:result:mira_DF}

\begin{figure*}	\includegraphics[width=\textwidth]{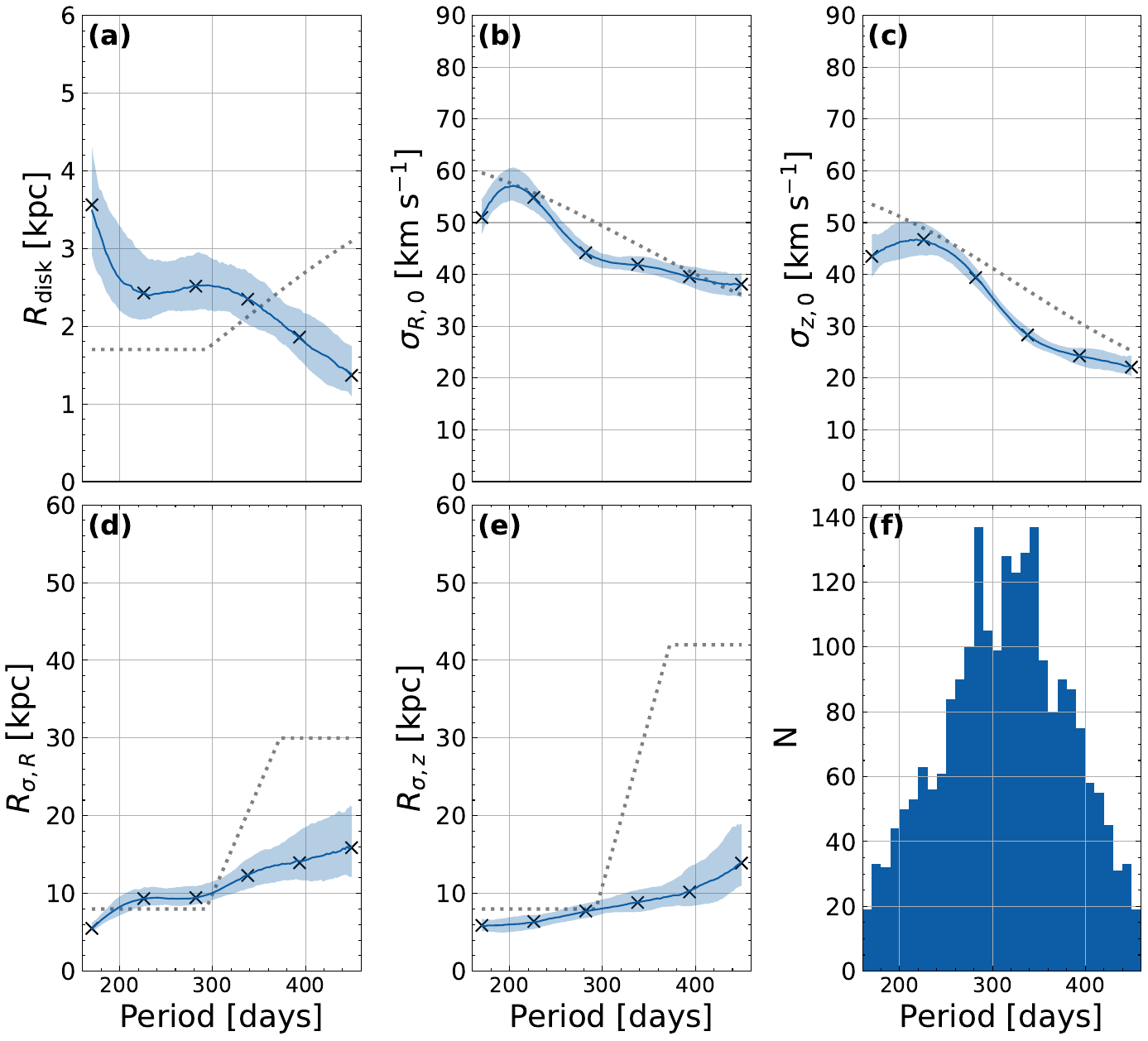}
    \caption{Posterior distributions of spline models describing the period dependence of the quasi-isothermal distribution-function parameters for Mira variables. Panels (a), (b), (c), (d) and (e) show the distribution-function parameters $R_\mathrm{disc}, \sigma_{R,0}, \sigma_{z,0}, R_{\sigma,R}$ and $R_{\sigma,z}$, respectively, while panel~(f) shows the stellar period distribution. The blue solid lines indicate the median of the fitting results, while the shaded blue regions represent the 1$\sigma$ confidence intervals. The grey dotted lines show the parameters used as proposal distribution function, $f^{\prime}_P$. Crosses denote the position of knots for the fitted spline function.}
    \label{fig:result:spline}
\end{figure*}
Fig.~\ref{fig:result:spline} shows the results of the distribution-function fitting for the Mira sample. Panels~(a)-(e) show the five distribution-function parameters as functions of pulsation period. 
The period dependence of each parameter is described by a cubic spline with knots denoted by crosses. The blue solid curves show the median of the posterior spline samples, and the blue shaded regions indicate the corresponding $16$th-$84$th percentile ranges.
The grey dotted curves show the proposal distribution functions, $f^\prime_P$, used for importance sampling. As described in Appendix~\ref{sec:IS}, the functional form of $f^\prime_\tau$ was fixed to match the age-dependent APOGEE red giant result \citep{2025MNRAS.543.2275F}. We converted this age-dependent
proposal function into a period-dependent one, $f^\prime_P$, by iteratively updating the period--age relation used in the transformation. In the final iteration, we adopted $\tau_0=17.0$~Gyr and $P_s=210$~days, which gave a proposal function close to the best-fit period--age relation presented in Section~\ref{sec:result:PAR}.
Panel~(f) shows the period distribution of the Mira sample used in the fit. 

Since the line-of-sight velocities are not used, we consider the most reliable kinematic tracer for the present analysis is the vertical velocity dispersion at the solar radius, $\sigma_{z,0}$. Panel~(c) shows that $\sigma_{z,0}$ decreased from approximately $45~\mathrm{km~s^{-1}}$ at short periods ($P\sim260$~days) to approximately $20~\mathrm{km~s^{-1}}$ at long periods ($P\sim560$~days). This indicates that shorter-period Miras are kinematically hotter than longer-period Miras in the vertical direction, consistent with the expectation that shorter-period Miras trace older stellar populations.

Panel~(b) shows that the radial velocity dispersion parameter, $\sigma_{R,0}$, also decreases toward longer periods. This trend is broadly consistent with that found for $\sigma_{z,0}$ \rev{in panel~(c)}, with shorter-period (older) Miras being kinematically hotter than longer-period (younger) Miras. 

Panel~(a) shows the disc scale length, $R_{\textrm{disc}}$. The inferred trend is different from the age-dependent behaviour found for APOGEE red giants by \citet{2025MNRAS.543.2275F}. In the APOGEE red giant sample, older stars have smaller disc scale lengths, while younger stars have larger ones. In contrast, the short-period Miras, which are associated with older ages, favour larger disc scale lengths, whereas the long-period Miras' ones decrease with period. \rev{The signal constraining $R_{\textrm{disc}}$ is subtle and relies on the angular-momentum distribution, which in the absence of line-of-sight velocities is only partially constrained by proper motions, and which may also be affected by non-axisymmetric features such as resonances that our axisymmetric model does not capture.} 
In Appendix~\ref{sec:validation}, we evaluate the performance of the distribution function fitting with mock data, which shows the challenge of recovering $R_\mathrm{disc}$ with the Mira data used in this study.
\rev{We therefore do not draw physical conclusions from the period dependence of $R_{\textrm{disc}}$.}

Panels~(d) and~(e) show the radial scale lengths of the radial and vertical velocity dispersions, $R_{\sigma,R}$ and $R_{\sigma,z}$, respectively. Both parameters increase toward longer periods, but only weakly, from approximately 6~kpc  to 15~kpc. This gradual increase differs from the behaviour found for APOGEE red giants by \citet{2025MNRAS.543.2275F}, where the scale lengths rise sharply for younger stars, reaching values of $\sim30$~kpc for $R_{\sigma,R}$ and above $\sim40$~kpc for $R_{\sigma,z}$. The Mira sample does not show such a rapid increase. From the mock data verification, Appendix~\ref{sec:validation} shows that the current Mira data provide poor constraints on these parameters.

Finally, panel~(f) shows the period distribution of the Mira sample. The sample is concentrated around $P\simeq 330~\mathrm{days}$ and becomes sparse toward both ends of the period range. Consequently, the fitted spline curves are most strongly constrained near $P\simeq 330~\mathrm{days}$ and less well constrained near the edges. This is reflected in the fitted spline curves particularly in $\sigma_{z,0}$: the posterior uncertainties are smallest near the centre of the period distribution and become broader toward the edges, where fewer stars are available.

\begin{table*}
\centering
\caption{
Our best-fit values (medians of the posterior distributions) for the distribution-function parameters at the knots shown in Fig.~\ref{fig:result:spline} together with their uncertainties, which correspond to the 16th and 84th percentiles of the posterior distributions.
}\label{tab:mira_df_parameters}
\begin{tabular}{c|ccccc}
\hline
$P$~[days] & $R_{\mathrm{disc}}$~[kpc] & $\sigma_{R,0}$~[km~$\mathrm{s^{-1}}$] & $\sigma_{z,0}$~[km~$\mathrm{s^{-1}}$] & $R_{\sigma,R}$~[kpc] & $R_{\sigma,z}$~[kpc] \\ \hline
170 & $3.56^{+0.79}_{-0.61}$ & $50.99^{+3.78}_{-3.17}$ & $43.49^{+3.75}_{-3.73}$ & $5.51^{+0.69}_{-0.51}$ & $5.89^{+0.86}_{-0.71}$ \\
226 & $2.43^{+0.48}_{-0.32}$ & $54.85^{+2.78}_{-2.72}$ & $46.73^{+3.02}_{-3.19}$ & $9.34^{+1.40}_{-0.99}$ & $6.39^{+0.90}_{-0.85}$ \\
282 & $2.52^{+0.35}_{-0.30}$ & $44.16^{+1.61}_{-1.84}$ & $39.38^{+1.92}_{-1.79}$ & $9.46^{+1.43}_{-0.96}$ & $7.74^{+1.08}_{-0.88}$ \\
338 & $2.35^{+0.38}_{-0.30}$ & $41.92^{+1.80}_{-1.56}$ & $28.31^{+1.47}_{-1.39}$ & $12.31^{+2.02}_{-1.75}$ & $8.83^{+1.62}_{-1.18}$ \\
394 & $1.86^{+0.37}_{-0.28}$ & $39.57^{+1.94}_{-1.85}$ & $24.27^{+1.56}_{-1.57}$ & $13.95^{+3.84}_{-2.37}$ & $10.19^{+2.53}_{-1.86}$ \\
450 & $1.37^{+0.36}_{-0.27}$ & $38.10^{+2.57}_{-2.50}$ & $22.10^{+2.12}_{-1.83}$ & $15.89^{+5.56}_{-3.90}$ & $13.90^{+5.31}_{-3.20}$ \\
\hline
\end{tabular}

\end{table*}

The numerical values of the fitted spline parameters are listed in Table~\ref{tab:mira_df_parameters}. For each distribution-function parameter and each spline knot, we report the posterior median and the uncertainty estimated from the $16$th and $84$th percentiles. In Appendix \ref{sec:ap:best}, we compare the best-fit distribution function model with the observed velocity distributions, which show good agreement between the model and the observational data.

\subsection{Period--age relation}\label{sec:result:PAR}

In the previous section, we showed that the vertical velocity dispersion of Miras at the solar radius, $\sigma_{z,0}$, decreases systematically with increasing period and provides a reliable kinematic tracer for our period--age calibration. The next step is to find the mapping between period and age that makes the Miras' period-$\sigma_{z,0}$ relation consistent with the APOGEE red giants' age-$\sigma_{z,0}$ relation obtained by \citet{2025MNRAS.543.2275F}, as described in Section~\ref{sec:pa}.

\begin{figure}	\includegraphics[width=0.5\textwidth]{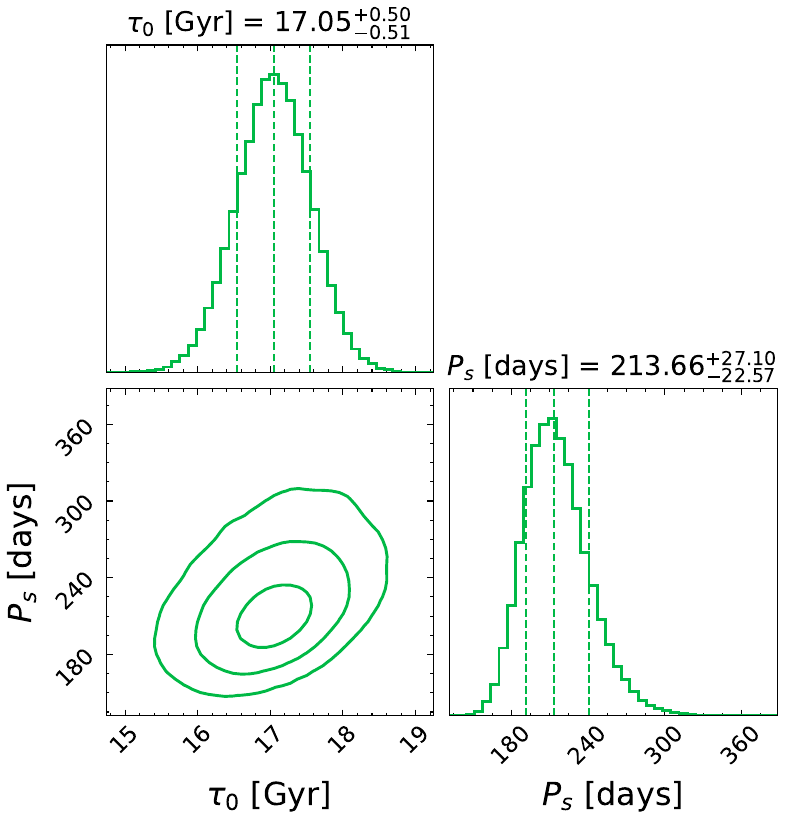}
    \caption{Corner plot of the posterior distributions of the period--age relation parameters, $\tau_0$ and $P_s$. The diagonal panels show the one-dimensional posterior distributions with dashed vertical lines showing the 16th, 50th and 84th percentiles, while the bottom left panel shows the joint posterior probability distributions of the two parameters. Contours enclose 39.3\%, 86.5\% and 98.9\% of the posterior probability, corresponding approximately to 1$\sigma$, 2$\sigma$ and 3$\sigma$ regions for a two-dimensional Gaussian distribution. The labels above the diagonal panels indicate the median values with the 16th and 84th percentiles.}
    \label{fig:result:corner}
\end{figure}

Fig.~\ref{fig:result:corner} shows the posterior distribution of the parameters of the period--age relation of equation~\eqref{eq:period_age_model}. The parameters $\tau_0$ and $P_s$ control the normalisation and transition width of $\tanh$ function of the period--age relation, respectively. The posterior is well constrained, with median values and uncertainties corresponding to the 16th and 84th percentiles: 
\begin{equation}
\tau_0 = 17.05^{+0.50}_{-0.51}~\mathrm{Gyr},
\qquad
P_s = 213.63^{+26.84}_{-22.44}~\mathrm{days}.
\end{equation}

\begin{figure*}	\includegraphics[width=\textwidth]{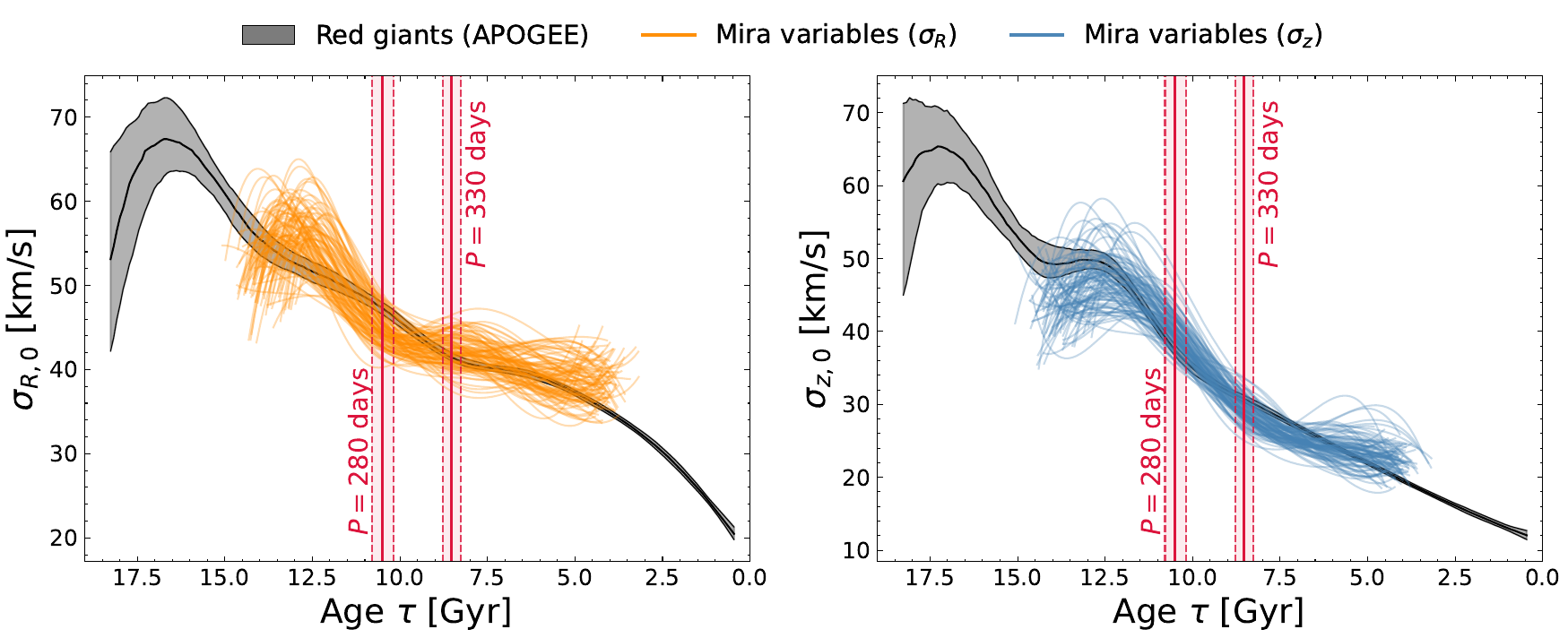}
    \caption{Comparison between the red giants' and \rev{Miras'} age--velocity-dispersion relations after applying the inferred period--age relation to the Miras' period--velocity-dispersion relations. The left panel shows $\sigma_{R,0}$, and the right panel shows $\sigma_{z,0}$ as a function of the age. The black curves show the median APOGEE red giant relations from \citet{2025MNRAS.543.2275F}, and the grey shaded regions indicate their $16$th-$84$th percentile ranges. The coloured curves show 100 posterior realisations of the Mira velocity-dispersion relations as a function of age. Each curve combines a posterior spline draw from the Mira distribution-function fit with a posterior draw of the period--age parameters. The spread of the curves therefore reflects both the uncertainty in the Miras' period-dependent kinematics and the uncertainty in the period--age relation. The vertical red lines mark the ages corresponding to $P=280~\mathrm{days}$ and $P=330~\mathrm{days}$ under the inferred period--age relation with 1~$\sigma$ uncertainty range shown with red shaded area and dashed lines. These periods correspond, respectively, to the onset of the rise in the NSD star-formation rate and to its peak in the star-formation history inferred by \citet{2024MNRAS.530.2972S}, and are  used here as markers of the bar formation epoch.}
    \label{fig:result:s_vs_age}
\end{figure*}

Fig.~\ref{fig:result:s_vs_age} shows \rev{Miras'} age-velocity-dispersion relations after applying the inferred period--age relation to the \rev{Miras'} period--velocity-dispersion relations. The black curves and shaded regions show the APOGEE red giant relations and their uncertainties. The coloured curves show posterior realisations of the Miras' relations mapped from period to age. Each realisation is constructed by combining one posterior spline draw from the Miras' distribution-function fit with one posterior draw of the period--age parameters, $\tau_0$ and $P_s$. Thus, the spread of the coloured curves reflects both the uncertainty in the inferred Miras' period-dependent velocity-dispersion relation and the uncertainty in the period--age relation. 

The right panel shows that the \rev{Miras'} age--$\sigma_{z,0}$ relation agrees well with the APOGEE red giant's age--$\sigma_{z,0}$ relation after applying the best-fit period--age relation. This agreement is expected because $\sigma_{z,0}$ is the quantity used to calibrate the period--age relation. The left panel shows the same comparison for $\sigma_{R,0}$. Although $\sigma_{R,0}$ is not used for the fitting of the period--age relation, the Mira age--radial velocity-dispersion relation is broadly consistent with the APOGEE red giant trend over the age range where the Mira data provide useful constraints.

\begin{figure}	\includegraphics[width=0.48\textwidth]{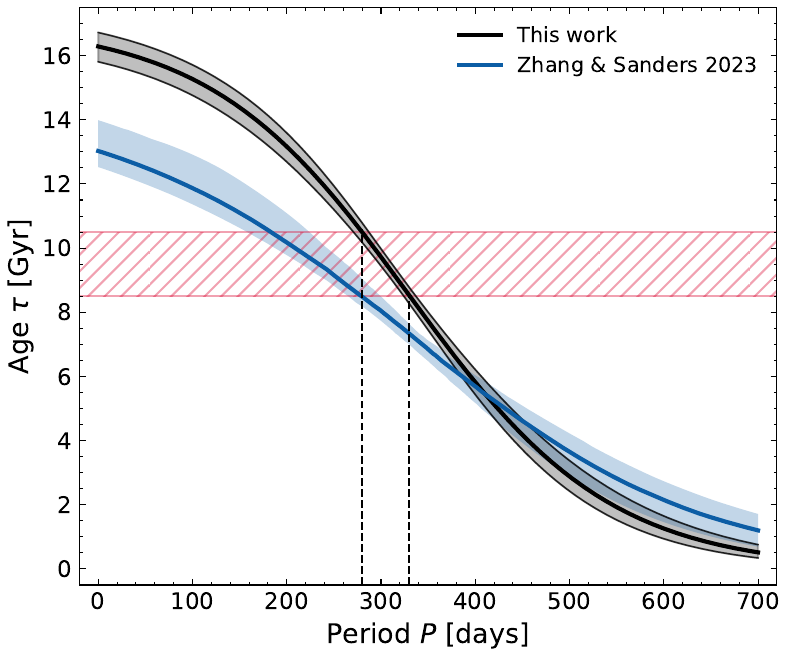}
    \caption{Inferred Miras' period--age relation. The black curve shows the posterior median relation, and the grey shaded region indicates the $16$th-$84$th percentile range. The vertical dashed lines mark the reference periods of bar formation, $P=280$ and $330~\mathrm{days}$, and the red hatched region shows their corresponding age range $8.5<\tau<10.5$~Gyr under our inferred period--age relation. The blue curve and shaded region show the period-age relation reported by \citet{2023MNRAS.521.1462Z}, together with its $1\sigma$ uncertainty. This relation was calibrated using isochrone ages in \citet{2018MNRAS.475.1093Y} and is therefore on an age scale different from the APOGEE/BINGO age scale targeted in this work.}
    \label{fig:result:p-a}
\end{figure}

Fig.~\ref{fig:result:p-a} shows our best-fit period--age relation whose parameters' posterior probability distribution is shown in Fig.~\ref{fig:result:corner}.
 The black curve shows the posterior median relation, and the grey shaded region indicates the 16th-84th percentile range. As imposed by the adopted functional form for the period--age relation in equation~\eqref{eq:period_age_model}, the relation maps shorter periods to older ages and longer periods to younger ages.
Fig.~\ref{fig:result:p-a} also shows the period-age relation suggested by \citet[][with $\tau_0=14.7\pm0.7$~Gyr and $P_s=308\pm54$~days]{2023MNRAS.521.1462Z}, \rev{which was calibrated using isochrone ages \citep{2018MNRAS.475.1093Y}. Our relation assigns systematically older ages to Mira variables with $P \lesssim 400$ days, by up to $\sim2$~Gyr around $P \approx 300$ days. This difference reflects the different age scales used in the two calibrations, and} demonstrates the importance of placing age estimates on a consistent age scale when comparing the timing of evolutionary events.

\section{Discussion}\label{sec:discussion}

\subsection{Bar formation epoch on the APOGEE/BINGO age scale}\label{sec:discussion:bar-formation}

In Section~\ref{sec:result}, we calibrated the Mira period--age relation by matching the period dependence of $\sigma_{z,0}$ for Miras to the age dependence of $\sigma_{z,0}$ for APOGEE red giants. This calibration placed Mira periods on the APOGEE/BINGO age scale used 
by \citet{2024MNRAS.528L.122C} and \citet{2025MNRAS.543.2275F}.
\rev{\citet{2024MNRAS.530.2972S} found that the star formation rate of the NSD, traced by Mira variables, began to rise at $P \approx 280$ days and peaked at $P \approx 330$ days. Because the NSD is expected to form from gas driven inwards by the bar \citep[e.g.][]{2020MNRAS.492.4500B}, we adopt this period range as a proxy for the bar formation epoch. Our calibrated period-–age relation converts this range into $8.5 < \tau < 10.5$ Gyr on the APOGEE/BINGO age scale, as indicated by the vertical red lines in Fig.~\ref{fig:result:s_vs_age} and the hatched region in Fig.~\ref{fig:result:p-a}. With the \citet{2023MNRAS.521.1462Z} relation, the same periods correspond to $7.5-8.5~\mathrm{Gyr}$ ago, consistent with the $\sim8$~Gyr reported by \citet{2024MNRAS.530.2972S}. We do not interpret this range as defining the precise onset or completion of bar formation, but rather as tracing a broad period during which the Galactic bar may have developed. Below, we compare this epoch with the chemical, structural and kinematic evolution of the disc inferred from APOGEE red giants.}

\begin{figure*}	\includegraphics[width=\textwidth]{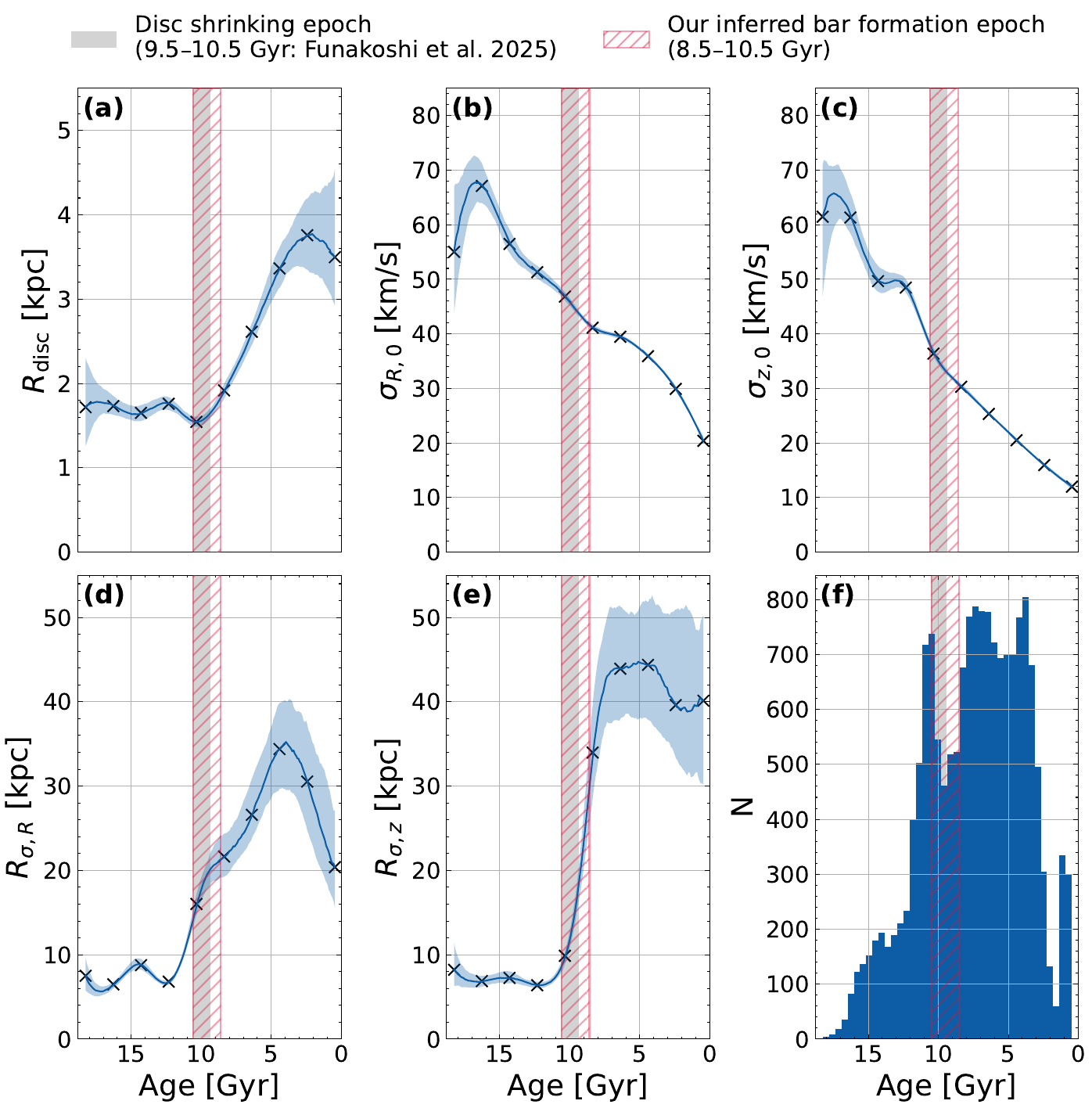}
    \caption{A comparison of the bar formation epoch inferred in this work and the age-dependent spline posterior distributions of the best-fit disc distribution function model parameters of the APOGEE red giants obtained in \citet{2025MNRAS.543.2275F}. Panels (a), (b), (c), (d) and (e) show the age-dependent distribution-function parameters $R_\mathrm{disc}, \sigma_{R,0}, \sigma_{z,0}, R_{\sigma,R}$ and $R_{\sigma,z}$, respectively. The blue solid lines indicate the median of the fitting results, while the shaded blue regions represent the 1$\sigma$ confidence intervals. Crosses denote the position of knots for the fitted spline function. 
    Panel~(f) shows the stellar age distribution of the APOGEE red giants. In all panels, the gray vertical shaded regions show ages around the dip in $R_\mathrm{disc}$ found by \citet{2025MNRAS.543.2275F}.
    The red vertical hatched regions show the bar formation epoch corresponding to Mira periods of $P=280-330$~days, as inferred from the derived period--age relation obtained in Section~\ref{sec:discussion:bar-formation}.}
    \label{fig:dis:spline_f25}
\end{figure*}

\subsection{Connection to the GSE merger and the disc evolution}\label{sec:discussion:GSE-disc}

Fig.~\ref{fig:dis:spline_f25} compares the bar formation epoch inferred in this work (Fig.~\ref{fig:result:p-a}) with the age-dependent disc structure obtained by \citet{2025MNRAS.543.2275F} using the APOGEE red giant samples described in Section~\ref{sec:data:apogee}. The figure shows the age-dependent distribution-function parameters of the APOGEE red giants, inferred using the same quasi-isothermal disc distribution function and Galactic potential as in Section~\ref{sec:DF}. In contrast to the Mira analysis, where the distribution-function parameters were modelled as functions of pulsation period, \citet{2025MNRAS.543.2275F} modelled the same parameters as functions of stellar age inferred with BINGO \citep{2021MNRAS.503.2814C,2024MNRAS.528L.122C}. Panels (a)-(e) show the five fitted distribution-function parameters, whose age dependences are represented by cubic splines with evenly-spaced ten knots. The solid blue curves show the posterior medians of the sampled spline curves, and the blue shaded regions indicate the corresponding $16$th-$84$th percentile ranges. Panel (f) shows the age distribution of the APOGEE red giant sample used in the fit. Further details of the sample selection (see also Sec.~\ref{sec:data:apogee}), distribution function model and fitting procedure are given in \citet{2025MNRAS.543.2275F}.

The red hatched region shows \rev{our inferred bar formation epoch.}
The grey shaded region marks the age around the dip in $R_{\textrm{disc}}$ identified by \citet{2025MNRAS.543.2275F}. $R_{\textrm{disc}}$ describes the present-day radial scale length of stars with a given age, and the dip in $R_{\textrm{disc}}$ indicates that the stars born at that epoch are more centrally concentrated than both older and younger populations. \citet{2025MNRAS.543.2275F} interpreted this feature as the fossil signature of a temporary shrinking of the star-forming gas disc at this epoch. In this interpretation, the star-forming gas disc became compact at that time and formed stars with a compact radial birth distribution. Although subsequent orbital evolution may have modified their present-day radii, so-called radial migration, the imprint of this compact birth distribution is still reflected in the dip in $R_{\textrm{disc}}$ for stars of that age. 
Moreover in the analysis by \citet{2025MNRAS.543.2275F}, stars older than the dip were found to form a compact and kinematically hot population, consistent with the known thick disc structure, whereas stars younger than the dip were found to be kinematically colder and to have a scale length that increases towards younger ages, as expected for inside-out formation of the thin disc. The dip therefore lies at the transition from thick disc population formation, through a phase of gas disc shrinking, to subsequent inside-out thin disc formation. 
Our bar formation epoch overlaps with this broad transition. It includes the epoch of the $R_{\textrm{disc}}$ dip within its older part and extends towards younger ages, where the disc scale length begins to increase. This temporal overlap suggests that \rev{the development of the Galactic bar may have occurred during the transition from the thick-disc phase to the thin-disc phase, although the available age resolution does not allow us to determine the precise timing of these processes relative to one another.}

\begin{figure}	\includegraphics[width=\columnwidth]{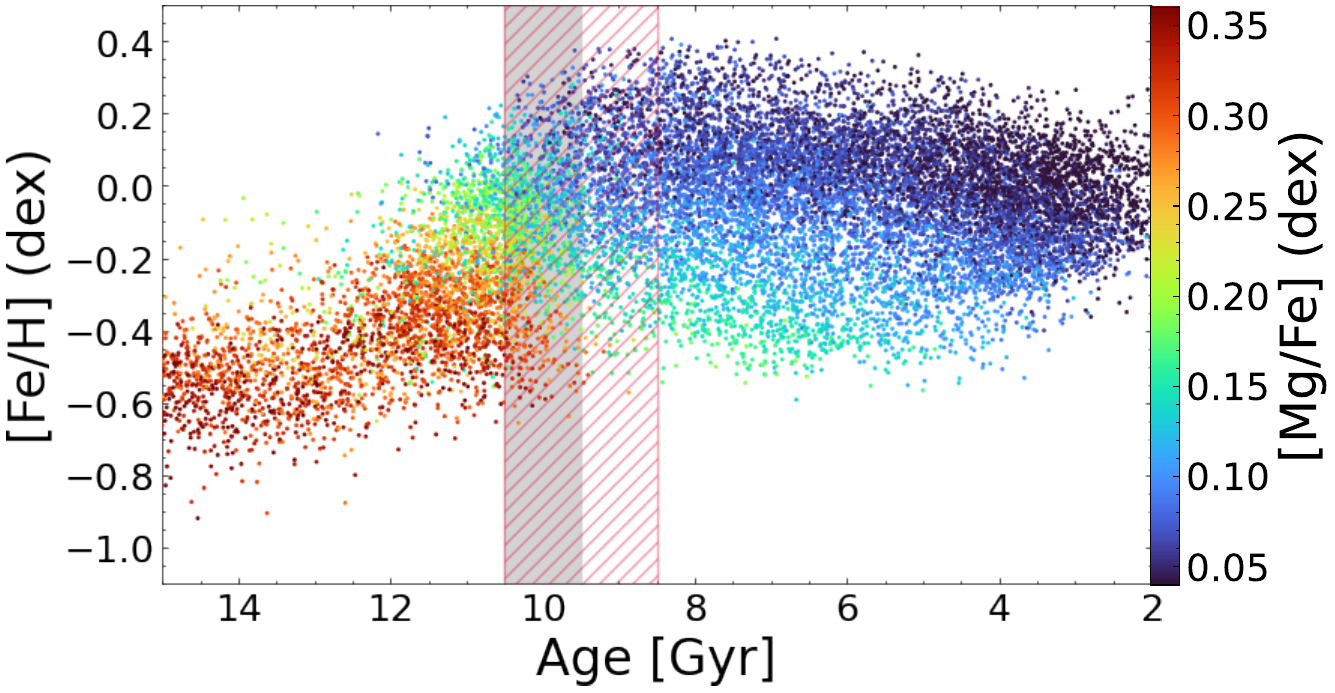}
    \caption{The age--metallicity relation of the APOGEE red giants described in Section~\ref{sec:data:apogee}, which were used in \citet{2025MNRAS.543.2275F}, coloured by [Mg/Fe]. The region shaded in grey shows the same age range shown in Fig.~\ref{fig:result:spline}, around the dip in $R_\mathrm{disc}$.
    The red vertical hatched regions indicate the bar formation epoch inferred from the Mira period range of $P=280-330$~days using the period-age relation derived in Section~\ref{sec:discussion:bar-formation}.}
    \label{fig:dis:age_metallicity}
\end{figure}

Fig.~\ref{fig:dis:age_metallicity} compares our bar formation epoch with the APOGEE age-metallicity relation. Each dot in the figure shows the APOGEE red giant described in Section~\ref{sec:data:apogee} in the age-$\mathrm{[Fe/H]}$ plane, colour-coded by $\mathrm{[Mg/Fe]}$. Stars older than $\sim 12~\mathrm{Gyr}$ are predominantly metal-poor ($\mathrm{[Fe/H]}\sim-0.5$) and $\mathrm{[Mg/Fe]}$-enhanced. In contrast, stars younger than $\sim 10~\mathrm{Gyr}$ are generally more metal-rich and have lower $\mathrm{[Mg/Fe]}$.
\citet{2024MNRAS.528L.122C} identified the epoch, approximately $10<\tau<12~\mathrm{Gyr}$ on their age scale, as the GGS phase. During this phase, the metallicity increases rapidly towards younger ages, while $\mathrm{[Mg/Fe]}$ decreases. By comparing the APOGEE age-metallicity relation with \texttt{Auriga} cosmological simulations, \citet{2024MNRAS.528L.122C} argued that this chemical transition was driven by the gas-rich GSE merger. In this interpretation, gas brought in by the merger triggered an intense starburst in the Galactic disc, followed by rapid chemical enrichment and a decline in $\mathrm{[Mg/Fe]}$. \rev{The slightly enhanced $\mathrm{[Mg/Fe]}$ values at $\tau\sim12~\mathrm{Gyr}$ (shown by the darker red distribution in Fig.~\ref{fig:dis:age_metallicity}) are consistent with an increased contribution from core-collapse supernovae during the starburst. The subsequent decrease in $\mathrm{[Mg/Fe]}$, accompanied by an increase in $\mathrm{[Fe/H]}$, may reflect the delayed contribution of Type I$\mathrm{\alpha}$ supernovae, which progressively enrich the interstellar medium with iron.}
This signature is particularly clear in the inner Galactic disc, suggesting that the merger had a strong impact on the central and inner regions of the Milky Way \citep{2024MNRAS.528L.122C}. 
The red hatched region shows our bar formation epoch, which lies near the final stage of the GGS phase and overlaps with the epoch of gas disc shrinking identified from the dip in $R_{\textrm{disc}}$ shown with the grey shaded region as also shown in Fig.~\ref{fig:dis:spline_f25}. This comparison suggests that the bar formation occurred during the transition from the early, metal-poor, $\mathrm{[Mg/Fe]}$-enhanced thick disc population to the more metal-rich, lower-$\mathrm{[Mg/Fe]}$ thin disc population. Interestingly, our inferred bar formation epoch coincides with the later phase of the rapid chemical enrichment of GGS, which has been linked to the GSE merger \citep{2024MNRAS.528L.122C}. This may suggest that the bar formation is triggered by the GSE merger.

\citet{2024MNRAS.531.1520M} analysed the \texttt{Auriga} cosmological simulation Au~18, which contains both a GSE-like merger history and a present-day barred Milky-Way-like galaxy. In their simulation, \rev{the first pericentric passage of the GSE-like progenitor marks the onset of strong tidal perturbations}
on the main galaxy.
The perturbation of the disc seeded the formation of a bar and induced angular-momentum loss in the gas disc, driving gas from the outer disc towards the central region. This gas inflow, together with the gas supply from the infalling satellite, triggered a starburst, rapidly increasing the star-formation rate, particularly within the inner $4~\mathrm{kpc}$. This enhanced star formation lasted for approximately 1 Gyr. The combination of the gas inflow and gas supply caused the baryonic component to become more dominant in the central region. This made the inner disc more susceptible to bar formation and created conditions in which a stellar bar could grow and remain stable. As a result, the stable bar was established shortly after this phase and formed approximately $1~\mathrm{Gyr}$ after the first pericentric passage of the GSE-like progenitor.

This sequence in Au~18 provides a useful comparison with the observational picture discussed above. On the APOGEE/BINGO age scale, \citet{2024MNRAS.528L.122C} placed the GGS phase at approximately $10<\tau<12~\mathrm{Gyr}$. \citet{2025MNRAS.543.2275F} identified the gas disc shrinking signature near the final stage of this phase, at $\tau\simeq 10~\mathrm{Gyr}$. In this work, the bar formation epoch spans a similar epoch between $\tau\approx10.5$~Gyr and $\tau\approx8.5$~Gyr.
A qualitatively similar sequence appears in the simulation of \citet{2024MNRAS.531.1520M}. In Au~18, the GSE-like event is associated with a starburst that began at a look-back time of $t_\textrm{lookback}\simeq 9~\mathrm{Gyr}$. Shortly afterwards, the first pericentric passage occurred at $t_\textrm{lookback}\simeq 8.9~\mathrm{Gyr}$, when the tidal force was strongest, driving gas inward motion from the outer disc and seeding bar formation. The merger was completed at $t_\textrm{lookback}\simeq 8.5~\mathrm{Gyr}$ and a stable bar formed by $t_\textrm{lookback}\simeq 8-7.5~\mathrm{Gyr}$. 
Taken together, our observational results and the scenarios shown by \citet{2024MNRAS.531.1520M}, \citet{2024MNRAS.528L.122C} and \citet{2025MNRAS.543.2275F} suggest that the gas-rich GSE merger triggered the starburst (GGS), star-forming gas disc shrinking and bar formation. 
In this picture, the Galactic bar formed as a natural consequence of the GSE-driven perturbation of the Galactic disc, with the redistribution of disc gas (possibly associated with gas-disc contraction).

\rev{Several processes may have contributed to the shrinking of the star-forming gas disc. The tidal perturbation from the GSE-like progenitor may have driven gas inflow by redistributing angular momentum \citep{2024MNRAS.531.1520M}, while gas consumption by the associated starburst may have reduced the gas disc and made star-forming gas disc smaller \citep{2025MNRAS.543.2275F}. A temporary reduction in gas supply due to the switch of the cold-to-hot gas accretion mode \citep{2005MNRAS.363....2K,2018Natur.559..585N,2026OJAp....963764N,2026arXiv261008931R} may also have contributed to both disc shrinking and the transition from thick-disc to thin-disc formation \citep{2020MNRAS.497.1603G,2026arXiv260522806M}}

\rev{A gas-rich merger may increase the gas fraction of a galaxy, potentially promoting the initial formation of a bar while also hindering its long-term survival \citep{2000MNRAS.312..194N,2024ApJ...968...86B,2026arXiv260903275B}. This picture may be consistent with the evolution seen in the Au~18 simulation, where the tidal force from GSE-like gas-rich merger seeds a bar, but a stable bar develops only after a further $\sim1~\mathrm{Gyr}$ \citep{2024MNRAS.531.1520M}. The merger-driven gas inflow and the associated high gas content may thus have raised the gas fraction and promoted the initial formation (a seed) of an unstable bar, while subsequent gas depletion marked by the shrinking of the star-forming disc may have favoured the development of a more stable bar. Although this interpretation is consistent with the simulation, our observations do not directly constrain the gas fraction at the time.}

\rev{
We stress that the temporal overlap between the GGS, the shrinking of the star-forming gas disc, and the inferred bar formation epoch does not establish a causal connection and that the scenario proposed above represents one plausible interpretation from observational constraints and theoretical model predictions, rather than a unique evolutionary pathway.
Given the limited time resolution of the age-dependent disc model and the systematic uncertainties in the age scale, we cannot determine whether the GGS, the dip in $R_{\textrm{disc}}$ near $\tau\sim10~\mathrm{Gyr}$, and the inferred bar formation epoch at $\tau\sim8.5$–$10.5~\mathrm{Gyr}$ correspond to distinct stages of the GSE merger or establish their precise temporal ordering. We therefore regard their temporal overlap as evidence that these events may be associated with the same broad evolutionary phase.
}

\label{sec:discussion:chemical_plane}
\begin{figure*}	\includegraphics[width=2\columnwidth]{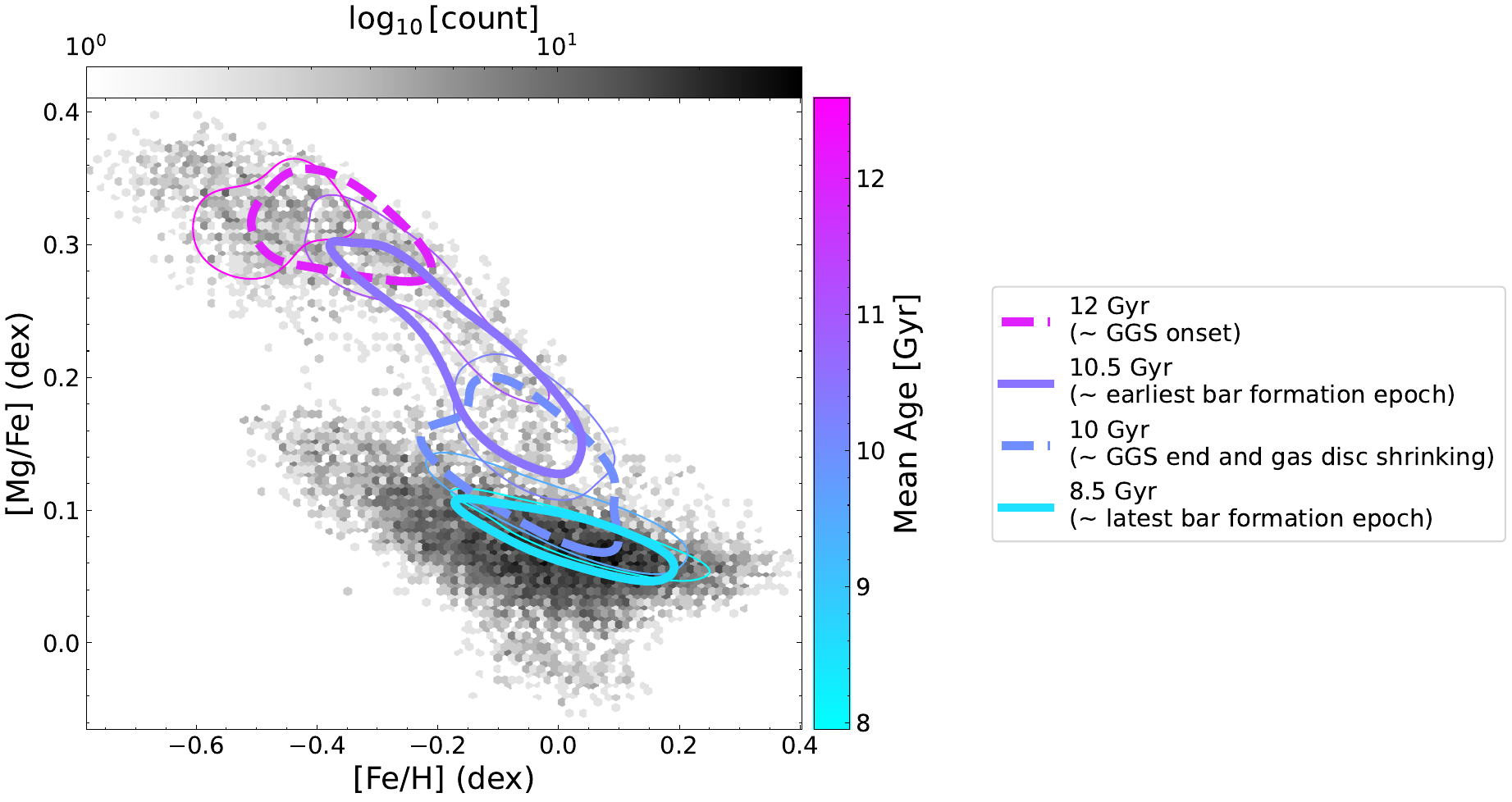}
    \caption{Distribution of the APOGEE red giants in the $\mathrm{[Fe/H]}$-$\mathrm{[Mg/Fe]}$ plane, using the sample described in Section~\ref{sec:data:apogee}. The background hexbin map shows the stellar number density, with the greyscale indicating the base-10 logarithm of the number of stars in each bin. The contours show the approximate $1\sigma$ regions of stars in different age bins, with the mean age indicated by the vertical colour bar. The dashed thick contours highlight the onset of GGS ($\tau\sim12$~Gyr) and the end of GGS which is at a similar epoch to the gas disc shrinking ($\tau\sim10$~Gyr).
    The solid thick contours show the earliest ($\tau\sim10.5$~Gyr) and the latest ($\tau\sim8.5$~Gyr) ranges of the inferred bar formation epoch in this study. The solid thin contours represent other age bins to illustrate the age-dependent shift of the stellar distribution across the $\mathrm{[Fe/H]}-\mathrm{[Mg/Fe]}$ plane.
}
    \label{fig:dis:FeH_MgFe}
\end{figure*}

Fig.~\ref{fig:dis:FeH_MgFe} provides a complementary chemical view of the transition epoch discussed above, showing where stellar populations of different ages lie in the $\mathrm{[Fe/H]}$-$\mathrm{[Mg/Fe]}$ plane. The underlying distribution is constructed from the APOGEE red giant sample described in Section~\ref{sec:data:apogee}, and is shown as a hexbin map with the greyscale indicating the base-10 logarithm of the star count in each bin. The overlaid solid thick, dashed thick and solid thin contours show distribution of stars in selected age bins. Each contour encloses approximately 39.3\% of stars around the density peak in the corresponding age bin, and its colour indicates the mean age of the population. The dashed thick contours trace the age bins associated with the GGS phase, while the solid thick contours trace those associated with the bar formation epoch. The solid thin contours show the other age bins and are included to illustrate how the stellar distribution shifts across the [Fe/H]-[Mg/Fe] plane with age.
This allows the transition epoch associated with the GGS, gas disc shrinking and bar formation to be placed in the chemical abundance plane. Here, it is important to note that the contours do not represent the chemical distribution of stars born in nor currently residing in the bar.

The onset of the GGS phase ($\tau\sim12$~Gyr) is located on the high-$\mathrm{[Mg/Fe]}$ sequence,
at relatively low metallicity ($\mathrm{[Fe/H]}\sim-0.4$~dex) and high $\mathrm{[Mg/Fe]}$ ($\mathrm{[Mg/Fe]}\sim0.3$~dex). The subsequent age contours move rapidly from this region towards higher $\mathrm{[Fe/H]}$ and lower $\mathrm{[Mg/Fe]}$, showing that the GGS phase corresponds to a rapid chemical transition from the high-$\mathrm{[Mg/Fe]}$ population towards the low-$\mathrm{[Mg/Fe]}$ disc \citep[see also][]{2021MNRAS.503.2814C,2026enap....4...38K}. 
The contour marking the end of the GGS phase ($\tau\sim10$~Gyr) lies in the upper part of the low-$\mathrm{[Mg/Fe]}$ sequence ($\mathrm{[Fe/H]}\sim-0.05$~dex and $\mathrm{[Mg/Fe]}\sim0.15$~dex) indicating that much of the chemical shift to low-$\mathrm{[Mg/Fe]}$ had already occurred by the end of the starburst phase and star-forming gas disc shrinking also occurred around this epoch.

Our inferred bar formation epoch spans approximately $\tau\sim10.5-8.5$~Gyr and the older and younger end of this interval are marked as the two solid thick contours. The older contour ($\tau\sim10.5$~Gyr) lies within the GGS phase, at the transition region between the high- and low-[Mg/Fe] sequences. The younger contour ($\tau\sim8.5$~Gyr) is located in the low-[Mg/Fe] sequence at slightly higher metallicity and lower [Mg/Fe] than the end of the GGS phase ([Fe/H]$~\sim0.0$ and [Mg/Fe]$~\sim0.1$), indicating further chemical evolution after the GGS phase. 
The distribution of these contours suggests that bar formation occurred during the later stages of the chemical transition from the high-$\mathrm{[Mg/Fe]}$ population, generally associated with a kinematically hotter disc, towards the low-$\mathrm{[Mg/Fe]}$ population associated with the kinematically colder thin disc.

\subsection{Caveats}\label{sec:dis:caveat}
The comparisons above should be interpreted with several caveats. First, the period range adopted here to represent the bar formation epoch, $P=280-330~\mathrm{days}$, is based on the NSD star formation history inferred from Mira variables. However, the time resolution and uncertainties associated with this star formation history have not been explicitly quantified in this study. In addition, the relation between the epoch of enhanced star formation traced by NSD Mira stars and the actual bar formation is not precisely established.
Therefore, the corresponding age range should not be interpreted as defining the precise onset or completion time of bar formation. Instead, it should be regarded as a broad epoch associated with the formation of the Galactic bar.

Second, the ages used in this work are tied to the APOGEE/BINGO age scale used by \citet{2024MNRAS.528L.122C} and \citet{2025MNRAS.543.2275F}. This choice is useful because it allows the bar formation epoch inferred from the Mira period--age relation to be placed on the same timeline as the chemical, structural and kinematic evolution traced by APOGEE red giants, including the GSE-related GGS phase \citep{2024MNRAS.528L.122C}, the gas disc shrinking epoch and the thick-to-thin disc transition \citep{2025MNRAS.543.2275F}. However, this age scale should be regarded primarily as a relative age scale rather than an absolute age calibration. In particular, APOGEE/BINGO ages older than $\sim10~\mathrm{Gyr}$ may be stretched relative to other age estimates, as shown in \citet{2023A&A...678A.158A}. Time intervals inferred in this old-age regime may therefore overestimate the corresponding physical time intervals. \rev{Our inferred bar formation epoch spans $8.5$–$10.5~\mathrm{Gyr}$, and the older end of the range may be more strongly affected by age stretching than the younger end. Thus, while the inferred epoch provides a useful reference on the APOGEE/BINGO relative age scale, its absolute timing remains subject to systematic uncertainty.}

A further limitation concerns the inferred timing of the $R_{\textrm{disc}}$ dip in \citet{2025MNRAS.543.2275F}, which is limited in precision by the time resolution of the age-dependent spline model. The spline knots were spaced by approximately $2~\mathrm{Gyr}$ on the APOGEE/BINGO age scale (see Fig.~\ref{fig:dis:spline_f25}), and the epoch at which $R_{\textrm{disc}}$ reaches its minimum cannot be localised much more finely than this. The dip should therefore not be interpreted as a precise timestamp of gas disc shrinking, but rather as indicating a broad epoch over which the disc structure changed significantly.

In addition, the mock validation discussed in Appendix~\ref{sec:validation} indicates that the modelling procedure tends to overestimate stellar ages by approximately $0.5~\mathrm{Gyr}$ on average. The inferred bar formation epoch may therefore be shifted towards younger ages by a similar amount. However, this systematic offset is small compared to the overall uncertainty in the timing of these evolutionary events.

\rev{Finally, the metallicities of our Mira variables are unknown. Mira variables are found in environments spanning a wide metallicity range (e.g. the Large and Small Magellanic Clouds, Sagittarius dwarf galaxy and globular clusters). Hence, the period–age relation may depend on metallicity. Our calibration is derived for disc Mira variables at $R > 6$~kpc, whereas it is applied to NSD Mira variables, which are likely more metal-rich on average. Any metallicity dependence would therefore introduce a systematic offset in the inferred bar formation epoch, whose size we cannot currently quantify.}

Taking these limitations together, the apparent alignment between \rev{the epoch associated with bar formation}, the $R_{\textrm{disc}}$ dip and the end of the GGS phase should be interpreted as a broad temporal coincidence, rather than as evidence that these events occurred at exactly the same time or with precisely determined timing. 
\rev{Their overlap is consistent with a scenario in which the GSE merger contributed to disc restructuring and the development of the Galactic bar through tidal perturbations and associated gas inflow, alongside star formation and changes in gas accretion mode.
However, the present analysis cannot establish the relative importance of these processes or demonstrate a unique causal sequence. Our main conclusion is therefore that bar formation and disc restructuring were likely connected to the same broad evolutionary phase of the Milky Way, as illustrated by the physically motivated simulation of \citet{2024MNRAS.531.1520M}.}

\section{Summary}\label{sec:summary}
In this study, we derived a Mira period-age relation by connecting the period-dependent kinematics of Galactic disc \rev{Mira variables} to the age-dependent kinematics of APOGEE red giants. The aim was to place \rev{Mira variable} periods on the APOGEE/BINGO age scale, so that the bar formation epoch inferred from NSD \rev{Mira variables} could be compared with the chemical, structural and kinematic evolution of the Milky Way disc traced by APOGEE red giants.

We fitted the \rev{Mira variable} sample with a period-dependent quasi-isothermal distribution function, using the same Galactic potential and distribution-function framework as in \citet{2025MNRAS.543.2275F}, who studied the age-dependent distribution function using the APOGEE red giants. Because reliable line-of-sight velocities were not available for the \rev{Mira variable} sample, the most reliable kinematic tracer was the vertical velocity dispersion at the solar radius, $\sigma_{z,0}$. We therefore calibrated the period-age relation by matching the period--vertical velocity-dispersion relation for \rev{Mira variables}, $\sigma_{z,0}(P)$, to the age--vertical velocity-dispersion relation for APOGEE red giants, $\sigma_{z,0}(\tau)$.

By applying the inferred period--age relation to the \rev{Mira variable} periods, we placed the bar formation epoch suggested by \citet{2024MNRAS.530.2972S} on the APOGEE/BINGO age scale. We found that the inferred bar formation epoch overlaps with the final stage of the GGS phase, the rapid chemical enrichment likely induced by the starburst caused by GSE gas-rich merger identified by \citet{2024MNRAS.528L.122C} and the epoch of gas disc shrinking inferred by \citet{2025MNRAS.543.2275F}. In the $\mathrm{[Fe/H]}$-$\mathrm{[Mg/Fe]}$ plane, the GGS phase lies between the high-$\mathrm{[Mg/Fe]}$ population and the low-$\mathrm{[Mg/Fe]}$ disc. This suggests that the bar formation may have started during the transition from the older kinematically hotter disc to the colder thin low-$\mathrm{[Mg/Fe]}$ disc sequence or just after the phase transition. These results support a picture in which the Galactic bar formed during the disc restructuring that followed the gas-rich GSE merger event. This interpretation is qualitatively consistent with simulations \citep{2024MNRAS.531.1520M}, in which a GSE-like merger supplies gas to the inner region of a galaxy and induces a starburst. Its tidal perturbations seed the bar formation and drive the disc gas inward. The combination of merger-driven gas supply and inward gas transport increase the baryonic dominance of the inner galaxy, making conditions more favourable for stable bar formation.

This study demonstrates the value of Mira variables as probes of Galactic archaeology, while also showing that their period--age relation remains difficult to calibrate precisely. In the present analysis, the main limitations are the lack of reliable line-of-sight velocities, the limited size of the current Mira sample, and the limited radial coverage available to constrain the period-dependent distribution function. These limitations restrict how strongly the radial structure parameters of the period-dependent distribution function can be constrained. In addition, possible metallicity dependence of the Mira period-age relation remains an important astrophysical systematic uncertainty when applying the calibration to different Galactic environments.

Future \Gaia data releases, especially \Gaia DR4, should substantially improve these limitations. A larger and better-characterised sample of O-rich Miras, together with improved astrometric measurements, would strengthen the constraints on the spatial and kinematic properties of the Mira population. If reliable time-series line-of-sight velocity measurements become available, the full three-dimensional velocity information could be used more directly, reducing the reliance on proper motions alone. Such improvements would lead to a more robust Mira period--age calibration, making Miras an even more powerful tracer of the connection between the Galactic bar, bulge and NSD and the broader evolutionary history of the Milky Way.

\section*{Acknowledgements}
This work is a part of MWGaiaDN, a Horizon Europe Marie Sk\l{}odowska-Curie Actions Doctoral Network funded under grant agreement no. 101072454 and also funded by UK Research and Innovation (EP/X031756/1). This work was also partly supported by the UK's Science \& Technology Facilities Council (STFC grant ST/S000216/1, ST/W001136/1). JLS acknowledges the support of the Royal Society (URF\textbackslash R1\textbackslash191555; URF\textbackslash R\textbackslash 241030). RJJG is supported by an STFC Ernest Rutherford Fellowship (ST/W003643/1). BT and NF acknowledge financial support from the Wenner-Gren Foundation (grant Nos.\ \mbox{FT2025-0009} and \mbox{WUP2026-0005}).
BT acknowledges the Royal Physiographic Society in Lund through the Stiftelsen Walter Gyllenbergs, M\"{a}rta och Erik Holmbergs donations.

This work has made use of data from the European Space Agency (ESA) mission \Gaia  (\url{https://www.cosmos.esa.int/gaia}), processed by the \Gaia Data Processing and Analysis Consortium (DPAC, \url{https://www.cosmos.esa.int/web/gaia/dpac/consortium}). Funding for the DPAC has been provided by national institutions, in particular the institutions participating in the \Gaia Multilateral Agreement.

This work has made use of data from the Apache Point Observatory Galactic Evolution Experiment (APOGEE), which is part of the Sloan Digital Sky Survey (SDSS). Funding for the Sloan Digital Sky Survey IV has been provided by the Alfred P. Sloan Foundation, the U.S. Department of Energy Office of Science, and the Participating Institutions. SDSS acknowledges support and resources from the Center for High-Performance Computing at the University of Utah. The SDSS web site is \url{www.sdss4.org}.

SDSS is managed by the Astrophysical Research Consortium for the Participating Institutions of the SDSS Collaboration including the Brazilian Participation Group, the Carnegie Institution for Science, Carnegie Mellon University, Center for Astrophysics | Harvard \& Smithsonian (CfA), the Chilean Participation Group, the French Participation Group, Instituto de Astrofísica de Canarias, The Johns Hopkins University, Kavli Institute for the Physics and Mathematics of the Universe (IPMU) / University of Tokyo, the Korean Participation Group, Lawrence Berkeley National Laboratory, Leibniz Institut für Astrophysik Potsdam (AIP), Max-Planck-Institut für Astronomie (MPIA Heidelberg), Max-Planck-Institut für Astrophysik (MPA Garching), Max-Planck-Institut für Extraterrestrische Physik (MPE), National Astronomical Observatories of China, New Mexico State University, New York University, University of Notre Dame, Observatório Nacional / MCTI, The Ohio State University, Pennsylvania State University, Shanghai Astronomical Observatory, United Kingdom Participation Group, Universidad Nacional Autónoma de México, University of Arizona, University of Colorado Boulder, University of Oxford, University of Portsmouth, University of Utah, University of Virginia, University of Washington, University of Wisconsin, Vanderbilt University, and Yale University.

The authors declare no conflicts of interest.

\section*{Data Availability}
The samples used in this work are based on publicly available survey data. The red giant ages used in this work will be shared by the corresponding author upon reasonable request.



\bibliographystyle{mnras}
\bibliography{example} 

@ARTICLE{2017JOSS....2..388P,
       author = {{Price-Whelan}, Adrian M.},
        title = "{Gala: A Python package for galactic dynamics}",
      journal = {The Journal of Open Source Software},
         year = 2017,
        month = oct,
       volume = {2},
          eid = {388},
        pages = {388},
          doi = {10.21105/joss.00388},
       adsurl = {https://ui.adsabs.harvard.edu/abs/2017JOSS....2..388P}
}

@ARTICLE{2004ARA&A..42..603K,
       author = {{Kormendy}, John and {Kennicutt}, Jr., Robert C.},
        title = "{Secular Evolution and the Formation of Pseudobulges in Disk Galaxies}",
      journal = {\araa},
         year = 2004,
        month = sep,
       volume = {42},
       number = {1},
        pages = {603-683},
          doi = {10.1146/annurev.astro.42.053102.134024},
archivePrefix = {arXiv},
       eprint = {astro-ph/0407343},
 primaryClass = {astro-ph},
       adsurl = {https://ui.adsabs.harvard.edu/abs/2004ARA&A..42..603K}
}

@ARTICLE{2013MNRAS.429.1949A,
       author = {{Athanassoula}, E. and {Machado}, Rubens E.~G. and {Rodionov}, S.~A.},
        title = "{Bar formation and evolution in disc galaxies with gas and a triaxial halo: morphology, bar strength and halo properties}",
      journal = {\mnras},
         year = 2013,
        month = mar,
       volume = {429},
       number = {3},
        pages = {1949-1969},
          doi = {10.1093/mnras/sts452},
archivePrefix = {arXiv},
       eprint = {1211.6754},
 primaryClass = {astro-ph.CO},
       adsurl = {https://ui.adsabs.harvard.edu/abs/2013MNRAS.429.1949A}
}

@ARTICLE{2002ApJ...569L..83A,
       author = {{Athanassoula}, E.},
        title = "{Bar-Halo Interaction and Bar Growth}",
      journal = {\apjl},
         year = 2002,
        month = apr,
       volume = {569},
       number = {2},
        pages = {L83-L86},
          doi = {10.1086/340784},
archivePrefix = {arXiv},
       eprint = {astro-ph/0203368},
 primaryClass = {astro-ph},
       adsurl = {https://ui.adsabs.harvard.edu/abs/2002ApJ...569L..83A}
}

@ARTICLE{1982MNRAS.199.1069E,
       author = {{Efstathiou}, G. and {Lake}, G. and {Negroponte}, J.},
        title = "{The stability and masses of disc galaxies.}",
      journal = {\mnras},
         year = 1982,
        month = jun,
       volume = {199},
        pages = {1069-1088},
          doi = {10.1093/mnras/199.4.1069},
       adsurl = {https://ui.adsabs.harvard.edu/abs/1982MNRAS.199.1069E}
}

@ARTICLE{2025arXiv251209987Z,
       author = {{Zhang}, HanYuan and {Belokurov}, Vasily and {Sanders}, Jason L. and {Evans}, N. Wyn and {Chemaly}, David and {Kawata}, Daisuke and {Funakoshi}, Natsuki and {Frankel}, Neige and {Kane}, Sarah G. and {Koposov}, Sergey E.},
        title = "{Orbital migration and heating history of the Galactic disc: a transition between the bimodal discs}",
      journal = {arXiv e-prints},
         year = 2025,
        month = dec,
          eid = {arXiv:2512.09987},
        pages = {arXiv:2512.09987},
          doi = {10.48550/arXiv.2512.09987},
archivePrefix = {arXiv},
       eprint = {2512.09987},
 primaryClass = {astro-ph.GA},
       adsurl = {https://ui.adsabs.harvard.edu/abs/2025arXiv251209987Z}
}

@ARTICLE{2025A&ARv..33....7S,
       author = {{Schultheis}, Mathias and {Sormani}, Mattia C. and {Gadotti}, Dimitri A.},
        title = "{Nuclear stellar discs}",
      journal = {\aapr},
         year = 2025,
        month = nov,
       volume = {33},
       number = {1},
          eid = {7},
        pages = {7},
          doi = {10.1007/s00159-025-00163-6},
archivePrefix = {arXiv},
       eprint = {2509.04562},
 primaryClass = {astro-ph.GA},
       adsurl = {https://ui.adsabs.harvard.edu/abs/2025A&ARv..33....7S}
}

@INPROCEEDINGS{2026enap....4...38K,
       author = {{Kawata}, Daisuke and {Grand}, Robert J.~J. and {Hunt}, Jason A.~S. and {Ciuc{\u{a}}}, Ioana},
        title = "{Milky Way disk}",
    booktitle = {Encyclopedia of Astrophysics, Volume 4},
         year = 2026,
       volume = {4},
        month = jan,
        pages = {38-60},
          doi = {10.1016/B978-0-443-21439-4.00064-X},
archivePrefix = {arXiv},
       eprint = {2412.12252},
 primaryClass = {astro-ph.GA},
       adsurl = {https://ui.adsabs.harvard.edu/abs/2026enap....4...38K}
}

@ARTICLE{2020A&A...636A.115D,
       author = {{Di Matteo}, P. and {Spite}, M. and {Haywood}, M. and {Bonifacio}, P. and {G{\'o}mez}, A. and {Spite}, F. and {Caffau}, E.},
        title = "{Reviving old controversies: is the early Galaxy flat or round?. Investigations into the early phases of the Milky Way's formation through stellar kinematics and chemical abundances}",
      journal = {\aap},
         year = 2020,
        month = apr,
       volume = {636},
          eid = {A115},
        pages = {A115},
          doi = {10.1051/0004-6361/201937016},
archivePrefix = {arXiv},
       eprint = {1910.13769},
 primaryClass = {astro-ph.GA},
       adsurl = {https://ui.adsabs.harvard.edu/abs/2020A&A...636A.115D}
}

@ARTICLE{2020MNRAS.494.3880B,
       author = {{Belokurov}, Vasily and {Sanders}, Jason L. and {Fattahi}, Azadeh and {Smith}, Martin C. and {Deason}, Alis J. and {Evans}, N. Wyn and {Grand}, Robert J.~J.},
        title = "{The biggest splash}",
      journal = {\mnras},
         year = 2020,
        month = may,
       volume = {494},
       number = {3},
        pages = {3880-3898},
          doi = {10.1093/mnras/staa876},
archivePrefix = {arXiv},
       eprint = {1909.04679},
 primaryClass = {astro-ph.GA},
       adsurl = {https://ui.adsabs.harvard.edu/abs/2020MNRAS.494.3880B}
}

@ARTICLE{2020NatAs...4..382C,
       author = {{Chaplin}, William J. and {Serenelli}, Aldo M. and {Miglio}, Andrea and {Morel}, Thierry and {Mackereth}, J. Ted and {Vincenzo}, Fiorenzo and {Kjeldsen}, Hans and {Basu}, Sarbani and {Ball}, Warrick H. and {Stokholm}, Amalie and {Verma}, Kuldeep and {Mosumgaard}, Jakob R{\o}rsted and {Silva Aguirre}, Victor and {Mazumdar}, Anwesh and {Ranadive}, Pritesh and {Antia}, H.~M. and {Lebreton}, Yveline and {Ong}, Joel and {Appourchaux}, Thierry and {Bedding}, Timothy R. and {Christensen-Dalsgaard}, J{\o}rgen and {Creevey}, Orlagh and {Garc{\'\i}a}, Rafael A. and {Handberg}, Rasmus and {Huber}, Daniel and {Kawaler}, Steven D. and {Lund}, Mikkel N. and {Metcalfe}, Travis S. and {Stassun}, Keivan G. and {Bazot}, Mich{\"a}el and {Beck}, Paul G. and {Bell}, Keaton J. and {Bergemann}, Maria and {Buzasi}, Derek L. and {Benomar}, Othman and {Bossini}, Diego and {Bugnet}, Lisa and {Campante}, Tiago L. and {Orhan}, Zeynep {\c{c}}elik and {Corsaro}, Enrico and {Gonz{\'a}lez-Cuesta}, Luc{\'\i}a and {Davies}, Guy R. and {Di Mauro}, Maria Pia and {Egeland}, Ricky and {Elsworth}, Yvonne P. and {Gaulme}, Patrick and {Ghasemi}, Hamed and {Guo}, Zhao and {Hall}, Oliver J. and {Hasanzadeh}, Amir and {Hekker}, Saskia and {Howe}, Rachel and {Jenkins}, Jon M. and {Jim{\'e}nez}, Antonio and {Kiefer}, Ren{\'e} and {Kuszlewicz}, James S. and {Kallinger}, Thomas and {Latham}, David W. and {Lundkvist}, Mia S. and {Mathur}, Savita and {Montalb{\'a}n}, Josefina and {Mosser}, Benoit and {Bed{\'o}n}, Andres Moya and {Nielsen}, Martin Bo and {{\"O}rtel}, Sibel and {Rendle}, Ben M. and {Ricker}, George R. and {Rodrigues}, Tha{\'\i}se S. and {Roxburgh}, Ian W. and {Safari}, Hossein and {Schofield}, Mathew and {Seager}, Sara and {Smalley}, Barry and {Stello}, Dennis and {Szab{\'o}}, R{\'o}bert and {Tayar}, Jamie and {Theme{\ss}l}, Nathalie and {Thomas}, Alexandra E.~L. and {Vanderspek}, Roland K. and {van Rossem}, Walter E. and {Vrard}, Mathieu and {Weiss}, Achim and {White}, Timothy R. and {Winn}, Joshua N. and {Y{\i}ld{\i}z}, Mutlu},
        title = "{Age dating of an early Milky Way merger via asteroseismology of the naked-eye star {\ensuremath{\nu}} Indi}",
      journal = {Nature Astronomy},
         year = 2020,
        month = jan,
       volume = {4},
        pages = {382-389},
          doi = {10.1038/s41550-019-0975-9},
archivePrefix = {arXiv},
       eprint = {2001.04653},
 primaryClass = {astro-ph.GA},
       adsurl = {https://ui.adsabs.harvard.edu/abs/2020NatAs...4..382C}
}

@ARTICLE{2010A&A...511L..10N,
       author = {{Nissen}, P.~E. and {Schuster}, W.~J.},
        title = "{Two distinct halo populations in the solar neighborhood. Evidence from stellar abundance ratios and kinematics}",
      journal = {\aap},
         year = 2010,
        month = feb,
       volume = {511},
          eid = {L10},
        pages = {L10},
          doi = {10.1051/0004-6361/200913877},
archivePrefix = {arXiv},
       eprint = {1002.4514},
 primaryClass = {astro-ph.GA},
       adsurl = {https://ui.adsabs.harvard.edu/abs/2010A&A...511L..10N}
}

@ARTICLE{2025MNRAS.538.1587F,
       author = {{Fragkoudi}, Francesca and {Grand}, Robert J.~J. and {Pakmor}, R{\"u}diger and {G{\'o}mez}, Facundo and {Marinacci}, Federico and {Springel}, Volker},
        title = "{Bar formation and evolution in the cosmological context: inputs from the Auriga simulations}",
      journal = {\mnras},
         year = 2025,
        month = apr,
       volume = {538},
       number = {3},
        pages = {1587-1608},
          doi = {10.1093/mnras/staf389},
archivePrefix = {arXiv},
       eprint = {2406.09453},
 primaryClass = {astro-ph.GA},
       adsurl = {https://ui.adsabs.harvard.edu/abs/2025MNRAS.538.1587F}
}

@ARTICLE{2025NewAR.10001721H,
       author = {{Hunt}, Jason A.~S. and {Vasiliev}, Eugene},
        title = "{Milky Way dynamics in light of Gaia}",
      journal = {\nar},
         year = 2025,
        month = jun,
       volume = {100},
          eid = {101721},
        pages = {101721},
          doi = {10.1016/j.newar.2024.101721},
archivePrefix = {arXiv},
       eprint = {2501.04075},
 primaryClass = {astro-ph.GA},
       adsurl = {https://ui.adsabs.harvard.edu/abs/2025NewAR.10001721H}
}

@ARTICLE{2012MNRAS.427..127B,
       author = {{Bressan}, Alessandro and {Marigo}, Paola and {Girardi}, L{\'e}o. and {Salasnich}, Bernardo and {Dal Cero}, Claudia and {Rubele}, Stefano and {Nanni}, Ambra},
        title = "{PARSEC: stellar tracks and isochrones with the PAdova and TRieste Stellar Evolution Code}",
      journal = {\mnras},
         year = 2012,
        month = nov,
       volume = {427},
       number = {1},
        pages = {127-145},
          doi = {10.1111/j.1365-2966.2012.21948.x},
archivePrefix = {arXiv},
       eprint = {1208.4498},
 primaryClass = {astro-ph.SR},
       adsurl = {https://ui.adsabs.harvard.edu/abs/2012MNRAS.427..127B}
}

@ARTICLE{2021A&A...650A.191S,
       author = {{Schultheis}, M. and {Fritz}, T.~K. and {Nandakumar}, G. and {Rojas-Arriagada}, A. and {Nogueras-Lara}, F. and {Feldmeier-Krause}, A. and {Gerhard}, O. and {Neumayer}, N. and {Patrick}, L.~R. and {Prieto}, M.~A. and {Sch{\"o}del}, R. and {Mastrobuono-Battisti}, A. and {Sormani}, M.~C.},
        title = "{The nuclear stellar disc of the Milky Way: A dynamically cool and metal-rich component possibly formed from the central molecular zone}",
      journal = {\aap},
         year = 2021,
        month = jun,
       volume = {650},
          eid = {A191},
        pages = {A191},
          doi = {10.1051/0004-6361/202140499},
archivePrefix = {arXiv},
       eprint = {2104.10439},
 primaryClass = {astro-ph.GA},
       adsurl = {https://ui.adsabs.harvard.edu/abs/2021A&A...650A.191S}
}

@ARTICLE{2002A&A...384..112L,
       author = {{Launhardt}, R. and {Zylka}, R. and {Mezger}, P.~G.},
        title = "{The nuclear bulge of the Galaxy. III. Large-scale physical characteristics of stars and interstellar matter}",
      journal = {\aap},
         year = 2002,
        month = mar,
       volume = {384},
        pages = {112-139},
          doi = {10.1051/0004-6361:20020017},
archivePrefix = {arXiv},
       eprint = {astro-ph/0201294},
 primaryClass = {astro-ph},
       adsurl = {https://ui.adsabs.harvard.edu/abs/2002A&A...384..112L}
}

@ARTICLE{2018MNRAS.475.1093Y,
       author = {{Yu}, Jincheng and {Liu}, Chao},
        title = "{The age-velocity dispersion relation of the Galactic discs from LAMOST-Gaia data}",
      journal = {\mnras},
         year = 2018,
        month = mar,
       volume = {475},
       number = {1},
        pages = {1093-1103},
          doi = {10.1093/mnras/stx3204},
archivePrefix = {arXiv},
       eprint = {1712.03965},
 primaryClass = {astro-ph.GA},
       adsurl = {https://ui.adsabs.harvard.edu/abs/2018MNRAS.475.1093Y}
}

@ARTICLE{1972ApJ...178..623T,
       author = {{Toomre}, Alar and {Toomre}, Juri},
        title = "{Galactic Bridges and Tails}",
      journal = {\apj},
         year = 1972,
        month = dec,
       volume = {178},
        pages = {623-666},
          doi = {10.1086/151823},
       adsurl = {https://ui.adsabs.harvard.edu/abs/1972ApJ...178..623T}
}

@ARTICLE{2018ApJ...857....6L,
       author = {{{\L}okas}, Ewa L.},
        title = "{Formation of Tidally Induced Bars in Galactic Flybys: Prograde versus Retrograde Encounters}",
      journal = {\apj},
         year = 2018,
        month = apr,
       volume = {857},
       number = {1},
          eid = {6},
        pages = {6},
          doi = {10.3847/1538-4357/aab4ff},
archivePrefix = {arXiv},
       eprint = {1803.09465},
 primaryClass = {astro-ph.GA},
       adsurl = {https://ui.adsabs.harvard.edu/abs/2018ApJ...857....6L}
}

@ARTICLE{2025A&A...702A...7L,
       author = {{{\L}okas}, Ewa L.},
        title = "{Tidally induced bar-like galaxies in simulated clusters}",
      journal = {\aap},
         year = 2025,
        month = sep,
       volume = {702},
          eid = {A7},
        pages = {A7},
          doi = {10.1051/0004-6361/202555503},
archivePrefix = {arXiv},
       eprint = {2509.03184},
 primaryClass = {astro-ph.GA},
       adsurl = {https://ui.adsabs.harvard.edu/abs/2025A&A...702A...7L}
}

@ARTICLE{1998ApJ...499..149M,
       author = {{Miwa}, Toshinobu and {Noguchi}, Masafumi},
        title = "{Dynamical Properties of Tidally Induced Galactic Bars}",
      journal = {\apj},
         year = 1998,
        month = may,
       volume = {499},
       number = {1},
        pages = {149-166},
          doi = {10.1086/305611},
       adsurl = {https://ui.adsabs.harvard.edu/abs/1998ApJ...499..149M}
}

@ARTICLE{2025ApJ...978...37A,
       author = {{Ansar}, Sioree and {Pearson}, Sarah and {Sanderson}, Robyn E. and {Arora}, Arpit and {Hopkins}, Philip F. and {Wetzel}, Andrew and {Cunningham}, Emily C. and {Quinn}, Jamie},
        title = "{Bar Formation and Destruction in the FIRE-2 Simulations}",
      journal = {\apj},
         year = 2025,
        month = jan,
       volume = {978},
       number = {1},
          eid = {37},
        pages = {37},
          doi = {10.3847/1538-4357/ad8b45},
archivePrefix = {arXiv},
       eprint = {2309.16811},
 primaryClass = {astro-ph.GA},
       adsurl = {https://ui.adsabs.harvard.edu/abs/2025ApJ...978...37A}
}

@ARTICLE{2012ApJ...757...60K,
       author = {{Kraljic}, Katarina and {Bournaud}, Fr{\'e}d{\'e}ric and {Martig}, Marie},
        title = "{The Two-phase Formation History of Spiral Galaxies Traced by the Cosmic Evolution of the Bar Fraction}",
      journal = {\apj},
         year = 2012,
        month = sep,
       volume = {757},
       number = {1},
          eid = {60},
        pages = {60},
          doi = {10.1088/0004-637X/757/1/60},
archivePrefix = {arXiv},
       eprint = {1207.0351},
 primaryClass = {astro-ph.GA},
       adsurl = {https://ui.adsabs.harvard.edu/abs/2012ApJ...757...60K}
}

@ARTICLE{2019MNRAS.483.2721P,
       author = {{Peschken}, Nicolas and {{\L}okas}, Ewa L.},
        title = "{Tidally induced bars in Illustris galaxies}",
      journal = {\mnras},
         year = 2019,
        month = feb,
       volume = {483},
       number = {2},
        pages = {2721-2735},
          doi = {10.1093/mnras/sty3277},
archivePrefix = {arXiv},
       eprint = {1804.06241},
 primaryClass = {astro-ph.GA},
       adsurl = {https://ui.adsabs.harvard.edu/abs/2019MNRAS.483.2721P}
}

@ARTICLE{2016ApJ...826..227L,
       author = {{{\L}okas}, Ewa L. and {Ebrov{\'a}}, Ivana and {del Pino}, Andr{\'e}s and {Sybilska}, Agnieszka and {Athanassoula}, E. and {Semczuk}, Marcin and {Gajda}, Grzegorz and {Fouquet}, Sylvain},
        title = "{Tidally Induced Bars of Galaxies in Clusters}",
      journal = {\apj},
         year = 2016,
        month = aug,
       volume = {826},
       number = {2},
          eid = {227},
        pages = {227},
          doi = {10.3847/0004-637X/826/2/227},
archivePrefix = {arXiv},
       eprint = {1601.07433},
 primaryClass = {astro-ph.GA},
       adsurl = {https://ui.adsabs.harvard.edu/abs/2016ApJ...826..227L}
}

@ARTICLE{2019MNRAS.482.1983F,
       author = {{Fujii}, M.~S. and {B{\'e}dorf}, J. and {Baba}, J. and {Portegies Zwart}, S.},
        title = "{Modelling the Milky Way as a dry Galaxy}",
      journal = {\mnras},
         year = 2019,
        month = jan,
       volume = {482},
       number = {2},
        pages = {1983-2015},
          doi = {10.1093/mnras/sty2747},
archivePrefix = {arXiv},
       eprint = {1807.10019},
 primaryClass = {astro-ph.GA},
       adsurl = {https://ui.adsabs.harvard.edu/abs/2019MNRAS.482.1983F}
}

@ARTICLE{2020ApJ...897L..18B,
       author = {{Bonaca}, Ana and {Conroy}, Charlie and {Cargile}, Phillip A. and {Naidu}, Rohan P. and {Johnson}, Benjamin D. and {Zaritsky}, Dennis and {Ting}, Yuan-Sen and {Caldwell}, Nelson and {Han}, Jiwon Jesse and {van Dokkum}, Pieter},
        title = "{Timing the Early Assembly of the Milky Way with the H3 Survey}",
      journal = {\apjl},
         year = 2020,
        month = jul,
       volume = {897},
       number = {1},
          eid = {L18},
        pages = {L18},
          doi = {10.3847/2041-8213/ab9caa},
archivePrefix = {arXiv},
       eprint = {2004.11384},
 primaryClass = {astro-ph.GA},
       adsurl = {https://ui.adsabs.harvard.edu/abs/2020ApJ...897L..18B}
}

@ARTICLE{2019MNRAS.490.4740B,
       author = {{Bovy}, Jo and {Leung}, Henry W. and {Hunt}, Jason A.~S. and {Mackereth}, J. Ted and {Garc{\'\i}a-Hern{\'a}ndez}, Domingo A. and {Roman-Lopes}, Alexandre},
        title = "{Life in the fast lane: a direct view of the dynamics, formation, and evolution of the Milky Way's bar}",
      journal = {\mnras},
         year = 2019,
        month = dec,
       volume = {490},
       number = {4},
        pages = {4740-4747},
          doi = {10.1093/mnras/stz2891},
archivePrefix = {arXiv},
       eprint = {1905.11404},
 primaryClass = {astro-ph.GA},
       adsurl = {https://ui.adsabs.harvard.edu/abs/2019MNRAS.490.4740B}
}

@ARTICLE{2015MNRAS.450.4050W,
       author = {{Wegg}, Christopher and {Gerhard}, Ortwin and {Portail}, Matthieu},
        title = "{The structure of the Milky Way's bar outside the bulge}",
      journal = {\mnras},
         year = 2015,
        month = jul,
       volume = {450},
       number = {4},
        pages = {4050-4069},
          doi = {10.1093/mnras/stv745},
archivePrefix = {arXiv},
       eprint = {1504.01401},
 primaryClass = {astro-ph.GA},
       adsurl = {https://ui.adsabs.harvard.edu/abs/2015MNRAS.450.4050W}
}

@ARTICLE{2013MNRAS.435.1874W,
       author = {{Wegg}, Christopher and {Gerhard}, Ortwin},
        title = "{Mapping the three-dimensional density of the Galactic bulge with VVV red clump stars}",
      journal = {\mnras},
         year = 2013,
        month = nov,
       volume = {435},
       number = {3},
        pages = {1874-1887},
          doi = {10.1093/mnras/stt1376},
archivePrefix = {arXiv},
       eprint = {1308.0593},
 primaryClass = {astro-ph.GA},
       adsurl = {https://ui.adsabs.harvard.edu/abs/2013MNRAS.435.1874W}
}

@ARTICLE{2014RvMP...86....1S,
       author = {{Sellwood}, J.~A.},
        title = "{Secular evolution in disk galaxies}",
      journal = {Reviews of Modern Physics},
         year = 2014,
        month = jan,
       volume = {86},
       number = {1},
        pages = {1-46},
          doi = {10.1103/RevModPhys.86.1},
archivePrefix = {arXiv},
       eprint = {1310.0403},
 primaryClass = {astro-ph.GA},
       adsurl = {https://ui.adsabs.harvard.edu/abs/2014RvMP...86....1S}
}

@ARTICLE{1995ApJ...445..716D,
       author = {{Dwek}, E. and {Arendt}, R.~G. and {Hauser}, M.~G. and {Kelsall}, T. and {Lisse}, C.~M. and {Moseley}, S.~H. and {Silverberg}, R.~F. and {Sodroski}, T.~J. and {Weiland}, J.~L.},
        title = "{Morphology, Near-Infrared Luminosity, and Mass of the Galactic Bulge from COBE DIRBE Observations}",
      journal = {\apj},
         year = 1995,
        month = jun,
       volume = {445},
        pages = {716},
          doi = {10.1086/175734},
       adsurl = {https://ui.adsabs.harvard.edu/abs/1995ApJ...445..716D}
}

@ARTICLE{1991ApJ...379..631B,
       author = {{Blitz}, Leo and {Spergel}, David N.},
        title = "{Direct Evidence for a Bar at the Galactic Center}",
      journal = {\apj},
         year = 1991,
        month = oct,
       volume = {379},
        pages = {631},
          doi = {10.1086/170535},
       adsurl = {https://ui.adsabs.harvard.edu/abs/1991ApJ...379..631B}
}

@ARTICLE{1991MNRAS.252..210B,
       author = {{Binney}, James and {Gerhard}, Ortwin E. and {Stark}, Antony A. and {Bally}, John and {Uchida}, Keven I.},
        title = "{Understanding the kinematics of Galactic Centre gas.}",
      journal = {\mnras},
         year = 1991,
        month = sep,
       volume = {252},
        pages = {210},
          doi = {10.1093/mnras/252.2.210},
       adsurl = {https://ui.adsabs.harvard.edu/abs/1991MNRAS.252..210B}
}

@ARTICLE{2019MNRAS.488.4552S,
       author = {{Sanders}, Jason L. and {Smith}, Leigh and {Evans}, N. Wyn},
        title = "{The pattern speed of the Milky Way bar from transverse velocities}",
      journal = {\mnras},
         year = 2019,
        month = oct,
       volume = {488},
       number = {4},
        pages = {4552-4564},
          doi = {10.1093/mnras/stz1827},
archivePrefix = {arXiv},
       eprint = {1903.02009},
 primaryClass = {astro-ph.GA},
       adsurl = {https://ui.adsabs.harvard.edu/abs/2019MNRAS.488.4552S}
}

@ARTICLE{1987MNRAS.228..635N,
       author = {{Noguchi}, Masafumi},
        title = "{Close encounter between galaxies - II. Tidal deformation of a disc galaxy stabilized by massive halo.}",
      journal = {\mnras},
         year = 1987,
        month = oct,
       volume = {228},
        pages = {635-651},
          doi = {10.1093/mnras/228.3.635},
       adsurl = {https://ui.adsabs.harvard.edu/abs/1987MNRAS.228..635N}
}

@ARTICLE{2024MNRAS.531.1520M,
       author = {{Merrow}, Alex and {Grand}, Robert J.~J. and {Fragkoudi}, Francesca and {Martig}, Marie},
        title = "{Did the Gaia Enceladus/Sausage merger form the Milky Way's bar?}",
      journal = {\mnras},
         year = 2024,
        month = jun,
       volume = {531},
       number = {1},
        pages = {1520-1533},
          doi = {10.1093/mnras/stae1250},
archivePrefix = {arXiv},
       eprint = {2312.02318},
 primaryClass = {astro-ph.GA},
       adsurl = {https://ui.adsabs.harvard.edu/abs/2024MNRAS.531.1520M}
}

@ARTICLE{2018Natur.559..585N,
       author = {{Noguchi}, Masafumi},
        title = "{The formation of solar-neighbourhood stars in two generations separated by 5 billion years}",
      journal = {\nat},
         year = 2018,
        month = jul,
       volume = {559},
       number = {7715},
        pages = {585-588},
          doi = {10.1038/s41586-018-0329-2},
archivePrefix = {arXiv},
       eprint = {1809.02299},
 primaryClass = {astro-ph.GA},
       adsurl = {https://ui.adsabs.harvard.edu/abs/2018Natur.559..585N}
}

@ARTICLE{2005MNRAS.363....2K,
       author = {{Kere{\v{s}}}, Du{\v{s}}an and {Katz}, Neal and {Weinberg}, David H. and {Dav{\'e}}, Romeel},
        title = "{How do galaxies get their gas?}",
      journal = {\mnras},
         year = 2005,
        month = oct,
       volume = {363},
       number = {1},
        pages = {2-28},
          doi = {10.1111/j.1365-2966.2005.09451.x},
archivePrefix = {arXiv},
       eprint = {astro-ph/0407095},
 primaryClass = {astro-ph},
       adsurl = {https://ui.adsabs.harvard.edu/abs/2005MNRAS.363....2K}
}

@ARTICLE{2010MNRAS.401.2318B,
       author = {{Binney}, James},
        title = "{Distribution functions for the Milky Way}",
      journal = {\mnras},
         year = 2010,
        month = feb,
       volume = {401},
       number = {4},
        pages = {2318-2330},
          doi = {10.1111/j.1365-2966.2009.15845.x},
archivePrefix = {arXiv},
       eprint = {0910.1512},
 primaryClass = {astro-ph.GA},
       adsurl = {https://ui.adsabs.harvard.edu/abs/2010MNRAS.401.2318B}
}

@ARTICLE{2018MNRAS.474.3629G,
       author = {{Grand}, Robert J.~J. and {Bustamante}, Sebasti{\'a}n and {G{\'o}mez}, Facundo A. and {Kawata}, Daisuke and {Marinacci}, Federico and {Pakmor}, R{\"u}diger and {Rix}, Hans-Walter and {Simpson}, Christine M. and {Sparre}, Martin and {Springel}, Volker},
        title = "{Origin of chemically distinct discs in the Auriga cosmological simulations}",
      journal = {\mnras},
         year = 2018,
        month = mar,
       volume = {474},
       number = {3},
        pages = {3629-3639},
          doi = {10.1093/mnras/stx3025},
archivePrefix = {arXiv},
       eprint = {1708.07834},
 primaryClass = {astro-ph.GA},
       adsurl = {https://ui.adsabs.harvard.edu/abs/2018MNRAS.474.3629G}
}

@ARTICLE{2011arXiv1111.4246H,
       author = {{Hoffman}, Matthew D. and {Gelman}, Andrew},
        title = "{The No-U-Turn Sampler: Adaptively Setting Path Lengths in Hamiltonian Monte Carlo}",
      journal = {arXiv e-prints},
         year = 2011,
        month = nov,
          eid = {arXiv:1111.4246},
        pages = {arXiv:1111.4246},
          doi = {10.48550/arXiv.1111.4246},
archivePrefix = {arXiv},
       eprint = {1111.4246},
 primaryClass = {stat.CO},
       adsurl = {https://ui.adsabs.harvard.edu/abs/2011arXiv1111.4246H}
}

@misc{jax2018github,
  author = {James Bradbury and Roy Frostig and Peter Hawkins and Matthew James Johnson and Chris Leary and Dougal Maclaurin and George Necula and Adam Paszke and Jake Vander{P}las and Skye Wanderman-{M}ilne and Qiao Zhang},
  title = {{JAX}: composable transformations of {P}ython+{N}um{P}y programs},
  url = {http://github.com/jax-ml/jax},
  version = {0.3.13},
  year = {2018},
}

@ARTICLE{2024MNRAS.532.1814G,
       author = {{Grand}, Robert J.~J. and {Fragkoudi}, Francesca and {G{\'o}mez}, Facundo A. and {Jenkins}, Adrian and {Marinacci}, Federico and {Pakmor}, R{\"u}diger and {Springel}, Volker},
        title = "{Overview and public data release of the augmented Auriga Project: cosmological simulations of dwarf and Milky Way-mass galaxies}",
      journal = {\mnras},
         year = 2024,
        month = aug,
       volume = {532},
       number = {2},
        pages = {1814-1831},
          doi = {10.1093/mnras/stae1598},
archivePrefix = {arXiv},
       eprint = {2401.08750},
 primaryClass = {astro-ph.GA},
       adsurl = {https://ui.adsabs.harvard.edu/abs/2024MNRAS.532.1814G}
}

@ARTICLE{2011MNRAS.413.1889B,
       author = {{Binney}, James and {McMillan}, Paul},
        title = "{Models of our Galaxy - II}",
      journal = {\mnras},
         year = 2011,
        month = may,
       volume = {413},
       number = {3},
        pages = {1889-1898},
          doi = {10.1111/j.1365-2966.2011.18268.x},
archivePrefix = {arXiv},
       eprint = {1101.0747},
 primaryClass = {astro-ph.GA},
       adsurl = {https://ui.adsabs.harvard.edu/abs/2011MNRAS.413.1889B}
}

@ARTICLE{2019PASP..131e5001W,
       author = {{Wilson}, J.~C. and {Hearty}, F.~R. and {Skrutskie}, M.~F. and {Majewski}, S.~R. and {Holtzman}, J.~A. and {Eisenstein}, D. and {Gunn}, J. and {Blank}, B. and {Henderson}, C. and {Smee}, S. and {Nelson}, M. and {Nidever}, D. and {Arns}, J. and {Barkhouser}, R. and {Barr}, J. and {Beland}, S. and {Bershady}, M.~A. and {Blanton}, M.~R. and {Brunner}, S. and {Burton}, A. and {Carey}, L. and {Carr}, M. and {Colque}, J.~P. and {Crane}, J. and {Damke}, G.~J. and {Davidson}, Jr., J.~W. and {Dean}, J. and {Di Mille}, F. and {Don}, K.~W. and {Ebelke}, G. and {Evans}, M. and {Fitzgerald}, G. and {Gillespie}, B. and {Hall}, M. and {Harding}, A. and {Harding}, P. and {Hammond}, R. and {Hancock}, D. and {Harrison}, C. and {Hope}, S. and {Horne}, T. and {Karakla}, J. and {Lam}, C. and {Leger}, F. and {MacDonald}, N. and {Maseman}, P. and {Matsunari}, J. and {Melton}, S. and {Mitcheltree}, T. and {O'Brien}, T. and {O'Connell}, R.~W. and {Patten}, A. and {Richardson}, W. and {Rieke}, G. and {Rieke}, M. and {Roman-Lopes}, A. and {Schiavon}, R.~P. and {Sobeck}, J.~S. and {Stolberg}, T. and {Stoll}, R. and {Tembe}, M. and {Trujillo}, J.~D. and {Uomoto}, A. and {Vernieri}, M. and {Walker}, E. and {Weinberg}, D.~H. and {Young}, E. and {Anthony-Brumfield}, B. and {Bizyaev}, D. and {Breslauer}, B. and {De Lee}, N. and {Downey}, J. and {Halverson}, S. and {Huehnerhoff}, J. and {Klaene}, M. and {Leon}, E. and {Long}, D. and {Mahadevan}, S. and {Malanushenko}, E. and {Nguyen}, D.~C. and {Owen}, R. and {S{\'a}nchez-Gallego}, J.~R. and {Sayres}, C. and {Shane}, N. and {Shectman}, S.~A. and {Shetrone}, M. and {Skinner}, D. and {Stauffer}, F. and {Zhao}, B.},
        title = "{The Apache Point Observatory Galactic Evolution Experiment (APOGEE) Spectrographs}",
      journal = {\pasp},
         year = 2019,
        month = may,
       volume = {131},
       number = {999},
        pages = {055001},
          doi = {10.1088/1538-3873/ab0075},
archivePrefix = {arXiv},
       eprint = {1902.00928},
 primaryClass = {astro-ph.IM},
       adsurl = {https://ui.adsabs.harvard.edu/abs/2019PASP..131e5001W}
}

@ARTICLE{2015AJ....150..173N,
       author = {{Nidever}, David L. and {Holtzman}, Jon A. and {Allende Prieto}, Carlos and {Beland}, Stephane and {Bender}, Chad and {Bizyaev}, Dmitry and {Burton}, Adam and {Desphande}, Rohit and {Fleming}, Scott W. and {Garc{\'\i}a P{\'e}rez}, Ana E. and {Hearty}, Fred R. and {Majewski}, Steven R. and {M{\'e}sz{\'a}ros}, Szabolcs and {Muna}, Demitri and {Nguyen}, Duy and {Schiavon}, Ricardo P. and {Shetrone}, Matthew and {Skrutskie}, Michael F. and {Sobeck}, Jennifer S. and {Wilson}, John C.},
        title = "{The Data Reduction Pipeline for the Apache Point Observatory Galactic Evolution Experiment}",
      journal = {\aj},
         year = 2015,
        month = dec,
       volume = {150},
       number = {6},
          eid = {173},
        pages = {173},
          doi = {10.1088/0004-6256/150/6/173},
archivePrefix = {arXiv},
       eprint = {1501.03742},
 primaryClass = {astro-ph.IM},
       adsurl = {https://ui.adsabs.harvard.edu/abs/2015AJ....150..173N}
}

@ARTICLE{2013ApJ...779..115B,
       author = {{Bovy}, Jo and {Rix}, Hans-Walter},
        title = "{A Direct Dynamical Measurement of the Milky Way's Disk Surface Density Profile, Disk Scale Length, and Dark Matter Profile at 4 kpc <\raisebox{-0.5ex}\textasciitilde R <\raisebox{-0.5ex}\textasciitilde 9 kpc}",
      journal = {\apj},
         year = 2013,
        month = dec,
       volume = {779},
       number = {2},
          eid = {115},
        pages = {115},
          doi = {10.1088/0004-637X/779/2/115},
archivePrefix = {arXiv},
       eprint = {1309.0809},
 primaryClass = {astro-ph.GA},
       adsurl = {https://ui.adsabs.harvard.edu/abs/2013ApJ...779..115B}
}

@ARTICLE{2023A&A...678A.158A,
       author = {{Anders}, F. and {Gispert}, P. and {Ratcliffe}, B. and {Chiappini}, C. and {Minchev}, I. and {Nepal}, S. and {Queiroz}, A.~B.~A. and {Amarante}, J.~A.~S. and {Antoja}, T. and {Casali}, G. and {Casamiquela}, L. and {Khalatyan}, A. and {Miglio}, A. and {Perottoni}, H. and {Schultheis}, M.},
        title = "{Spectroscopic age estimates for APOGEE red-giant stars: Precise spatial and kinematic trends with age in the Galactic disc}",
      journal = {\aap},
         year = 2023,
        month = oct,
       volume = {678},
          eid = {A158},
        pages = {A158},
          doi = {10.1051/0004-6361/202346666},
archivePrefix = {arXiv},
       eprint = {2304.08276},
 primaryClass = {astro-ph.GA},
       adsurl = {https://ui.adsabs.harvard.edu/abs/2023A&A...678A.158A}
}

@ARTICLE{2024MNRAS.529.1035A,
       author = {{Almannaei}, Aisha S. and {Kawata}, Daisuke and {Baba}, Junichi and {Hunt}, Jason A.~S. and {Seabroke}, George and {Yan}, Ziyang},
        title = "{Impacts of the Local arm on the local circular velocity inferred from the Gaia DR3 young stars in the Milky Way}",
      journal = {\mnras},
         year = 2024,
        month = apr,
       volume = {529},
       number = {2},
        pages = {1035-1046},
          doi = {10.1093/mnras/stae158},
archivePrefix = {arXiv},
       eprint = {2310.06831},
 primaryClass = {astro-ph.GA},
       adsurl = {https://ui.adsabs.harvard.edu/abs/2024MNRAS.529.1035A}
}

@ARTICLE{2019MNRAS.482.1417B,
       author = {{Bennett}, Morgan and {Bovy}, Jo},
        title = "{Vertical waves in the solar neighbourhood in Gaia DR2}",
      journal = {\mnras},
         year = 2019,
        month = jan,
       volume = {482},
       number = {1},
        pages = {1417-1425},
          doi = {10.1093/mnras/sty2813},
archivePrefix = {arXiv},
       eprint = {1809.03507},
 primaryClass = {astro-ph.GA},
       adsurl = {https://ui.adsabs.harvard.edu/abs/2019MNRAS.482.1417B}
}

@ARTICLE{2003ApJ...585L.125B,
       author = {{Brook}, Chris B. and {Kawata}, Daisuke and {Gibson}, Brad K. and {Flynn}, Chris},
        title = "{Galactic Halo Stars in Phase Space: A Hint of Satellite Accretion?}",
      journal = {\apjl},
         year = 2003,
        month = mar,
       volume = {585},
       number = {2},
        pages = {L125-L129},
          doi = {10.1086/374306},
archivePrefix = {arXiv},
       eprint = {astro-ph/0301596},
 primaryClass = {astro-ph},
       adsurl = {https://ui.adsabs.harvard.edu/abs/2003ApJ...585L.125B}
}

@ARTICLE{2021A&A...647A..59G,
       author = {{GRAVITY Collaboration} and {Abuter}, R. and {Amorim}, A. and {Baub{\"o}ck}, M. and {Berger}, J.~P. and {Bonnet}, H. and {Brandner}, W. and {Cl{\'e}net}, Y. and {Davies}, R. and {de Zeeuw}, P.~T. and {Dexter}, J. and {Dallilar}, Y. and {Drescher}, A. and {Eckart}, A. and {Eisenhauer}, F. and {F{\"o}rster Schreiber}, N.~M. and {Garcia}, P. and {Gao}, F. and {Gendron}, E. and {Genzel}, R. and {Gillessen}, S. and {Habibi}, M. and {Haubois}, X. and {Hei{\ss}el}, G. and {Henning}, T. and {Hippler}, S. and {Horrobin}, M. and {Jim{\'e}nez-Rosales}, A. and {Jochum}, L. and {Jocou}, L. and {Kaufer}, A. and {Kervella}, P. and {Lacour}, S. and {Lapeyr{\`e}re}, V. and {Le Bouquin}, J. -B. and {L{\'e}na}, P. and {Lutz}, D. and {Nowak}, M. and {Ott}, T. and {Paumard}, T. and {Perraut}, K. and {Perrin}, G. and {Pfuhl}, O. and {Rabien}, S. and {Rodr{\'\i}guez-Coira}, G. and {Shangguan}, J. and {Shimizu}, T. and {Scheithauer}, S. and {Stadler}, J. and {Straub}, O. and {Straubmeier}, C. and {Sturm}, E. and {Tacconi}, L.~J. and {Vincent}, F. and {von Fellenberg}, S. and {Waisberg}, I. and {Widmann}, F. and {Wieprecht}, E. and {Wiezorrek}, E. and {Woillez}, J. and {Yazici}, S. and {Young}, A. and {Zins}, G.},
        title = "{Improved GRAVITY astrometric accuracy from modeling optical aberrations}",
      journal = {\aap},
         year = 2021,
        month = mar,
       volume = {647},
          eid = {A59},
        pages = {A59},
          doi = {10.1051/0004-6361/202040208},
archivePrefix = {arXiv},
       eprint = {2101.12098},
 primaryClass = {astro-ph.GA},
       adsurl = {https://ui.adsabs.harvard.edu/abs/2021A&A...647A..59G}
}

@ARTICLE{2019MNRAS.482.1525V,
       author = {{Vasiliev}, Eugene},
        title = "{AGAMA: action-based galaxy modelling architecture}",
      journal = {\mnras},
         year = 2019,
        month = jan,
       volume = {482},
       number = {2},
        pages = {1525-1544},
          doi = {10.1093/mnras/sty2672},
archivePrefix = {arXiv},
       eprint = {1802.08239},
 primaryClass = {astro-ph.GA},
       adsurl = {https://ui.adsabs.harvard.edu/abs/2019MNRAS.482.1525V}
}

@ARTICLE{2021A&A...645A..85M,
       author = {{Miglio}, A. and {Chiappini}, C. and {Mackereth}, J.~T. and {Davies}, G.~R. and {Brogaard}, K. and {Casagrande}, L. and {Chaplin}, W.~J. and {Girardi}, L. and {Kawata}, D. and {Khan}, S. and {Izzard}, R. and {Montalb{\'a}n}, J. and {Mosser}, B. and {Vincenzo}, F. and {Bossini}, D. and {Noels}, A. and {Rodrigues}, T. and {Valentini}, M. and {Mandel}, I.},
        title = "{Age dissection of the Milky Way discs: Red giants in the Kepler field}",
      journal = {\aap},
         year = 2021,
        month = jan,
       volume = {645},
          eid = {A85},
        pages = {A85},
          doi = {10.1051/0004-6361/202038307},
archivePrefix = {arXiv},
       eprint = {2004.14806},
 primaryClass = {astro-ph.GA},
       adsurl = {https://ui.adsabs.harvard.edu/abs/2021A&A...645A..85M}
}

@ARTICLE{2020MNRAS.497.1603G,
       author = {{Grand}, Robert J.~J. and {Kawata}, Daisuke and {Belokurov}, Vasily and {Deason}, Alis J. and {Fattahi}, Azadeh and {Fragkoudi}, Francesca and {G{\'o}mez}, Facundo A. and {Marinacci}, Federico and {Pakmor}, R{\"u}diger},
        title = "{The dual origin of the Galactic thick disc and halo from the gas-rich Gaia-Enceladus Sausage merger}",
      journal = {\mnras},
         year = 2020,
        month = sep,
       volume = {497},
       number = {2},
        pages = {1603-1618},
          doi = {10.1093/mnras/staa2057},
archivePrefix = {arXiv},
       eprint = {2001.06009},
 primaryClass = {astro-ph.GA},
       adsurl = {https://ui.adsabs.harvard.edu/abs/2020MNRAS.497.1603G}
}

@ARTICLE{2019NatAs...3..932G,
       author = {{Gallart}, Carme and {Bernard}, Edouard J. and {Brook}, Chris B. and {Ruiz-Lara}, Tom{\'a}s and {Cassisi}, Santi and {Hill}, Vanessa and {Monelli}, Matteo},
        title = "{Uncovering the birth of the Milky Way through accurate stellar ages with Gaia}",
      journal = {Nature Astronomy},
         year = 2019,
        month = jul,
       volume = {3},
        pages = {932-939},
          doi = {10.1038/s41550-019-0829-5},
archivePrefix = {arXiv},
       eprint = {1901.02900},
 primaryClass = {astro-ph.GA},
       adsurl = {https://ui.adsabs.harvard.edu/abs/2019NatAs...3..932G}
}

@ARTICLE{2021A&A...649A...2L,
       author = {{Lindegren}, L. and {Klioner}, S.~A. and {Hern{\'a}ndez}, J. and {Bombrun}, A. and {Ramos-Lerate}, M. and {Steidelm{\"u}ller}, H. and {Bastian}, U. and {Biermann}, M. and {de Torres}, A. and {Gerlach}, E. and {Geyer}, R. and {Hilger}, T. and {Hobbs}, D. and {Lammers}, U. and {McMillan}, P.~J. and {Stephenson}, C.~A. and {Casta{\~n}eda}, J. and {Davidson}, M. and {Fabricius}, C. and {Gracia-Abril}, G. and {Portell}, J. and {Rowell}, N. and {Teyssier}, D. and {Torra}, F. and {Bartolom{\'e}}, S. and {Clotet}, M. and {Garralda}, N. and {Gonz{\'a}lez-Vidal}, J.~J. and {Torra}, J. and {Abbas}, U. and {Altmann}, M. and {Anglada Varela}, E. and {Balaguer-N{\'u}{\~n}ez}, L. and {Balog}, Z. and {Barache}, C. and {Becciani}, U. and {Bernet}, M. and {Bertone}, S. and {Bianchi}, L. and {Bouquillon}, S. and {Brown}, A.~G.~A. and {Bucciarelli}, B. and {Busonero}, D. and {Butkevich}, A.~G. and {Buzzi}, R. and {Cancelliere}, R. and {Carlucci}, T. and {Charlot}, P. and {Cioni}, M. -R.~L. and {Crosta}, M. and {Crowley}, C. and {del Peloso}, E.~F. and {del Pozo}, E. and {Drimmel}, R. and {Esquej}, P. and {Fienga}, A. and {Fraile}, E. and {Gai}, M. and {Garcia-Reinaldos}, M. and {Guerra}, R. and {Hambly}, N.~C. and {Hauser}, M. and {Jan{\ss}en}, K. and {Jordan}, S. and {Kostrzewa-Rutkowska}, Z. and {Lattanzi}, M.~G. and {Liao}, S. and {Licata}, E. and {Lister}, T.~A. and {L{\"o}ffler}, W. and {Marchant}, J.~M. and {Masip}, A. and {Mignard}, F. and {Mints}, A. and {Molina}, D. and {Mora}, A. and {Morbidelli}, R. and {Murphy}, C.~P. and {Pagani}, C. and {Panuzzo}, P. and {Pe{\~n}alosa Esteller}, X. and {Poggio}, E. and {Re Fiorentin}, P. and {Riva}, A. and {Sagrist{\`a} Sell{\'e}s}, A. and {Sanchez Gimenez}, V. and {Sarasso}, M. and {Sciacca}, E. and {Siddiqui}, H.~I. and {Smart}, R.~L. and {Souami}, D. and {Spagna}, A. and {Steele}, I.~A. and {Taris}, F. and {Utrilla}, E. and {van Reeven}, W. and {Vecchiato}, A.},
        title = "{Gaia Early Data Release 3. The astrometric solution}",
      journal = {\aap},
         year = 2021,
        month = may,
       volume = {649},
          eid = {A2},
        pages = {A2},
          doi = {10.1051/0004-6361/202039709},
archivePrefix = {arXiv},
       eprint = {2012.03380},
 primaryClass = {astro-ph.IM},
       adsurl = {https://ui.adsabs.harvard.edu/abs/2021A&A...649A...2L}
}

@ARTICLE{2018ApJ...863..113H,
       author = {{Haywood}, M. and {Di Matteo}, P. and {Lehnert}, M.~D. and {Snaith}, O. and {Khoperskov}, S. and {G{\'o}mez}, A.},
        title = "{In Disguise or Out of Reach: First Clues about In Situ and Accreted Stars in the Stellar Halo of the Milky Way from Gaia DR2}",
      journal = {\apj},
         year = 2018,
        month = aug,
       volume = {863},
       number = {2},
          eid = {113},
        pages = {113},
          doi = {10.3847/1538-4357/aad235},
archivePrefix = {arXiv},
       eprint = {1805.02617},
 primaryClass = {astro-ph.GA},
       adsurl = {https://ui.adsabs.harvard.edu/abs/2018ApJ...863..113H}
}

@ARTICLE{2023MNRAS.521.1462Z,
       author = {{Zhang}, Hanyuan and {Sanders}, Jason L.},
        title = "{A kinematic calibration of the O-rich Mira variable period-age relation from Gaia}",
      journal = {\mnras},
         year = 2023,
        month = may,
       volume = {521},
       number = {1},
        pages = {1462-1478},
          doi = {10.1093/mnras/stad575},
archivePrefix = {arXiv},
       eprint = {2302.10024},
 primaryClass = {astro-ph.SR},
       adsurl = {https://ui.adsabs.harvard.edu/abs/2023MNRAS.521.1462Z}
}

@ARTICLE{2018arXiv180208255V,
       author = {{Vasiliev}, Eugene},
        title = "{Agama reference documentation}",
      journal = {arXiv e-prints},
         year = 2018,
        month = feb,
          eid = {arXiv:1802.08255},
        pages = {arXiv:1802.08255},
          doi = {10.48550/arXiv.1802.08255},
archivePrefix = {arXiv},
       eprint = {1802.08255},
 primaryClass = {astro-ph.IM},
       adsurl = {https://ui.adsabs.harvard.edu/abs/2018arXiv180208255V}
}

@ARTICLE{2023A&A...674A..37G,
       author = {{Gaia Collaboration} and {Drimmel}, R. and {Romero-G{\'o}mez}, M. and {Chemin}, L. and {Ramos}, P. and {Poggio}, E. and {Ripepi}, V. and {Andrae}, R. and {Blomme}, R. and {Cantat-Gaudin}, T. and {Castro-Ginard}, A. and {Clementini}, G. and {Figueras}, F. and {Fouesneau}, M. and {Fr{\'e}mat}, Y. and {Jardine}, K. and {Khanna}, S. and {Lobel}, A. and {Marshall}, D.~J. and {Muraveva}, T. and {Brown}, A.~G.~A. and {Vallenari}, A. and {Prusti}, T. and {de Bruijne}, J.~H.~J. and {Arenou}, F. and {Babusiaux}, C. and {Biermann}, M. and {Creevey}, O.~L. and {Ducourant}, C. and {Evans}, D.~W. and {Eyer}, L. and {Guerra}, R. and {Hutton}, A. and {Jordi}, C. and {Klioner}, S.~A. and {Lammers}, U.~L. and {Lindegren}, L. and {Luri}, X. and {Mignard}, F. and {Panem}, C. and {Pourbaix}, D. and {Randich}, S. and {Sartoretti}, P. and {Soubiran}, C. and {Tanga}, P. and {Walton}, N.~A. and {Bailer-Jones}, C.~A.~L. and {Bastian}, U. and {Jansen}, F. and {Katz}, D. and {Lattanzi}, M.~G. and {van Leeuwen}, F. and {Bakker}, J. and {Cacciari}, C. and {Casta{\~n}eda}, J. and {De Angeli}, F. and {Fabricius}, C. and {Galluccio}, L. and {Guerrier}, A. and {Heiter}, U. and {Masana}, E. and {Messineo}, R. and {Mowlavi}, N. and {Nicolas}, C. and {Nienartowicz}, K. and {Pailler}, F. and {Panuzzo}, P. and {Riclet}, F. and {Roux}, W. and {Seabroke}, G.~M. and {Sordo}, R. and {Th{\'e}venin}, F. and {Gracia-Abril}, G. and {Portell}, J. and {Teyssier}, D. and {Altmann}, M. and {Audard}, M. and {Bellas-Velidis}, I. and {Benson}, K. and {Berthier}, J. and {Burgess}, P.~W. and {Busonero}, D. and {Busso}, G. and {C{\'a}novas}, H. and {Carry}, B. and {Cellino}, A. and {Cheek}, N. and {Damerdji}, Y. and {Davidson}, M. and {de Teodoro}, P. and {Nu{\~n}ez Campos}, M. and {Delchambre}, L. and {Dell'Oro}, A. and {Esquej}, P. and {Fern{\'a}ndez-Hern{\'a}ndez}, J. and {Fraile}, E. and {Garabato}, D. and {Garc{\'\i}a-Lario}, P. and {Gosset}, E. and {Haigron}, R. and {Halbwachs}, J. -L. and {Hambly}, N.~C. and {Harrison}, D.~L. and {Hern{\'a}ndez}, J. and {Hestroffer}, D. and {Hodgkin}, S.~T. and {Holl}, B. and {Jan{\ss}en}, K. and {Jevardat de Fombelle}, G. and {Jordan}, S. and {Krone-Martins}, A. and {Lanzafame}, A.~C. and {L{\"o}ffler}, W. and {Marchal}, O. and {Marrese}, P.~M. and {Moitinho}, A. and {Muinonen}, K. and {Osborne}, P. and {Pancino}, E. and {Pauwels}, T. and {Recio-Blanco}, A. and {Reyl{\'e}}, C. and {Riello}, M. and {Rimoldini}, L. and {Roegiers}, T. and {Rybizki}, J. and {Sarro}, L.~M. and {Siopis}, C. and {Smith}, M. and {Sozzetti}, A. and {Utrilla}, E. and {van Leeuwen}, M. and {Abbas}, U. and {{\'A}brah{\'a}m}, P. and {Abreu Aramburu}, A. and {Aerts}, C. and {Aguado}, J.~J. and {Ajaj}, M. and {Aldea-Montero}, F. and {Altavilla}, G. and {{\'A}lvarez}, M.~A. and {Alves}, J. and {Anders}, F. and {Anderson}, R.~I. and {Anglada Varela}, E. and {Antoja}, T. and {Baines}, D. and {Baker}, S.~G. and {Balaguer-N{\'u}{\~n}ez}, L. and {Balbinot}, E. and {Balog}, Z. and {Barache}, C. and {Barbato}, D. and {Barros}, M. and {Barstow}, M.~A. and {Bartolom{\'e}}, S. and {Bassilana}, J. -L. and {Bauchet}, N. and {Becciani}, U. and {Bellazzini}, M. and {Berihuete}, A. and {Bernet}, M. and {Bertone}, S. and {Bianchi}, L. and {Binnenfeld}, A. and {Blanco-Cuaresma}, S. and {Boch}, T. and {Bombrun}, A. and {Bossini}, D. and {Bouquillon}, S. and {Bragaglia}, A. and {Bramante}, L. and {Breedt}, E. and {Bressan}, A. and {Brouillet}, N. and {Brugaletta}, E. and {Bucciarelli}, B. and {Burlacu}, A. and {Butkevich}, A.~G. and {Buzzi}, R. and {Caffau}, E. and {Cancelliere}, R. and {Carballo}, R. and {Carlucci}, T. and {Carnerero}, M.~I. and {Carrasco}, J.~M. and {Casamiquela}, L. and {Castellani}, M. and {Chaoul}, L. and {Charlot}, P. and {Chiaramida}, V. and {Chiavassa}, A. and {Chornay}, N. and {Comoretto}, G. and {Contursi}, G. and {Cooper}, W.~J. and {Cornez}, T. and {Cowell}, S. and {Crifo}, F. and {Cropper}, M. and {Crosta}, M. and {Crowley}, C. and {Dafonte}, C. and {Dapergolas}, A. and {David}, P. and {de Laverny}, P. and {De Luise}, F. and {De March}, R. and {De Ridder}, J. and {de Souza}, R. and {de Torres}, A. and {del Peloso}, E.~F. and {del Pozo}, E. and {Delbo}, M. and {Delgado}, A. and {Delisle}, J. -B. and {Demouchy}, C. and {Dharmawardena}, T.~E. and {Di Matteo}, P. and {Diakite}, S. and {Diener}, C. and {Distefano}, E. and {Dolding}, C. and {Enke}, H. and {Fabre}, C. and {Fabrizio}, M. and {Faigler}, S. and {Fedorets}, G. and {Fernique}, P. and {Fournier}, Y. and {Fouron}, C. and {Fragkoudi}, F. and {Gai}, M. and {Garcia-Gutierrez}, A. and {Garcia-Reinaldos}, M. and {Garc{\'\i}a-Torres}, M. and {Garofalo}, A. and {Gavel}, A. and {Gavras}, P. and {Gerlach}, E. and {Geyer}, R. and {Giacobbe}, P. and {Gilmore}, G. and {Girona}, S. and {Giuffrida}, G. and {Gomel}, R. and {Gomez}, A. and {Gonz{\'a}lez-N{\'u}{\~n}ez}, J. and {Gonz{\'a}lez-Santamar{\'\i}a}, I. and {Gonz{\'a}lez-Vidal}, J.~J. and {Granvik}, M. and {Guillout}, P. and {Guiraud}, J. and {Guti{\'e}rrez-S{\'a}nchez}, R. and {Guy}, L.~P. and {Hatzidimitriou}, D. and {Hauser}, M. and {Haywood}, M. and {Helmer}, A. and {Helmi}, A. and {Sarmiento}, M.~H. and {Hidalgo}, S.~L. and {H{\l}adczuk}, N. and {Hobbs}, D. and {Holland}, G. and {Huckle}, H.~E. and {Jasniewicz}, G. and {Jean-Antoine Piccolo}, A. and {Jim{\'e}nez-Arranz}, {\'O}. and {Juaristi Campillo}, J. and {Julbe}, F. and {Karbevska}, L. and {Kervella}, P. and {Kordopatis}, G. and {Korn}, A.~J. and {K{\'o}sp{\'a}l}, {\'A}. and {Kostrzewa-Rutkowska}, Z. and {Kruszy{\'n}ska}, K. and {Kun}, M. and {Laizeau}, P. and {Lambert}, S. and {Lanza}, A.~F. and {Lasne}, Y. and {Le Campion}, J. -F. and {Lebreton}, Y. and {Lebzelter}, T. and {Leccia}, S. and {Leclerc}, N. and {Lecoeur-Taibi}, I. and {Liao}, S. and {Licata}, E.~L. and {Lindstr{\o}m}, H.~E.~P. and {Lister}, T.~A. and {Livanou}, E. and {Lorca}, A. and {Loup}, C. and {Madrero Pardo}, P. and {Magdaleno Romeo}, A. and {Managau}, S. and {Mann}, R.~G. and {Manteiga}, M. and {Marchant}, J.~M. and {Marconi}, M. and {Marcos}, J. and {Marcos Santos}, M.~M.~S. and {Mar{\'\i}n Pina}, D. and {Marinoni}, S. and {Marocco}, F. and {Martin Polo}, L. and {Mart{\'\i}n-Fleitas}, J.~M. and {Marton}, G. and {Mary}, N. and {Masip}, A. and {Massari}, D. and {Mastrobuono-Battisti}, A. and {Mazeh}, T. and {McMillan}, P.~J. and {Messina}, S. and {Michalik}, D. and {Millar}, N.~R. and {Mints}, A. and {Molina}, D. and {Molinaro}, R. and {Moln{\'a}r}, L. and {Monari}, G. and {Mongui{\'o}}, M. and {Montegriffo}, P. and {Montero}, A. and {Mor}, R. and {Mora}, A. and {Morbidelli}, R. and {Morel}, T. and {Morris}, D. and {Murphy}, C.~P. and {Musella}, I. and {Nagy}, Z. and {Noval}, L. and {Oca{\~n}a}, F. and {Ogden}, A. and {Ordenovic}, C. and {Osinde}, J.~O. and {Pagani}, C. and {Pagano}, I. and {Palaversa}, L. and {Palicio}, P.~A. and {Pallas-Quintela}, L. and {Panahi}, A. and {Payne-Wardenaar}, S. and {Pe{\~n}alosa Esteller}, X. and {Penttil{\"a}}, A. and {Pichon}, B. and {Piersimoni}, A.~M. and {Pineau}, F. -X. and {Plachy}, E. and {Plum}, G. and {Pr{\v{s}}a}, A. and {Pulone}, L. and {Racero}, E. and {Ragaini}, S. and {Rainer}, M. and {Raiteri}, C.~M. and {Ramos-Lerate}, M. and {Re Fiorentin}, P. and {Regibo}, S. and {Richards}, P.~J. and {Rios Diaz}, C. and {Riva}, A. and {Rix}, H. -W. and {Rixon}, G. and {Robichon}, N. and {Robin}, A.~C. and {Robin}, C. and {Roelens}, M. and {Rogues}, H.~R.~O. and {Rohrbasser}, L. and {Rowell}, N. and {Royer}, F. and {Ruz Mieres}, D. and {Rybicki}, K.~A. and {Sadowski}, G. and {S{\'a}ez N{\'u}{\~n}ez}, A. and {Sagrist{\`a} Sell{\'e}s}, A. and {Sahlmann}, J. and {Salguero}, E. and {Samaras}, N. and {Sanchez Gimenez}, V. and {Sanna}, N. and {Santove{\~n}a}, R. and {Sarasso}, M. and {Schultheis}, M.~S. and {Sciacca}, E. and {Segol}, M. and {Segovia}, J.~C. and {S{\'e}gransan}, D. and {Semeux}, D. and {Shahaf}, S. and {Siddiqui}, H.~I. and {Siebert}, A. and {Siltala}, L. and {Silvelo}, A. and {Slezak}, E. and {Slezak}, I. and {Smart}, R.~L. and {Snaith}, O.~N. and {Solano}, E. and {Solitro}, F. and {Souami}, D. and {Souchay}, J. and {Spagna}, A. and {Spina}, L. and {Spoto}, F. and {Steele}, I.~A. and {Steidelm{\"u}ller}, H. and {Stephenson}, C.~A. and {S{\"u}veges}, M. and {Surdej}, J. and {Szabados}, L. and {Szegedi-Elek}, E. and {Taris}, F. and {Taylor}, M.~B. and {Teixeira}, R. and {Tolomei}, L. and {Tonello}, N. and {Torra}, F. and {Torra}, J. and {Torralba Elipe}, G. and {Trabucchi}, M. and {Tsounis}, A.~T. and {Turon}, C. and {Ulla}, A. and {Unger}, N. and {Vaillant}, M.~V. and {van Dillen}, E. and {van Reeven}, W. and {Vanel}, O. and {Vecchiato}, A. and {Viala}, Y. and {Vicente}, D. and {Voutsinas}, S. and {Weiler}, M. and {Wevers}, T. and {Wyrzykowski}, {\L}. and {Yoldas}, A. and {Yvard}, P. and {Zhao}, H. and {Zorec}, J. and {Zucker}, S. and {Zwitter}, T.},
        title = "{Gaia Data Release 3. Mapping the asymmetric disc of the Milky Way}",
      journal = {\aap},
         year = 2023,
        month = jun,
       volume = {674},
          eid = {A37},
        pages = {A37},
          doi = {10.1051/0004-6361/202243797},
archivePrefix = {arXiv},
       eprint = {2206.06207},
 primaryClass = {astro-ph.GA},
       adsurl = {https://ui.adsabs.harvard.edu/abs/2023A&A...674A..37G}
}

@ARTICLE{2021MNRAS.503.2814C,
       author = {{Ciuc{\u{a}}}, Ioana and {Kawata}, Daisuke and {Miglio}, Andrea and {Davies}, Guy R. and {Grand}, Robert J.~J.},
        title = "{Unveiling the distinct formation pathways of the inner and outer discs of the Milky Way with Bayesian Machine Learning}",
      journal = {\mnras},
         year = 2021,
        month = may,
       volume = {503},
       number = {2},
        pages = {2814-2824},
          doi = {10.1093/mnras/stab639},
archivePrefix = {arXiv},
       eprint = {2003.03316},
 primaryClass = {astro-ph.GA},
       adsurl = {https://ui.adsabs.harvard.edu/abs/2021MNRAS.503.2814C}
}

@ARTICLE{2024MNRAS.528L.122C,
       author = {{Ciuc{\u{a}}}, Ioana and {Kawata}, Daisuke and {Ting}, Yuan-Sen and {Grand}, Robert J.~J. and {Miglio}, Andrea and {Hayden}, Michael and {Baba}, Junichi and {Fragkoudi}, Francesca and {Monty}, Stephanie and {Buder}, Sven and {Freeman}, Ken},
        title = "{Chasing the impact of the Gaia-Sausage-Enceladus merger on the formation of the Milky Way thick disc}",
      journal = {\mnras},
         year = 2024,
        month = feb,
       volume = {528},
       number = {1},
        pages = {L122-L126},
          doi = {10.1093/mnrasl/slad033},
archivePrefix = {arXiv},
       eprint = {2211.01006},
 primaryClass = {astro-ph.GA},
       adsurl = {https://ui.adsabs.harvard.edu/abs/2024MNRAS.528L.122C}
}

@ARTICLE{2011A&A...528A.120C,
       author = {{Chiavassa}, A. and {Pasquato}, E. and {Jorissen}, A. and {Sacuto}, S. and {Babusiaux}, C. and {Freytag}, B. and {Ludwig}, H.-G. and {Cruzal{\`e}bes}, P. and {Rabbia}, Y. and {Spang}, A. and {Chesneau}, O.},
        title = "{Radiative hydrodynamic simulations of red supergiant stars. III. Spectro-photocentric variability, photometric variability, and consequences on Gaia measurements}",
      journal = {\aap},
         year = 2011,
        month = apr,
       volume = {528},
          eid = {A120},
        pages = {A120},
          doi = {10.1051/0004-6361/201015768},
archivePrefix = {arXiv},
       eprint = {1012.5234},
 primaryClass = {astro-ph.SR},
       adsurl = {https://ui.adsabs.harvard.edu/abs/2011A&A...528A.120C}
}

@ARTICLE{2018A&A...617L...1C,
       author = {{Chiavassa}, A. and {Freytag}, B. and {Schultheis}, M.},
        title = "{Heading Gaia to measure atmospheric dynamics in AGB stars}",
      journal = {\aap},
         year = 2018,
        month = sep,
       volume = {617},
          eid = {L1},
        pages = {L1},
          doi = {10.1051/0004-6361/201833844},
archivePrefix = {arXiv},
       eprint = {1808.02548},
 primaryClass = {astro-ph.SR},
       adsurl = {https://ui.adsabs.harvard.edu/abs/2018A&A...617L...1C}
}

@ARTICLE{2017A&A...600A.137F,
       author = {{Freytag}, B. and {Liljegren}, S. and {H{\"o}fner}, S.},
        title = "{Global 3D radiation-hydrodynamics models of AGB stars. Effects of convection and radial pulsations on atmospheric structures}",
      journal = {\aap},
         year = 2017,
        month = apr,
       volume = {600},
          eid = {A137},
        pages = {A137},
          doi = {10.1051/0004-6361/201629594},
archivePrefix = {arXiv},
       eprint = {1702.05433},
 primaryClass = {astro-ph.SR},
       adsurl = {https://ui.adsabs.harvard.edu/abs/2017A&A...600A.137F}
}

@ARTICLE{2003A&A...399.1167P,
       author = {{Pourbaix}, D. and {Platais}, I. and {Detournay}, S. and {Jorissen}, A. and {Knapp}, G. and {Makarov}, V.~V.},
        title = "{How many Hipparcos Variability-Induced Movers are genuine binaries?}",
      journal = {\aap},
         year = 2003,
        month = mar,
       volume = {399},
        pages = {1167-1175},
          doi = {10.1051/0004-6361:20021843},
archivePrefix = {arXiv},
       eprint = {astro-ph/0212201},
 primaryClass = {astro-ph},
       adsurl = {https://ui.adsabs.harvard.edu/abs/2003A&A...399.1167P}
}

@ARTICLE{2021A&A...649A..11R,
       author = {{Rowell}, N. and {Davidson}, M. and {Lindegren}, L. and {van Leeuwen}, F. and {Casta{\~n}eda}, J. and {Fabricius}, C. and {Bastian}, U. and {Hambly}, N.~C. and {Hern{\'a}ndez}, J. and {Bombrun}, A. and {Evans}, D.~W. and {De Angeli}, F. and {Riello}, M. and {Busonero}, D. and {Crowley}, C. and {Mora}, A. and {Lammers}, U. and {Gracia}, G. and {Portell}, J. and {Biermann}, M. and {Brown}, A.~G.~A.},
        title = "{Gaia Early Data Release 3. Modelling and calibration of Gaia's point and line spread functions}",
      journal = {\aap},
         year = 2021,
        month = may,
       volume = {649},
          eid = {A11},
        pages = {A11},
          doi = {10.1051/0004-6361/202039448},
archivePrefix = {arXiv},
       eprint = {2012.02069},
 primaryClass = {astro-ph.IM},
       adsurl = {https://ui.adsabs.harvard.edu/abs/2021A&A...649A..11R}
}

@ARTICLE{2019MNRAS.483.3022G,
       author = {{Grady}, J. and {Belokurov}, V. and {Evans}, N.~W.},
        title = "{Age gradients throughout the Galaxy with long-period variables}",
      journal = {\mnras},
         year = 2019,
        month = mar,
       volume = {483},
       number = {3},
        pages = {3022-3035},
          doi = {10.1093/mnras/sty3284},
archivePrefix = {arXiv},
       eprint = {1804.09186},
 primaryClass = {astro-ph.GA},
       adsurl = {https://ui.adsabs.harvard.edu/abs/2019MNRAS.483.3022G}
}

@ARTICLE{2019ApJ...877..116W,
       author = {{Wang}, Shu and {Chen}, Xiaodian},
        title = "{The Optical to Mid-infrared Extinction Law Based on the APOGEE, Gaia DR2, Pan-STARRS1, SDSS, APASS, 2MASS, and WISE Surveys}",
      journal = {\apj},
         year = 2019,
        month = jun,
       volume = {877},
       number = {2},
          eid = {116},
        pages = {116},
          doi = {10.3847/1538-4357/ab1c61},
archivePrefix = {arXiv},
       eprint = {1904.04575},
 primaryClass = {astro-ph.GA},
       adsurl = {https://ui.adsabs.harvard.edu/abs/2019ApJ...877..116W}
}

@ARTICLE{2009MNRAS.399.1709M,
       author = {{Matsunaga}, Noriyuki and {Kawadu}, Takahiro and {Nishiyama}, Shogo and {Nagayama}, Takahiro and {Hatano}, Hirofumi and {Tamura}, Motohide and {Glass}, Ian S. and {Nagata}, Tetsuya},
        title = "{A near-infrared survey of Miras and the distance to the Galactic Centre}",
      journal = {\mnras},
         year = 2009,
        month = nov,
       volume = {399},
       number = {4},
        pages = {1709-1729},
          doi = {10.1111/j.1365-2966.2009.15393.x},
archivePrefix = {arXiv},
       eprint = {0907.2761},
 primaryClass = {astro-ph.SR},
       adsurl = {https://ui.adsabs.harvard.edu/abs/2009MNRAS.399.1709M}
}

@ARTICLE{2006AJ....131.1163S,
       author = {{Skrutskie}, M.~F. and {Cutri}, R.~M. and {Stiening}, R. and {Weinberg}, M.~D. and {Schneider}, S. and {Carpenter}, J.~M. and {Beichman}, C. and {Capps}, R. and {Chester}, T. and {Elias}, J. and {Huchra}, J. and {Liebert}, J. and {Lonsdale}, C. and {Monet}, D.~G. and {Price}, S. and {Seitzer}, P. and {Jarrett}, T. and {Kirkpatrick}, J.~D. and {Gizis}, J.~E. and {Howard}, E. and {Evans}, T. and {Fowler}, J. and {Fullmer}, L. and {Hurt}, R. and {Light}, R. and {Kopan}, E.~L. and {Marsh}, K.~A. and {McCallon}, H.~L. and {Tam}, R. and {Van Dyk}, S. and {Wheelock}, S.},
        title = "{The Two Micron All Sky Survey (2MASS)}",
      journal = {\aj},
         year = 2006,
        month = feb,
       volume = {131},
       number = {2},
        pages = {1163-1183},
          doi = {10.1086/498708},
       adsurl = {https://ui.adsabs.harvard.edu/abs/2006AJ....131.1163S}
}

@ARTICLE{2023A&A...674A..15L,
       author = {{Lebzelter}, T. and {Mowlavi}, N. and {Lecoeur-Taibi}, I. and {Trabucchi}, M. and {Audard}, M. and {Garc{\'\i}a-Lario}, P. and {Gavras}, P. and {Holl}, B. and {Jevardat de Fombelle}, G. and {Nienartowicz}, K. and {Rimoldini}, L. and {Eyer}, L.},
        title = "{Gaia Data Release 3. The second Gaia catalogue of long-period variable candidates}",
      journal = {\aap},
         year = 2023,
        month = jun,
       volume = {674},
          eid = {A15},
        pages = {A15},
          doi = {10.1051/0004-6361/202244241},
archivePrefix = {arXiv},
       eprint = {2206.05745},
 primaryClass = {astro-ph.SR},
       adsurl = {https://ui.adsabs.harvard.edu/abs/2023A&A...674A..15L}
}

@ARTICLE{2025MNRAS.543.2275F,
       author = {{Funakoshi}, Natsuki and {Kawata}, Daisuke and {Sanders}, Jason L. and {Ciuc{\u{a}}}, Ioana and {Grand}, Robert J.~J. and {Zh{\={a}}ng}, HanYuan},
        title = "{Thick-to-thin disc transition and gas disc shrinking induced by the Gaia-Sausage-Enceladus merger}",
      journal = {\mnras},
         year = 2025,
        month = nov,
       volume = {543},
       number = {3},
        pages = {2275-2291},
          doi = {10.1093/mnras/staf1632},
archivePrefix = {arXiv},
       eprint = {2507.22979},
 primaryClass = {astro-ph.GA},
       adsurl = {https://ui.adsabs.harvard.edu/abs/2025MNRAS.543.2275F}
}

@ARTICLE{2013PASP..125..306F,
       author = {{Foreman-Mackey}, Daniel and {Hogg}, David W. and {Lang}, Dustin and {Goodman}, Jonathan},
        title = "{emcee: The MCMC Hammer}",
      journal = {\pasp},
         year = 2013,
        month = mar,
       volume = {125},
       number = {925},
        pages = {306},
          doi = {10.1086/670067},
archivePrefix = {arXiv},
       eprint = {1202.3665},
 primaryClass = {astro-ph.IM},
       adsurl = {https://ui.adsabs.harvard.edu/abs/2013PASP..125..306F}
}

@ARTICLE{2022MNRAS.512.2171C,
       author = {{Clarke}, Jonathan P. and {Gerhard}, Ortwin},
        title = "{The pattern speed of the Milky Way bar/bulge from VIRAC and Gaia}",
      journal = {\mnras},
         year = 2022,
        month = may,
       volume = {512},
       number = {2},
        pages = {2171-2188},
          doi = {10.1093/mnras/stac603},
archivePrefix = {arXiv},
       eprint = {2107.10875},
 primaryClass = {astro-ph.GA},
       adsurl = {https://ui.adsabs.harvard.edu/abs/2022MNRAS.512.2171C}
}

@ARTICLE{2017MNRAS.465.1621P,
       author = {{Portail}, Matthieu and {Gerhard}, Ortwin and {Wegg}, Christopher and {Ness}, Melissa},
        title = "{Dynamical modelling of the galactic bulge and bar: the Milky Way's pattern speed, stellar and dark matter mass distribution}",
      journal = {\mnras},
         year = 2017,
        month = feb,
       volume = {465},
       number = {2},
        pages = {1621-1644},
          doi = {10.1093/mnras/stw2819},
archivePrefix = {arXiv},
       eprint = {1608.07954},
 primaryClass = {astro-ph.GA},
       adsurl = {https://ui.adsabs.harvard.edu/abs/2017MNRAS.465.1621P}
}

@ARTICLE{2024MNRAS.530.2972S,
       author = {{Sanders}, Jason L. and {Kawata}, Daisuke and {Matsunaga}, Noriyuki and {Sormani}, Mattia C. and {Smith}, Leigh C. and {Minniti}, Dante and {Gerhard}, Ortwin},
        title = "{The epoch of the Milky Way's bar formation: dynamical modelling of Mira variables in the nuclear stellar disc}",
      journal = {\mnras},
         year = 2024,
        month = may,
       volume = {530},
       number = {3},
        pages = {2972-2993},
          doi = {10.1093/mnras/stae711},
archivePrefix = {arXiv},
       eprint = {2311.00035},
 primaryClass = {astro-ph.GA},
       adsurl = {https://ui.adsabs.harvard.edu/abs/2024MNRAS.530.2972S}
}

@ARTICLE{1971ApJ...168..343H,
       author = {{Hohl}, Frank},
        title = "{Numerical Experiments with a Disk of Stars}",
      journal = {\apj},
         year = 1971,
        month = sep,
       volume = {168},
        pages = {343},
          doi = {10.1086/151091},
       adsurl = {https://ui.adsabs.harvard.edu/abs/1971ApJ...168..343H}
}

@ARTICLE{2018MNRAS.473.2608Z,
       author = {{Zana}, Tommaso and {Dotti}, Massimo and {Capelo}, Pedro R. and {Bonoli}, Silvia and {Haardt}, Francesco and {Mayer}, Lucio and {Spinoso}, Daniele},
        title = "{External versus internal triggers of bar formation in cosmological zoom-in simulations}",
      journal = {\mnras},
         year = 2018,
        month = jan,
       volume = {473},
       number = {2},
        pages = {2608-2621},
          doi = {10.1093/mnras/stx2503},
archivePrefix = {arXiv},
       eprint = {1705.02348},
 primaryClass = {astro-ph.GA},
       adsurl = {https://ui.adsabs.harvard.edu/abs/2018MNRAS.473.2608Z}
}

@ARTICLE{1990A&A...230...37G,
       author = {{Gerin}, M. and {Combes}, F. and {Athanassoula}, E.},
        title = "{The influence of galaxy interactions on stellar bars.}",
      journal = {\aap},
         year = 1990,
        month = apr,
       volume = {230},
        pages = {37-54},
       adsurl = {https://ui.adsabs.harvard.edu/abs/1990A&A...230...37G}
}

@ARTICLE{2022A&A...659A..80W,
       author = {{Wylie}, Shola M. and {Clarke}, Jonathan P. and {Gerhard}, Ortwin E.},
        title = "{The Milky Way's middle-aged inner ring}",
      journal = {\aap},
         year = 2022,
        month = mar,
       volume = {659},
          eid = {A80},
        pages = {A80},
          doi = {10.1051/0004-6361/202142343},
archivePrefix = {arXiv},
       eprint = {2110.03658},
 primaryClass = {astro-ph.GA},
       adsurl = {https://ui.adsabs.harvard.edu/abs/2022A&A...659A..80W}
}

@ARTICLE{1973ApJ...186..467O,
       author = {{Ostriker}, J.~P. and {Peebles}, P.~J.~E.},
        title = "{A Numerical Study of the Stability of Flattened Galaxies: or, can Cold Galaxies Survive?}",
      journal = {\apj},
         year = 1973,
        month = dec,
       volume = {186},
        pages = {467-480},
          doi = {10.1086/152513},
       adsurl = {https://ui.adsabs.harvard.edu/abs/1973ApJ...186..467O}
}

@ARTICLE{2018MNRAS.478..611B,
       author = {{Belokurov}, V. and {Erkal}, D. and {Evans}, N.~W. and {Koposov}, S.~E. and {Deason}, A.~J.},
        title = "{Co-formation of the disc and the stellar halo}",
      journal = {\mnras},
         year = 2018,
        month = jul,
       volume = {478},
       number = {1},
        pages = {611-619},
          doi = {10.1093/mnras/sty982},
archivePrefix = {arXiv},
       eprint = {1802.03414},
 primaryClass = {astro-ph.GA},
       adsurl = {https://ui.adsabs.harvard.edu/abs/2018MNRAS.478..611B}
}

@ARTICLE{2018Natur.563...85H,
       author = {{Helmi}, Amina and {Babusiaux}, Carine and {Koppelman}, Helmer H. and {Massari}, Davide and {Veljanoski}, Jovan and {Brown}, Anthony G.~A.},
        title = "{The merger that led to the formation of the Milky Way's inner stellar halo and thick disk}",
      journal = {\nat},
         year = 2018,
        month = oct,
       volume = {563},
       number = {7729},
        pages = {85-88},
          doi = {10.1038/s41586-018-0625-x},
archivePrefix = {arXiv},
       eprint = {1806.06038},
 primaryClass = {astro-ph.GA},
       adsurl = {https://ui.adsabs.harvard.edu/abs/2018Natur.563...85H}
}

@ARTICLE{2019arXiv191211554P,
       author = {{Phan}, Du and {Pradhan}, Neeraj and {Jankowiak}, Martin},
        title = "{Composable Effects for Flexible and Accelerated Probabilistic Programming in NumPyro}",
      journal = {arXiv e-prints},
         year = 2019,
        month = dec,
          eid = {arXiv:1912.11554},
        pages = {arXiv:1912.11554},
          doi = {10.48550/arXiv.1912.11554},
archivePrefix = {arXiv},
       eprint = {1912.11554},
 primaryClass = {stat.ML},
       adsurl = {https://ui.adsabs.harvard.edu/abs/2019arXiv191211554P}
}

@ARTICLE{2016A&A...595A...1G,
       author = {{Gaia Collaboration} and {Prusti}, T. and {de Bruijne}, J.~H.~J. and {Brown}, A.~G.~A. and {Vallenari}, A. and {Babusiaux}, C. and {Bailer-Jones}, C.~A.~L. and {Bastian}, U. and {Biermann}, M. and {Evans}, D.~W. and {Eyer}, L. and {Jansen}, F. and {Jordi}, C. and {Klioner}, S.~A. and {Lammers}, U. and {Lindegren}, L. and {Luri}, X. and {Mignard}, F. and {Milligan}, D.~J. and {Panem}, C. and {Poinsignon}, V. and {Pourbaix}, D. and {Randich}, S. and {Sarri}, G. and {Sartoretti}, P. and {Siddiqui}, H.~I. and {Soubiran}, C. and {Valette}, V. and {van Leeuwen}, F. and {Walton}, N.~A. and {Aerts}, C. and {Arenou}, F. and {Cropper}, M. and {Drimmel}, R. and {H{\o}g}, E. and {Katz}, D. and {Lattanzi}, M.~G. and {O'Mullane}, W. and {Grebel}, E.~K. and {Holland}, A.~D. and {Huc}, C. and {Passot}, X. and {Bramante}, L. and {Cacciari}, C. and {Casta{\~n}eda}, J. and {Chaoul}, L. and {Cheek}, N. and {De Angeli}, F. and {Fabricius}, C. and {Guerra}, R. and {Hern{\'a}ndez}, J. and {Jean-Antoine-Piccolo}, A. and {Masana}, E. and {Messineo}, R. and {Mowlavi}, N. and {Nienartowicz}, K. and {Ord{\'o}{\~n}ez-Blanco}, D. and {Panuzzo}, P. and {Portell}, J. and {Richards}, P.~J. and {Riello}, M. and {Seabroke}, G.~M. and {Tanga}, P. and {Th{\'e}venin}, F. and {Torra}, J. and {Els}, S.~G. and {Gracia-Abril}, G. and {Comoretto}, G. and {Garcia-Reinaldos}, M. and {Lock}, T. and {Mercier}, E. and {Altmann}, M. and {Andrae}, R. and {Astraatmadja}, T.~L. and {Bellas-Velidis}, I. and {Benson}, K. and {Berthier}, J. and {Blomme}, R. and {Busso}, G. and {Carry}, B. and {Cellino}, A. and {Clementini}, G. and {Cowell}, S. and {Creevey}, O. and {Cuypers}, J. and {Davidson}, M. and {De Ridder}, J. and {de Torres}, A. and {Delchambre}, L. and {Dell'Oro}, A. and {Ducourant}, C. and {Fr{\'e}mat}, Y. and {Garc{\'\i}a-Torres}, M. and {Gosset}, E. and {Halbwachs}, J. -L. and {Hambly}, N.~C. and {Harrison}, D.~L. and {Hauser}, M. and {Hestroffer}, D. and {Hodgkin}, S.~T. and {Huckle}, H.~E. and {Hutton}, A. and {Jasniewicz}, G. and {Jordan}, S. and {Kontizas}, M. and {Korn}, A.~J. and {Lanzafame}, A.~C. and {Manteiga}, M. and {Moitinho}, A. and {Muinonen}, K. and {Osinde}, J. and {Pancino}, E. and {Pauwels}, T. and {Petit}, J. -M. and {Recio-Blanco}, A. and {Robin}, A.~C. and {Sarro}, L.~M. and {Siopis}, C. and {Smith}, M. and {Smith}, K.~W. and {Sozzetti}, A. and {Thuillot}, W. and {van Reeven}, W. and {Viala}, Y. and {Abbas}, U. and {Abreu Aramburu}, A. and {Accart}, S. and {Aguado}, J.~J. and {Allan}, P.~M. and {Allasia}, W. and {Altavilla}, G. and {{\'A}lvarez}, M.~A. and {Alves}, J. and {Anderson}, R.~I. and {Andrei}, A.~H. and {Anglada Varela}, E. and {Antiche}, E. and {Antoja}, T. and {Ant{\'o}n}, S. and {Arcay}, B. and {Atzei}, A. and {Ayache}, L. and {Bach}, N. and {Baker}, S.~G. and {Balaguer-N{\'u}{\~n}ez}, L. and {Barache}, C. and {Barata}, C. and {Barbier}, A. and {Barblan}, F. and {Baroni}, M. and {Barrado y Navascu{\'e}s}, D. and {Barros}, M. and {Barstow}, M.~A. and {Becciani}, U. and {Bellazzini}, M. and {Bellei}, G. and {Bello Garc{\'\i}a}, A. and {Belokurov}, V. and {Bendjoya}, P. and {Berihuete}, A. and {Bianchi}, L. and {Bienaym{\'e}}, O. and {Billebaud}, F. and {Blagorodnova}, N. and {Blanco-Cuaresma}, S. and {Boch}, T. and {Bombrun}, A. and {Borrachero}, R. and {Bouquillon}, S. and {Bourda}, G. and {Bouy}, H. and {Bragaglia}, A. and {Breddels}, M.~A. and {Brouillet}, N. and {Br{\"u}semeister}, T. and {Bucciarelli}, B. and {Budnik}, F. and {Burgess}, P. and {Burgon}, R. and {Burlacu}, A. and {Busonero}, D. and {Buzzi}, R. and {Caffau}, E. and {Cambras}, J. and {Campbell}, H. and {Cancelliere}, R. and {Cantat-Gaudin}, T. and {Carlucci}, T. and {Carrasco}, J.~M. and {Castellani}, M. and {Charlot}, P. and {Charnas}, J. and {Charvet}, P. and {Chassat}, F. and {Chiavassa}, A. and {Clotet}, M. and {Cocozza}, G. and {Collins}, R.~S. and {Collins}, P. and {Costigan}, G. and {Crifo}, F. and {Cross}, N.~J.~G. and {Crosta}, M. and {Crowley}, C. and {Dafonte}, C. and {Damerdji}, Y. and {Dapergolas}, A. and {David}, P. and {David}, M. and {De Cat}, P. and {de Felice}, F. and {de Laverny}, P. and {De Luise}, F. and {De March}, R. and {de Martino}, D. and {de Souza}, R. and {Debosscher}, J. and {del Pozo}, E. and {Delbo}, M. and {Delgado}, A. and {Delgado}, H.~E. and {di Marco}, F. and {Di Matteo}, P. and {Diakite}, S. and {Distefano}, E. and {Dolding}, C. and {Dos Anjos}, S. and {Drazinos}, P. and {Dur{\'a}n}, J. and {Dzigan}, Y. and {Ecale}, E. and {Edvardsson}, B. and {Enke}, H. and {Erdmann}, M. and {Escolar}, D. and {Espina}, M. and {Evans}, N.~W. and {Eynard Bontemps}, G. and {Fabre}, C. and {Fabrizio}, M. and {Faigler}, S. and {Falc{\~a}o}, A.~J. and {Farr{\`a}s Casas}, M. and {Faye}, F. and {Federici}, L. and {Fedorets}, G. and {Fern{\'a}ndez-Hern{\'a}ndez}, J. and {Fernique}, P. and {Fienga}, A. and {Figueras}, F. and {Filippi}, F. and {Findeisen}, K. and {Fonti}, A. and {Fouesneau}, M. and {Fraile}, E. and {Fraser}, M. and {Fuchs}, J. and {Furnell}, R. and {Gai}, M. and {Galleti}, S. and {Galluccio}, L. and {Garabato}, D. and {Garc{\'\i}a-Sedano}, F. and {Gar{\'e}}, P. and {Garofalo}, A. and {Garralda}, N. and {Gavras}, P. and {Gerssen}, J. and {Geyer}, R. and {Gilmore}, G. and {Girona}, S. and {Giuffrida}, G. and {Gomes}, M. and {Gonz{\'a}lez-Marcos}, A. and {Gonz{\'a}lez-N{\'u}{\~n}ez}, J. and {Gonz{\'a}lez-Vidal}, J.~J. and {Granvik}, M. and {Guerrier}, A. and {Guillout}, P. and {Guiraud}, J. and {G{\'u}rpide}, A. and {Guti{\'e}rrez-S{\'a}nchez}, R. and {Guy}, L.~P. and {Haigron}, R. and {Hatzidimitriou}, D. and {Haywood}, M. and {Heiter}, U. and {Helmi}, A. and {Hobbs}, D. and {Hofmann}, W. and {Holl}, B. and {Holland}, G. and {Hunt}, J.~A.~S. and {Hypki}, A. and {Icardi}, V. and {Irwin}, M. and {Jevardat de Fombelle}, G. and {Jofr{\'e}}, P. and {Jonker}, P.~G. and {Jorissen}, A. and {Julbe}, F. and {Karampelas}, A. and {Kochoska}, A. and {Kohley}, R. and {Kolenberg}, K. and {Kontizas}, E. and {Koposov}, S.~E. and {Kordopatis}, G. and {Koubsky}, P. and {Kowalczyk}, A. and {Krone-Martins}, A. and {Kudryashova}, M. and {Kull}, I. and {Bachchan}, R.~K. and {Lacoste-Seris}, F. and {Lanza}, A.~F. and {Lavigne}, J. -B. and {Le Poncin-Lafitte}, C. and {Lebreton}, Y. and {Lebzelter}, T. and {Leccia}, S. and {Leclerc}, N. and {Lecoeur-Taibi}, I. and {Lemaitre}, V. and {Lenhardt}, H. and {Leroux}, F. and {Liao}, S. and {Licata}, E. and {Lindstr{\o}m}, H.~E.~P. and {Lister}, T.~A. and {Livanou}, E. and {Lobel}, A. and {L{\"o}ffler}, W. and {L{\'o}pez}, M. and {Lopez-Lozano}, A. and {Lorenz}, D. and {Loureiro}, T. and {MacDonald}, I. and {Magalh{\~a}es Fernandes}, T. and {Managau}, S. and {Mann}, R.~G. and {Mantelet}, G. and {Marchal}, O. and {Marchant}, J.~M. and {Marconi}, M. and {Marie}, J. and {Marinoni}, S. and {Marrese}, P.~M. and {Marschalk{\'o}}, G. and {Marshall}, D.~J. and {Mart{\'\i}n-Fleitas}, J.~M. and {Martino}, M. and {Mary}, N. and {Matijevi{\v{c}}}, G. and {Mazeh}, T. and {McMillan}, P.~J. and {Messina}, S. and {Mestre}, A. and {Michalik}, D. and {Millar}, N.~R. and {Miranda}, B.~M.~H. and {Molina}, D. and {Molinaro}, R. and {Molinaro}, M. and {Moln{\'a}r}, L. and {Moniez}, M. and {Montegriffo}, P. and {Monteiro}, D. and {Mor}, R. and {Mora}, A. and {Morbidelli}, R. and {Morel}, T. and {Morgenthaler}, S. and {Morley}, T. and {Morris}, D. and {Mulone}, A.~F. and {Muraveva}, T. and {Musella}, I. and {Narbonne}, J. and {Nelemans}, G. and {Nicastro}, L. and {Noval}, L. and {Ord{\'e}novic}, C. and {Ordieres-Mer{\'e}}, J. and {Osborne}, P. and {Pagani}, C. and {Pagano}, I. and {Pailler}, F. and {Palacin}, H. and {Palaversa}, L. and {Parsons}, P. and {Paulsen}, T. and {Pecoraro}, M. and {Pedrosa}, R. and {Pentik{\"a}inen}, H. and {Pereira}, J. and {Pichon}, B. and {Piersimoni}, A.~M. and {Pineau}, F. -X. and {Plachy}, E. and {Plum}, G. and {Poujoulet}, E. and {Pr{\v{s}}a}, A. and {Pulone}, L. and {Ragaini}, S. and {Rago}, S. and {Rambaux}, N. and {Ramos-Lerate}, M. and {Ranalli}, P. and {Rauw}, G. and {Read}, A. and {Regibo}, S. and {Renk}, F. and {Reyl{\'e}}, C. and {Ribeiro}, R.~A. and {Rimoldini}, L. and {Ripepi}, V. and {Riva}, A. and {Rixon}, G. and {Roelens}, M. and {Romero-G{\'o}mez}, M. and {Rowell}, N. and {Royer}, F. and {Rudolph}, A. and {Ruiz-Dern}, L. and {Sadowski}, G. and {Sagrist{\`a} Sell{\'e}s}, T. and {Sahlmann}, J. and {Salgado}, J. and {Salguero}, E. and {Sarasso}, M. and {Savietto}, H. and {Schnorhk}, A. and {Schultheis}, M. and {Sciacca}, E. and {Segol}, M. and {Segovia}, J.~C. and {Segransan}, D. and {Serpell}, E. and {Shih}, I. -C. and {Smareglia}, R. and {Smart}, R.~L. and {Smith}, C. and {Solano}, E. and {Solitro}, F. and {Sordo}, R. and {Soria Nieto}, S. and {Souchay}, J. and {Spagna}, A. and {Spoto}, F. and {Stampa}, U. and {Steele}, I.~A. and {Steidelm{\"u}ller}, H. and {Stephenson}, C.~A. and {Stoev}, H. and {Suess}, F.~F. and {S{\"u}veges}, M. and {Surdej}, J. and {Szabados}, L. and {Szegedi-Elek}, E. and {Tapiador}, D. and {Taris}, F. and {Tauran}, G. and {Taylor}, M.~B. and {Teixeira}, R. and {Terrett}, D. and {Tingley}, B. and {Trager}, S.~C. and {Turon}, C. and {Ulla}, A. and {Utrilla}, E. and {Valentini}, G. and {van Elteren}, A. and {Van Hemelryck}, E. and {van Leeuwen}, M. and {Varadi}, M. and {Vecchiato}, A. and {Veljanoski}, J. and {Via}, T. and {Vicente}, D. and {Vogt}, S. and {Voss}, H. and {Votruba}, V. and {Voutsinas}, S. and {Walmsley}, G. and {Weiler}, M. and {Weingrill}, K. and {Werner}, D. and {Wevers}, T. and {Whitehead}, G. and {Wyrzykowski}, {\L}. and {Yoldas}, A. and {{\v{Z}}erjal}, M. and {Zucker}, S. and {Zurbach}, C. and {Zwitter}, T. and {Alecu}, A. and {Allen}, M. and {Allende Prieto}, C. and {Amorim}, A. and {Anglada-Escud{\'e}}, G. and {Arsenijevic}, V. and {Azaz}, S. and {Balm}, P. and {Beck}, M. and {Bernstein}, H. -H. and {Bigot}, L. and {Bijaoui}, A. and {Blasco}, C. and {Bonfigli}, M. and {Bono}, G. and {Boudreault}, S. and {Bressan}, A. and {Brown}, S. and {Brunet}, P. -M. and {Bunclark}, P. and {Buonanno}, R. and {Butkevich}, A.~G. and {Carret}, C. and {Carrion}, C. and {Chemin}, L. and {Ch{\'e}reau}, F. and {Corcione}, L. and {Darmigny}, E. and {de Boer}, K.~S. and {de Teodoro}, P. and {de Zeeuw}, P.~T. and {Delle Luche}, C. and {Domingues}, C.~D. and {Dubath}, P. and {Fodor}, F. and {Fr{\'e}zouls}, B. and {Fries}, A. and {Fustes}, D. and {Fyfe}, D. and {Gallardo}, E. and {Gallegos}, J. and {Gardiol}, D. and {Gebran}, M. and {Gomboc}, A. and {G{\'o}mez}, A. and {Grux}, E. and {Gueguen}, A. and {Heyrovsky}, A. and {Hoar}, J. and {Iannicola}, G. and {Isasi Parache}, Y. and {Janotto}, A. -M. and {Joliet}, E. and {Jonckheere}, A. and {Keil}, R. and {Kim}, D. -W. and {Klagyivik}, P. and {Klar}, J. and {Knude}, J. and {Kochukhov}, O. and {Kolka}, I. and {Kos}, J. and {Kutka}, A. and {Lainey}, V. and {LeBouquin}, D. and {Liu}, C. and {Loreggia}, D. and {Makarov}, V.~V. and {Marseille}, M.~G. and {Martayan}, C. and {Martinez-Rubi}, O. and {Massart}, B. and {Meynadier}, F. and {Mignot}, S. and {Munari}, U. and {Nguyen}, A. -T. and {Nordlander}, T. and {Ocvirk}, P. and {O'Flaherty}, K.~S. and {Olias Sanz}, A. and {Ortiz}, P. and {Osorio}, J. and {Oszkiewicz}, D. and {Ouzounis}, A. and {Palmer}, M. and {Park}, P. and {Pasquato}, E. and {Peltzer}, C. and {Peralta}, J. and {P{\'e}turaud}, F. and {Pieniluoma}, T. and {Pigozzi}, E. and {Poels}, J. and {Prat}, G. and {Prod'homme}, T. and {Raison}, F. and {Rebordao}, J.~M. and {Risquez}, D. and {Rocca-Volmerange}, B. and {Rosen}, S. and {Ruiz-Fuertes}, M.~I. and {Russo}, F. and {Sembay}, S. and {Serraller Vizcaino}, I. and {Short}, A. and {Siebert}, A. and {Silva}, H. and {Sinachopoulos}, D. and {Slezak}, E. and {Soffel}, M. and {Sosnowska}, D. and {Strai{\v{z}}ys}, V. and {ter Linden}, M. and {Terrell}, D. and {Theil}, S. and {Tiede}, C. and {Troisi}, L. and {Tsalmantza}, P. and {Tur}, D. and {Vaccari}, M. and {Vachier}, F. and {Valles}, P. and {Van Hamme}, W. and {Veltz}, L. and {Virtanen}, J. and {Wallut}, J. -M. and {Wichmann}, R. and {Wilkinson}, M.~I. and {Ziaeepour}, H. and {Zschocke}, S.},
        title = "{The Gaia mission}",
      journal = {\aap},
         year = 2016,
        month = nov,
       volume = {595},
          eid = {A1},
        pages = {A1},
          doi = {10.1051/0004-6361/201629272},
archivePrefix = {arXiv},
       eprint = {1609.04153},
 primaryClass = {astro-ph.IM},
       adsurl = {https://ui.adsabs.harvard.edu/abs/2016A&A...595A...1G}
}

@ARTICLE{2017AJ....154...94M,
       author = {{Majewski}, Steven R. and {Schiavon}, Ricardo P. and {Frinchaboy}, Peter M. and {Allende Prieto}, Carlos and {Barkhouser}, Robert and {Bizyaev}, Dmitry and {Blank}, Basil and {Brunner}, Sophia and {Burton}, Adam and {Carrera}, Ricardo and {Chojnowski}, S. Drew and {Cunha}, K{\'a}tia and {Epstein}, Courtney and {Fitzgerald}, Greg and {Garc{\'\i}a P{\'e}rez}, Ana E. and {Hearty}, Fred R. and {Henderson}, Chuck and {Holtzman}, Jon A. and {Johnson}, Jennifer A. and {Lam}, Charles R. and {Lawler}, James E. and {Maseman}, Paul and {M{\'e}sz{\'a}ros}, Szabolcs and {Nelson}, Matthew and {Nguyen}, Duy Coung and {Nidever}, David L. and {Pinsonneault}, Marc and {Shetrone}, Matthew and {Smee}, Stephen and {Smith}, Verne V. and {Stolberg}, Todd and {Skrutskie}, Michael F. and {Walker}, Eric and {Wilson}, John C. and {Zasowski}, Gail and {Anders}, Friedrich and {Basu}, Sarbani and {Beland}, Stephane and {Blanton}, Michael R. and {Bovy}, Jo and {Brownstein}, Joel R. and {Carlberg}, Joleen and {Chaplin}, William and {Chiappini}, Cristina and {Eisenstein}, Daniel J. and {Elsworth}, Yvonne and {Feuillet}, Diane and {Fleming}, Scott W. and {Galbraith-Frew}, Jessica and {Garc{\'\i}a}, Rafael A. and {Garc{\'\i}a-Hern{\'a}ndez}, D. An{\'\i}bal and {Gillespie}, Bruce A. and {Girardi}, L{\'e}o and {Gunn}, James E. and {Hasselquist}, Sten and {Hayden}, Michael R. and {Hekker}, Saskia and {Ivans}, Inese and {Kinemuchi}, Karen and {Klaene}, Mark and {Mahadevan}, Suvrath and {Mathur}, Savita and {Mosser}, Beno{\^\i}t and {Muna}, Demitri and {Munn}, Jeffrey A. and {Nichol}, Robert C. and {O'Connell}, Robert W. and {Parejko}, John K. and {Robin}, A.~C. and {Rocha-Pinto}, Helio and {Schultheis}, Matthias and {Serenelli}, Aldo M. and {Shane}, Neville and {Silva Aguirre}, Victor and {Sobeck}, Jennifer S. and {Thompson}, Benjamin and {Troup}, Nicholas W. and {Weinberg}, David H. and {Zamora}, Olga},
        title = "{The Apache Point Observatory Galactic Evolution Experiment (APOGEE)}",
      journal = {\aj},
         year = 2017,
        month = sep,
       volume = {154},
       number = {3},
          eid = {94},
        pages = {94},
          doi = {10.3847/1538-3881/aa784d},
archivePrefix = {arXiv},
       eprint = {1509.05420},
 primaryClass = {astro-ph.IM},
       adsurl = {https://ui.adsabs.harvard.edu/abs/2017AJ....154...94M}
}

@ARTICLE{2020MNRAS.492.4500B,
       author = {{Baba}, Junichi and {Kawata}, Daisuke},
        title = "{Age dating the Galactic bar with the nuclear stellar disc}",
      journal = {\mnras},
         year = 2020,
        month = mar,
       volume = {492},
       number = {3},
        pages = {4500-4511},
          doi = {10.1093/mnras/staa140},
archivePrefix = {arXiv},
       eprint = {1909.07548},
 primaryClass = {astro-ph.GA},
       adsurl = {https://ui.adsabs.harvard.edu/abs/2020MNRAS.492.4500B}
}

@ARTICLE{2023A&A...674A...1G,
       author = {{Gaia Collaboration} and {Vallenari}, A. and {Brown}, A.~G.~A. and {Prusti}, T. and {de Bruijne}, J.~H.~J. and {Arenou}, F. and {Babusiaux}, C. and {Biermann}, M. and {Creevey}, O.~L. and {Ducourant}, C. and {Evans}, D.~W. and {Eyer}, L. and {Guerra}, R. and {Hutton}, A. and {Jordi}, C. and {Klioner}, S.~A. and {Lammers}, U.~L. and {Lindegren}, L. and {Luri}, X. and {Mignard}, F. and {Panem}, C. and {Pourbaix}, D. and {Randich}, S. and {Sartoretti}, P. and {Soubiran}, C. and {Tanga}, P. and {Walton}, N.~A. and {Bailer-Jones}, C.~A.~L. and {Bastian}, U. and {Drimmel}, R. and {Jansen}, F. and {Katz}, D. and {Lattanzi}, M.~G. and {van Leeuwen}, F. and {Bakker}, J. and {Cacciari}, C. and {Casta{\~n}eda}, J. and {De Angeli}, F. and {Fabricius}, C. and {Fouesneau}, M. and {Fr{\'e}mat}, Y. and {Galluccio}, L. and {Guerrier}, A. and {Heiter}, U. and {Masana}, E. and {Messineo}, R. and {Mowlavi}, N. and {Nicolas}, C. and {Nienartowicz}, K. and {Pailler}, F. and {Panuzzo}, P. and {Riclet}, F. and {Roux}, W. and {Seabroke}, G.~M. and {Sordo}, R. and {Th{\'e}venin}, F. and {Gracia-Abril}, G. and {Portell}, J. and {Teyssier}, D. and {Altmann}, M. and {Andrae}, R. and {Audard}, M. and {Bellas-Velidis}, I. and {Benson}, K. and {Berthier}, J. and {Blomme}, R. and {Burgess}, P.~W. and {Busonero}, D. and {Busso}, G. and {C{\'a}novas}, H. and {Carry}, B. and {Cellino}, A. and {Cheek}, N. and {Clementini}, G. and {Damerdji}, Y. and {Davidson}, M. and {de Teodoro}, P. and {Nu{\~n}ez Campos}, M. and {Delchambre}, L. and {Dell'Oro}, A. and {Esquej}, P. and {Fern{\'a}ndez-Hern{\'a}ndez}, J. and {Fraile}, E. and {Garabato}, D. and {Garc{\'\i}a-Lario}, P. and {Gosset}, E. and {Haigron}, R. and {Halbwachs}, J. -L. and {Hambly}, N.~C. and {Harrison}, D.~L. and {Hern{\'a}ndez}, J. and {Hestroffer}, D. and {Hodgkin}, S.~T. and {Holl}, B. and {Jan{\ss}en}, K. and {Jevardat de Fombelle}, G. and {Jordan}, S. and {Krone-Martins}, A. and {Lanzafame}, A.~C. and {L{\"o}ffler}, W. and {Marchal}, O. and {Marrese}, P.~M. and {Moitinho}, A. and {Muinonen}, K. and {Osborne}, P. and {Pancino}, E. and {Pauwels}, T. and {Recio-Blanco}, A. and {Reyl{\'e}}, C. and {Riello}, M. and {Rimoldini}, L. and {Roegiers}, T. and {Rybizki}, J. and {Sarro}, L.~M. and {Siopis}, C. and {Smith}, M. and {Sozzetti}, A. and {Utrilla}, E. and {van Leeuwen}, M. and {Abbas}, U. and {{\'A}brah{\'a}m}, P. and {Abreu Aramburu}, A. and {Aerts}, C. and {Aguado}, J.~J. and {Ajaj}, M. and {Aldea-Montero}, F. and {Altavilla}, G. and {{\'A}lvarez}, M.~A. and {Alves}, J. and {Anders}, F. and {Anderson}, R.~I. and {Anglada Varela}, E. and {Antoja}, T. and {Baines}, D. and {Baker}, S.~G. and {Balaguer-N{\'u}{\~n}ez}, L. and {Balbinot}, E. and {Balog}, Z. and {Barache}, C. and {Barbato}, D. and {Barros}, M. and {Barstow}, M.~A. and {Bartolom{\'e}}, S. and {Bassilana}, J. -L. and {Bauchet}, N. and {Becciani}, U. and {Bellazzini}, M. and {Berihuete}, A. and {Bernet}, M. and {Bertone}, S. and {Bianchi}, L. and {Binnenfeld}, A. and {Blanco-Cuaresma}, S. and {Blazere}, A. and {Boch}, T. and {Bombrun}, A. and {Bossini}, D. and {Bouquillon}, S. and {Bragaglia}, A. and {Bramante}, L. and {Breedt}, E. and {Bressan}, A. and {Brouillet}, N. and {Brugaletta}, E. and {Bucciarelli}, B. and {Burlacu}, A. and {Butkevich}, A.~G. and {Buzzi}, R. and {Caffau}, E. and {Cancelliere}, R. and {Cantat-Gaudin}, T. and {Carballo}, R. and {Carlucci}, T. and {Carnerero}, M.~I. and {Carrasco}, J.~M. and {Casamiquela}, L. and {Castellani}, M. and {Castro-Ginard}, A. and {Chaoul}, L. and {Charlot}, P. and {Chemin}, L. and {Chiaramida}, V. and {Chiavassa}, A. and {Chornay}, N. and {Comoretto}, G. and {Contursi}, G. and {Cooper}, W.~J. and {Cornez}, T. and {Cowell}, S. and {Crifo}, F. and {Cropper}, M. and {Crosta}, M. and {Crowley}, C. and {Dafonte}, C. and {Dapergolas}, A. and {David}, M. and {David}, P. and {de Laverny}, P. and {De Luise}, F. and {De March}, R. and {De Ridder}, J. and {de Souza}, R. and {de Torres}, A. and {del Peloso}, E.~F. and {del Pozo}, E. and {Delbo}, M. and {Delgado}, A. and {Delisle}, J. -B. and {Demouchy}, C. and {Dharmawardena}, T.~E. and {Di Matteo}, P. and {Diakite}, S. and {Diener}, C. and {Distefano}, E. and {Dolding}, C. and {Edvardsson}, B. and {Enke}, H. and {Fabre}, C. and {Fabrizio}, M. and {Faigler}, S. and {Fedorets}, G. and {Fernique}, P. and {Fienga}, A. and {Figueras}, F. and {Fournier}, Y. and {Fouron}, C. and {Fragkoudi}, F. and {Gai}, M. and {Garcia-Gutierrez}, A. and {Garcia-Reinaldos}, M. and {Garc{\'\i}a-Torres}, M. and {Garofalo}, A. and {Gavel}, A. and {Gavras}, P. and {Gerlach}, E. and {Geyer}, R. and {Giacobbe}, P. and {Gilmore}, G. and {Girona}, S. and {Giuffrida}, G. and {Gomel}, R. and {Gomez}, A. and {Gonz{\'a}lez-N{\'u}{\~n}ez}, J. and {Gonz{\'a}lez-Santamar{\'\i}a}, I. and {Gonz{\'a}lez-Vidal}, J.~J. and {Granvik}, M. and {Guillout}, P. and {Guiraud}, J. and {Guti{\'e}rrez-S{\'a}nchez}, R. and {Guy}, L.~P. and {Hatzidimitriou}, D. and {Hauser}, M. and {Haywood}, M. and {Helmer}, A. and {Helmi}, A. and {Sarmiento}, M.~H. and {Hidalgo}, S.~L. and {Hilger}, T. and {H{\l}adczuk}, N. and {Hobbs}, D. and {Holland}, G. and {Huckle}, H.~E. and {Jardine}, K. and {Jasniewicz}, G. and {Jean-Antoine Piccolo}, A. and {Jim{\'e}nez-Arranz}, {\'O}. and {Jorissen}, A. and {Juaristi Campillo}, J. and {Julbe}, F. and {Karbevska}, L. and {Kervella}, P. and {Khanna}, S. and {Kontizas}, M. and {Kordopatis}, G. and {Korn}, A.~J. and {K{\'o}sp{\'a}l}, {\'A}. and {Kostrzewa-Rutkowska}, Z. and {Kruszy{\'n}ska}, K. and {Kun}, M. and {Laizeau}, P. and {Lambert}, S. and {Lanza}, A.~F. and {Lasne}, Y. and {Le Campion}, J. -F. and {Lebreton}, Y. and {Lebzelter}, T. and {Leccia}, S. and {Leclerc}, N. and {Lecoeur-Taibi}, I. and {Liao}, S. and {Licata}, E.~L. and {Lindstr{\o}m}, H.~E.~P. and {Lister}, T.~A. and {Livanou}, E. and {Lobel}, A. and {Lorca}, A. and {Loup}, C. and {Madrero Pardo}, P. and {Magdaleno Romeo}, A. and {Managau}, S. and {Mann}, R.~G. and {Manteiga}, M. and {Marchant}, J.~M. and {Marconi}, M. and {Marcos}, J. and {Marcos Santos}, M.~M.~S. and {Mar{\'\i}n Pina}, D. and {Marinoni}, S. and {Marocco}, F. and {Marshall}, D.~J. and {Martin Polo}, L. and {Mart{\'\i}n-Fleitas}, J.~M. and {Marton}, G. and {Mary}, N. and {Masip}, A. and {Massari}, D. and {Mastrobuono-Battisti}, A. and {Mazeh}, T. and {McMillan}, P.~J. and {Messina}, S. and {Michalik}, D. and {Millar}, N.~R. and {Mints}, A. and {Molina}, D. and {Molinaro}, R. and {Moln{\'a}r}, L. and {Monari}, G. and {Mongui{\'o}}, M. and {Montegriffo}, P. and {Montero}, A. and {Mor}, R. and {Mora}, A. and {Morbidelli}, R. and {Morel}, T. and {Morris}, D. and {Muraveva}, T. and {Murphy}, C.~P. and {Musella}, I. and {Nagy}, Z. and {Noval}, L. and {Oca{\~n}a}, F. and {Ogden}, A. and {Ordenovic}, C. and {Osinde}, J.~O. and {Pagani}, C. and {Pagano}, I. and {Palaversa}, L. and {Palicio}, P.~A. and {Pallas-Quintela}, L. and {Panahi}, A. and {Payne-Wardenaar}, S. and {Pe{\~n}alosa Esteller}, X. and {Penttil{\"a}}, A. and {Pichon}, B. and {Piersimoni}, A.~M. and {Pineau}, F. -X. and {Plachy}, E. and {Plum}, G. and {Poggio}, E. and {Pr{\v{s}}a}, A. and {Pulone}, L. and {Racero}, E. and {Ragaini}, S. and {Rainer}, M. and {Raiteri}, C.~M. and {Rambaux}, N. and {Ramos}, P. and {Ramos-Lerate}, M. and {Re Fiorentin}, P. and {Regibo}, S. and {Richards}, P.~J. and {Rios Diaz}, C. and {Ripepi}, V. and {Riva}, A. and {Rix}, H. -W. and {Rixon}, G. and {Robichon}, N. and {Robin}, A.~C. and {Robin}, C. and {Roelens}, M. and {Rogues}, H.~R.~O. and {Rohrbasser}, L. and {Romero-G{\'o}mez}, M. and {Rowell}, N. and {Royer}, F. and {Ruz Mieres}, D. and {Rybicki}, K.~A. and {Sadowski}, G. and {S{\'a}ez N{\'u}{\~n}ez}, A. and {Sagrist{\`a} Sell{\'e}s}, A. and {Sahlmann}, J. and {Salguero}, E. and {Samaras}, N. and {Sanchez Gimenez}, V. and {Sanna}, N. and {Santove{\~n}a}, R. and {Sarasso}, M. and {Schultheis}, M. and {Sciacca}, E. and {Segol}, M. and {Segovia}, J.~C. and {S{\'e}gransan}, D. and {Semeux}, D. and {Shahaf}, S. and {Siddiqui}, H.~I. and {Siebert}, A. and {Siltala}, L. and {Silvelo}, A. and {Slezak}, E. and {Slezak}, I. and {Smart}, R.~L. and {Snaith}, O.~N. and {Solano}, E. and {Solitro}, F. and {Souami}, D. and {Souchay}, J. and {Spagna}, A. and {Spina}, L. and {Spoto}, F. and {Steele}, I.~A. and {Steidelm{\"u}ller}, H. and {Stephenson}, C.~A. and {S{\"u}veges}, M. and {Surdej}, J. and {Szabados}, L. and {Szegedi-Elek}, E. and {Taris}, F. and {Taylor}, M.~B. and {Teixeira}, R. and {Tolomei}, L. and {Tonello}, N. and {Torra}, F. and {Torra}, J. and {Torralba Elipe}, G. and {Trabucchi}, M. and {Tsounis}, A.~T. and {Turon}, C. and {Ulla}, A. and {Unger}, N. and {Vaillant}, M.~V. and {van Dillen}, E. and {van Reeven}, W. and {Vanel}, O. and {Vecchiato}, A. and {Viala}, Y. and {Vicente}, D. and {Voutsinas}, S. and {Weiler}, M. and {Wevers}, T. and {Wyrzykowski}, {\L}. and {Yoldas}, A. and {Yvard}, P. and {Zhao}, H. and {Zorec}, J. and {Zucker}, S. and {Zwitter}, T.},
        title = "{Gaia Data Release 3. Summary of the content and survey properties}",
      journal = {\aap},
         year = 2023,
        month = jun,
       volume = {674},
          eid = {A1},
        pages = {A1},
          doi = {10.1051/0004-6361/202243940},
archivePrefix = {arXiv},
       eprint = {2208.00211},
 primaryClass = {astro-ph.GA},
       adsurl = {https://ui.adsabs.harvard.edu/abs/2023A&A...674A...1G}
}

@ARTICLE{2024A&A...690A.125B,
       author = {{B{\'e}guin}, E. and {Chiavassa}, A. and {Ahmad}, A. and {Freytag}, B. and {Uttenthaler}, S.},
        title = "{Retrieving stellar parameters and dynamics of AGB stars with Gaia parallax measurements and CO$^{5}$BOLD RHD simulations}",
      journal = {\aap},
         year = 2024,
        month = oct,
       volume = {690},
          eid = {A125},
        pages = {A125},
          doi = {10.1051/0004-6361/202450245},
archivePrefix = {arXiv},
       eprint = {2409.03422},
 primaryClass = {astro-ph.SR},
       adsurl = {https://ui.adsabs.harvard.edu/abs/2024A&A...690A.125B}
}

@ARTICLE{2026arXiv260903275B,
       author = {{Bland-Hawthorn}, Joss and {Tepper-Garcia}, Thor and {Hamilton}, Chris and {Villa-Alatorre}, Johan M. and {Agertz}, Oscar and {Tsukui}, Takafumi and {Freeman}, Ken},
        title = "{Turbulent gas-rich discs at high redshift: the origin of early massive stellar bars}",
      journal = {arXiv e-prints},
         year = 2026,
        month = sep,
          eid = {arXiv:2609.03275},
        pages = {arXiv:2609.03275},
          doi = {10.48550/arXiv.2609.03275},
archivePrefix = {arXiv},
       eprint = {2609.03275},
 primaryClass = {astro-ph.GA},
       adsurl = {https://ui.adsabs.harvard.edu/abs/2026arXiv260903275B}
}

@ARTICLE{2023MNRAS.523.2369S,
       author = {{Sanders}, Jason L.},
        title = "{The period-luminosity relation for Mira variables in the Milky Way using Gaia DR3: a further distance anchor for H$_{0}$}",
      journal = {\mnras},
         year = 2023,
        month = aug,
       volume = {523},
       number = {2},
        pages = {2369-2398},
          doi = {10.1093/mnras/stad1431},
archivePrefix = {arXiv},
       eprint = {2304.01671},
 primaryClass = {astro-ph.GA},
       adsurl = {https://ui.adsabs.harvard.edu/abs/2023MNRAS.523.2369S}
}

@ARTICLE{2024ApJ...968...86B,
       author = {{Bland-Hawthorn}, Joss and {Tepper-Garcia}, Thor and {Agertz}, Oscar and {Federrath}, Christoph},
        title = "{Turbulent Gas-rich Disks at High Redshift: Bars and Bulges in a Radial Shear Flow}",
      journal = {\apj},
         year = 2024,
        month = jun,
       volume = {968},
       number = {2},
          eid = {86},
        pages = {86},
          doi = {10.3847/1538-4357/ad4118},
archivePrefix = {arXiv},
       eprint = {2402.06060},
 primaryClass = {astro-ph.GA},
       adsurl = {https://ui.adsabs.harvard.edu/abs/2024ApJ...968...86B}
}

@ARTICLE{2026MNRAS.548f2154O,
       author = {{Orkney}, Matthew D.~A. and {Laporte}, Chervin F.~P.},
        title = "{Build-up and survival of the disc: from numerical models of galaxy formation to the Milky Way}",
      journal = {\mnras},
         year = 2026,
        month = jun,
       volume = {548},
       number = {4},
          eid = {staf2154},
        pages = {staf2154},
          doi = {10.1093/mnras/staf2154},
archivePrefix = {arXiv},
       eprint = {2509.09576},
 primaryClass = {astro-ph.GA},
       adsurl = {https://ui.adsabs.harvard.edu/abs/2026MNRAS.548f2154O}
}

@ARTICLE{2020NatAs...4..377N,
       author = {{Nogueras-Lara}, Francisco and {Sch{\"o}del}, Rainer and {Gallego-Calvente}, Aurelia Teresa and {Gallego-Cano}, Eulalia and {Shahzamanian}, Banafsheh and {Dong}, Hui and {Neumayer}, Nadine and {Hilker}, Michael and {Najarro}, Francisco and {Nishiyama}, Shogo and {Feldmeier-Krause}, Anja and {Girard}, Julien H.~V. and {Cassisi}, Santi},
        title = "{Early formation and recent starburst activity in the nuclear disk of the Milky Way}",
      journal = {Nature Astronomy},
         year = 2020,
        month = jan,
       volume = {4},
        pages = {377-381},
          doi = {10.1038/s41550-019-0967-9},
archivePrefix = {arXiv},
       eprint = {1910.06968},
 primaryClass = {astro-ph.GA},
       adsurl = {https://ui.adsabs.harvard.edu/abs/2020NatAs...4..377N}
}

@ARTICLE{2024A&A...681L...8N,
       author = {{Nepal}, S. and {Chiappini}, C. and {Guiglion}, G. and {Steinmetz}, M. and {P{\'e}rez-Villegas}, A. and {Queiroz}, A.~B. and {Miglio}, A. and {Dohme}, P. and {Khalatyan}, A.},
        title = "{Insights from super-metal-rich stars: Is the Milky Way bar young?}",
      journal = {\aap},
         year = 2024,
        month = jan,
       volume = {681},
          eid = {L8},
        pages = {L8},
          doi = {10.1051/0004-6361/202348365},
archivePrefix = {arXiv},
       eprint = {2311.16993},
 primaryClass = {astro-ph.GA},
       adsurl = {https://ui.adsabs.harvard.edu/abs/2024A&A...681L...8N}
}

@ARTICLE{2026A&A...705A..92R,
       author = {{Ruiz-Lara}, Tom{\'a}s and {Mirabal}, David and {Gallart}, Carme and {Grand}, Robert and {Fragkoudi}, Francesca and {P{\'e}rez}, Isabel and {Cassisi}, Santi and {Fern{\'a}ndez-Alvar}, Emma and {Queiroz}, Anna B. and {Aznar-Menargues}, Guillem and {Gonz{\'a}lez-Koda}, Yllari K. and {Rivero}, Alicia and {Surot}, Francisco and {Thomas}, Guillaume F. and {Bieri}, Rebekka and {Gomez}, Facundo A. and {Pakmor}, R{\"u}diger and {van de Voort}, Freeke},
        title = "{Chronology of our Galaxy from Gaia colour-magnitude diagram fitting (ChronoGal): IV. The inner Milky Way stellar age distribution}",
      journal = {\aap},
         year = 2026,
        month = jan,
       volume = {705},
          eid = {A92},
        pages = {A92},
          doi = {10.1051/0004-6361/202556468},
archivePrefix = {arXiv},
       eprint = {2510.02238},
 primaryClass = {astro-ph.GA},
       adsurl = {https://ui.adsabs.harvard.edu/abs/2026A&A...705A..92R}
}

@ARTICLE{2026OJAp....963764N,
       author = {{Noguchi}, Masafumi},
        title = "{Dual-disk galaxies and thermal states of circumgalactic medium}",
      journal = {The Open Journal of Astrophysics},
         year = 2026,
        month = jun,
       volume = {9},
        pages = {63764},
          doi = {10.33232/001c.163764},
archivePrefix = {arXiv},
       eprint = {2606.24158},
 primaryClass = {astro-ph.GA},
       adsurl = {https://ui.adsabs.harvard.edu/abs/2026OJAp....963764N}
}

@ARTICLE{2026arXiv260522806M,
       author = {{Myrtaj}, Olti and {Bullock}, James S. and {Boylan-Kolchin}, Michael and {Chandra}, Vedant and {Faucher-Gigu{\`e}re}, Claude-Andr{\'e} and {Feldmann}, Robert and {Mercado}, Francisco J. and {Moreno}, Jorge and {Stern}, Jonathan and {Wetzel}, Andrew and {Gandhi}, Pratik J. and {Xia}, Luke},
        title = "{From protogalaxy through thick and thin: Why did the Milky Way evolve in three kinematic phases?}",
      journal = {arXiv e-prints},
         year = 2026,
        month = may,
          eid = {arXiv:2605.22806},
        pages = {arXiv:2605.22806},
          doi = {10.48550/arXiv.2605.22806},
archivePrefix = {arXiv},
       eprint = {2605.22806},
 primaryClass = {astro-ph.GA},
       adsurl = {https://ui.adsabs.harvard.edu/abs/2026arXiv260522806M}
}

@ARTICLE{2026arXiv261008931R,
       author = {{Rufo-Pastor}, V{\'\i}ctor and {Agertz}, Oscar and {Roca-F{\'a}brega}, Santi and {Rey}, Martin P. and {Bland-Hawthorn}, Joss and {Read}, Justin I. and {Tepper-Garc{\'\i}a}, Thor and {Cadiou}, Corentin and {Storck}, Anatole},
        title = "{Early onset of the hot circumgalactic medium around Milky Way galaxies}",
      journal = {arXiv e-prints},
         year = 2026,
        month = oct,
          eid = {arXiv:2610.08931},
        pages = {arXiv:2610.08931},
          doi = {10.48550/arXiv.2610.08931},
archivePrefix = {arXiv},
       eprint = {2610.08931},
 primaryClass = {astro-ph.GA},
       adsurl = {https://ui.adsabs.harvard.edu/abs/2026arXiv261008931R}
}

@ARTICLE{2000MNRAS.312..194N,
       author = {{Noguchi}, Masafumi},
        title = "{Secular evolution of late-type disc galaxies: formation of bulges and the origin of bar dichotomy}",
      journal = {\mnras},
         year = 2000,
        month = feb,
       volume = {312},
       number = {1},
        pages = {194-206},
          doi = {10.1046/j.1365-8711.2000.03130.x},
       adsurl = {https://ui.adsabs.harvard.edu/abs/2000MNRAS.312..194N}
}




\appendix

\section{Importance Sampling}\label{sec:IS}
As shown in equations~\eqref{eq:propto_s5} and \eqref{eq:propto_s3}, evaluating the equation~\eqref{eq:probabiliry} requires integrals over velocity volume, making it computationally expensive. Following the approach used in \citet{2023MNRAS.521.1462Z}, we applied importance sampling separately to the numerator and denominator of equation~\eqref{eq:probabiliry} to reduce this computational cost. In the numerator, the integral is over the unknown line-of-sight velocity, $v_{\|}$. For each star with observed quantities ($\ell,b,m,\bm{\mu}$, $P$), we sampled $v_{\|}$ from a sampling function, $G_P(v_{\|}\mid \ell,b,m,\mu)$, for weighted Monte Carlo integration. In the denominator, the integral is over the full three-dimensional velocity, $\bm{v}$. For each star with observed quantities ($\ell,b,m$) and $P$, we sampled $\bm{v}$ from a sampling function, $G_P(\bm{v}\mid \ell,b,m)$. We defined the period-dependent proposal distribution function using a quasi-isothermal distribution function, $f^\prime_P(\bm{J})$, which was fixed in advance at each $P$. The sampling functions were then defined as
\begin{align}
G_P\left(v_{\|} \mid \ell, b, m, \boldsymbol{\mu}\right)&=\frac{p\left(\ell, b, m, \boldsymbol{\mu}, v_{\|},P\right)}{\int \mathrm{d} v_{\|}~ p\left(\ell, b, m, \boldsymbol{\mu}, v_{\|},P\right)}=\frac{f_P^{\prime}(\boldsymbol{J})}{\int \mathrm{d}v_{\|}~f^\prime_P(\bm{J})},\label{eq:Gv_para}\\ 
G_P(\bm{v}\mid \ell,b,m)&=\frac{p\left(\ell, b, m, \bm{v},P\right)}{\int \mathrm{d}^3 \bm{v}~p\left(\ell, b, m, \bm{v},P\right)}=\frac{f_P^{\prime}(\boldsymbol{J})}{\int \mathrm{d}\bm{v}~f^\prime_P(\bm{J})}.\label{eq:Gv}
\end{align}

We generated the importance-sampling points separately for the numerator and denominator of equation~\eqref{eq:probabiliry} using equations~\eqref{eq:Gv_para} and~\eqref{eq:Gv}, respectively. For the numerator, we generated 1\,280 samples of $v_\|$ for each star. Specifically, for the $k$-th star, we drew 1\,280 $v_\|$ from the sampling function in equation~\eqref{eq:Gv_para} at the given observed position, proper motion and period $(\ell_k,b_k,m_k,\bm{\mu}_k,P_k)$ using inverse transform sampling.
For the denominator, we generated 1\,280 samples of the three-dimensional velocity $\bm{v}$ for each star. We sampled 1\,280 $\bm{v}$ from the sampling function in equation~\eqref{eq:Gv} at the given observed position of the $k$-th star, $(\ell_k,b_k,m_k)$, using Markov-Chain Monte Carlo (MCMC) sampling with \texttt{emcee} \citep{2013PASP..125..306F}.
For each sampled value, we stored the sampled line-of-sight velocity and three-dimensional velocity, the corresponding value of $f^\prime_{P_k}(\bm{J})$, and the normalization integral in the denominator of equations~\eqref{eq:Gv_para} and \eqref{eq:Gv}.

Using the precomputed importance-sampling points, we evaluated the numerator and denominator of equation~\eqref{eq:probabiliry} while accounting for the observational uncertainties through Monte Carlo sampling. Here, we denote the index of a star by $k$ and the index of a Monte Carlo sample by $i$.
For the numerator, we used the same number of samples as for the importance sampling of $v_\|$, $N=1\,280$ samples per star, to marginalise over the observational uncertainties. For the $k$-th star, we drew $m_i$, $\bm{\mu}_i$ and $P_i$ from Gaussian distributions centred on the observed values, with variance given by the squared uncertainties, while $v_{\|i}$ was drawn from the sampling distribution in equation~\eqref{eq:Gv_para}. We then approximated the numerator as
\begin{equation}
\begin{aligned}\label{eq:plbmmu}
&p(\ell_k, b_k, m_k, \bm{\mu}_k, P_k) \\&\quad \approx \frac{A_{\|k}}{N} \sum_i^{\substack{
m_i \text{ from } \mathcal{N} (m_k,\sigma^2_{m_k}) \\
\bm{\mu}_i \text{ from } \mathcal{N} (\bm{\mu}_k,\sigma^2_{\bm{\mu}_k}) \\
v_{\|i} \text { from } G\left(v_{\|} \mid \cdots\right)\\
P_i \text { from }\mathcal{N} (P_k,\sigma^2_{P_k})
}} s_i^5 \cos b_k \frac{f_{P_i}\left(\boldsymbol{J}(\ell_k,b_k,m_i,\bm{\mu}_i, v_{\|i})\right)}{f_{P_k}^{\prime}\left(\boldsymbol{J}(\ell_k,b_k,m_k,\bm{\mu}_k, v_{\|i})\right)},
\end{aligned}
\end{equation}
where
\begin{align}
A_{\|k} & =\int \mathrm{d} v_{\|} f_{P_k}^{\prime}\left(\bm{J}\left(\ell_k, b_k, m_k, \bm{\mu}_k, v_{\|}\right)\right) \nonumber \\
& =\sum_i^{v_{\|i} \text{ from } G\left(v_{\|} \mid \cdots\right)} f_{P_k}^{\prime}\left(\bm{J}\left(\ell_k, b_k, m_k, \bm{\mu}_k, v_{\|i}\right)\right).
\end{align}
$A_{\|k}$ is the normalization constant for the proposal distribution.
For the denominator of equation~\eqref{eq:probabiliry}, we used $N=1\,280$ samples per star, matching the number of importance-sampling points used for the three-dimensional velocity integral. For the $k$-th star, we drew $m_i$ and $P_i$ from Gaussian distributions centred on the observed values, while $\bm{v}_i$ was sampled from the proposal distribution in equation~\eqref{eq:Gv}. The denominator was then approximated as
\begin{equation}\begin{aligned}
\label{eq:plbm}
&p(\ell_k, b_k, m_k, P_k) \\
&\quad \approx \frac{A_k}{N} \sum_i^{\substack{
m_i \text{ from } \mathcal{N} (m_k,\sigma^2_{m_k}) \\
\bm{v}_i \text { from } G\left(\bm{v} \mid \cdots\right)\\
P_i \text { from }\mathcal{N} (P_k,\sigma^2_{P_k})
}} s_i^3 \cos b_k \frac{f_{P_i}\left(\boldsymbol{J}(\ell_k,b_k,m_i, \bm{v}_{i})\right)}{f_{P_k}^{\prime}\left(\boldsymbol{J}(\ell_k,b_k,m_k, \bm{v}_{i})\right)},
\end{aligned}
\end{equation}
where
\begin{align}
A_k & =\int \mathrm{d}^3 \bm{v}f_{P_k}^{\prime}\left(\bm{J}\left(\ell_k, b_k, m_k, \bm{v}\right)\right) \nonumber \\
& =\sum_i^{\bm{v}_{i} \text{ from } G\left(\bm{v} \mid \cdots\right)} f_{P_k}^{\prime}\left(\bm{J}\left(\ell_k, b_k, m_k, \bm{v}_i\right)\right).
\end{align}
$A_k$ is the normalization constant for the proposal distribution.

Note that, in both the numerator and denominator, the proposal distribution function, $f^\prime_{P_k}$, and the sampled distances, velocities and periods in equations~\eqref{eq:plbmmu} and~\eqref{eq:plbm} were fixed throughout the MCMC fitting of the Mira kinematics. The actions, $\bm{J}$, at the sampled phase-space points were also precomputed. However, the period-dependent parameters of the target distribution function, $f_P$, were updated at each MCMC step. Since the normalization constants $A_{\|k}$ and $A_k$ were evaluated only once for each star before the MCMC fitting, the numerical integration of equation~\eqref{eq:probabiliry} during the MCMC was reduced to simple summations over the precomputed samples.

As discussed by \citet{2025MNRAS.543.2275F}, the functional form adopted for the proposal distribution function, $f^\prime$, must be sufficiently close to the true age-dependent, and therefore period-dependent, distribution function especially when the goal is to measure how the disc structure changes with period. If the proposed distribution function differs substantially from the true distribution, importance sampling can introduce biases, making it difficult to recover the true structure. Since the APOGEE red giants data include radial velocity measurements and therefore provide stronger constraints on disc modelling, we considered the best-fit age-dependent distribution function obtained by \citet{2025MNRAS.543.2275F} to be the reference model for the true distribution function as a function of the age. We therefore adopted the same functional form for the proposal age-dependent distribution function as that used by \citet{2025MNRAS.543.2275F}, matching their best-fit function:
\begin{align}
&R_{\mathrm{disc}}/\mathrm{kpc}=
\begin{cases}-0.25 \times \tau +4.2 & \text { if } \tau \leq 10, \\
1.7& \text { otherwise },\end{cases}\notag \\
&\sigma_{R,0}/\mathrm{km~s^{-1}}=2.5\times\tau+25,\notag\\
&\sigma_{z,0}/\mathrm{km~s^{-1}}=3\times\tau+12,\notag\\
&R_{\sigma,R}/\mathrm{kpc}=\begin{cases}30 & \text { if } \tau \leq 7, \\
-\frac{22}{3}\times\tau+\frac{244}{3}& \text { if } 7 <\tau \leq 10,\\
8& \text { if } \tau>10,\end{cases} \notag\\
&R_{\sigma,z}/\mathrm{kpc}=\begin{cases}42 & \text { if } \tau \leq 7, \\
-\frac{34}{3}\times\tau+\frac{364}{3}& \text { if } 7 <\tau \leq 10,\\
8& \text { if } \tau>10,\end{cases}  \label{eq:f_tau}
\end{align}
where the age, $\tau$, is in Gyr. Although $\sigma_{R,0}$ and $\sigma_{z,0}$ were chosen to be slightly higher than those in the best-fit distribution function of \citet{2025MNRAS.543.2275F}, this choice helps to minimise bias in the Monte Carlo integration without substantially sacrificing precision, because it samples a wider range of velocities.
 
The Mira sample used here has pulsation periods instead of direct age measurements. We therefore needed to assume a period--age relation to transform the age-dependent proposal distribution function $f^\prime_\tau$ used for the APOGEE red giants into a period-dependent proposal distribution function, $f^\prime_P$, required for the Miras. To construct an appropriate $f^\prime_P$, we iterated the procedure of fitting the period-dependent disc structure, deriving the period--age relation. Then, adjusting $f^\prime_P$ using the best-fit period--age relation, we fitted the period-dependent disc structure again. Through this iteration, we obtained the best period--age relation that allowed equation~\eqref{eq:f_tau} to be converted into a period-dependent proposal distribution, $f^\prime_P$. The resulting $f^\prime_P$ was sufficiently close to the true period-dependent disc structure of the Mira sample inferred from the APOGEE red giants in \citet{2025MNRAS.543.2275F} and provides more precise integrations of equation~\eqref{eq:probabiliry}.

\section{Mock Data Validation}\label{sec:validation}
We validated the analysis pipeline used to measure the Mira period-dependent distribution function and to derive the period-age relation, using mock data created with the known parameters of the distribution function and an assumed period-age relation. This validation was designed to test the two main steps of the analysis. First, we tested how well the period-dependent distribution-function fitting described in Section~\ref{sec:DF} could recover the period dependence of the input parameters. Second, we tested how well the subsequent period-age calibration recovered the input period-age relation and whether or not there is any systematic offsets.

We constructed mock Mira catalogues that mimic the observational properties of the real Mira sample. The mock catalogues were generated from known period-dependent distribution-function parameters and period-age relation parameters using \texttt{AGAMA} \citep{2019MNRAS.482.1525V}. For each star in the observed Mira sample, we created a corresponding mock star with the same sky position, observed distance modulus and pulsation period, but with generated proper motions from a given distribution function. The mock catalogues therefore reproduce the observed spatial and period distributions of the real sample.

To define the proper motions of the mock Mira stars, we first specified the distribution-function parameters as functions of stellar age using the age-dependent forms in equation~\eqref{eq:f_tau}. We then converted these age-dependent functions into period-dependent ones using the period--age relation defined in equation~\eqref{eq:period_age_model}, adopting $(\tau_0,P_s)=(17.0~\mathrm{Gyr},210~\mathrm{days})$. These values were chosen so that the input period-age relation approximately matches the best-fit relation derived from the real Mira sample as described in Section~\ref{sec:result:PAR}.
Given these period-dependent input parameters, we drew a mock stellar velocity from the velocity distribution function at the true position of each star from the quasi-isothermal distribution function described in equation~\eqref{eq:DF}. The same Galactic potential and solar position and motion as in the main analysis were adopted when generating the mock star data.

When generating the mock kinematics, we also included the effect of distance uncertainty by perturbing the true distance modulus of each star before sampling its three-dimensional velocity in Cartesian coordinate. For each star, we drew a true distance modulus, $m_k^{\prime}$, as
\begin{equation}\label{eq:vali:dmerror}
m^{\prime}_k
\sim
\mathcal{N}
\left(
m_k,\sigma^2_{m_k}
\right),
\end{equation}
where $m_k$ and $\sigma_{m_k}$ are the observed distance modulus and its uncertainty of $k$-th Mira star used in this study. We then used the true position defined by $(\ell_k,b_k,m^{\prime}_k)$ to sample a velocity from the given quasi-isothermal distribution function at the observed period of the star. 
The sampled velocities were assigned to the mock stars and converted into true mock proper motions, $(\mu_{\ell_k}^{\prime},\mu_{b_k}^{\prime})$, using the sampling position, $(\ell_k,b_k,m^{\prime}_k)$. 

We then added observational errors to the proper motions. We drew
\begin{align}\label{eq:vali:pmerror}
\bm{\mu}_{k}^{\rm mock}
\sim
\mathcal{N}
\left(
\bm{\mu}_{k}^{\prime},
\sigma^2_{\bm{\mu}_{k}}
\right),
\end{align}
where $\bm{\mu}_k=(\mu_{\ell_k},\mu_{b_k})$ and $\sigma_{\bm{\mu}_k}$ denotes the observed uncertainties in the proper motions of our Mira data. The resulting $\bm{\mu}^{\rm mock}_k$ was used as the observed proper motion together with the observed position of $(\ell_k,b_k,m_k)$ in the mock catalogue.

The observed position of the mock star used in the subsequent fit was kept identical to the observed position of our Miras, $(\ell_k,b_k,m_k)$ though the errors by the observational uncertainties are added to the sampled true distance and proper motions. In this way, the mock catalogue preserved the observed spatial selection of the real Mira sample, while allowing the generated kinematics to reflect the distance and proper-motion uncertainties of the observed stars. When constructing the mock catalogues, we did not add observational noise to the periods for simplicity. The observed catalogue periods, $P_k$, were adopted as the true periods of the $k$-th mock stars when we selected the distribution-function parameters.

We repeated this mock generation procedure to produce 25 independent mock Mira catalogues.
The period-dependent distribution-function parameters, the period-age relation parameters, the Galactic potential and the solar position and motion were kept fixed across all realisations. The mock stars were tied one-to-one to the observed Mira sample, so that their sky positions, periods and observed distance moduli are the same in every realisation. The realisations differ only in the independently drawn true distance moduli, the random velocity draws from the velocity distribution function at the true position and period and the observational random noises applied to the sampled proper motions. This ensemble allowed us to quantify the mock-to-mock scatter arising from the stochastic elements of the mock construction for the limited number of stars and to examine possible systematic trends in both the recovered period-dependent kinematics and the calibrated period-age relation.

\subsection{Recovery of the period-dependent distribution-function parameters}\label{sec:vali:df}
We first tested whether the distribution-function fitting procedure described in Appendix~\ref{sec:result:mira_DF} could recover the period dependence imposed on the input distribution-function parameters. This test used the 25 mock Mira catalogues described above and focused on the first step of the analysis pipeline, the recovery of the period-dependent distribution function from the Mira observables.

We fitted each mock catalogue using the same likelihood, importance-sampling scheme and MCMC setup as in the analysis of the observed Mira sample described in Section~\ref{sec:result:mira_DF}. The observational uncertainties were treated in the same way as in the main analysis described in Section~\ref{sec:modelling}: in the likelihood evaluation, we marginalised over the uncertainties in distance modulus, proper motion and period using Monte Carlo sampling. As in the real Mira analysis, line-of-sight velocities were not used in the fit.

For computational efficiency, we evaluated equation~\eqref{eq:plbmmu} using 100 Monte Carlo samples, rather than the 1\,280 samples used for the real data analysis in Appendix~\ref{sec:IS}. We verified that this reduction did not significantly affect the recovered distribution-function parameters using several examples of the mock data.

\begin{figure*}	\includegraphics[width=\textwidth]{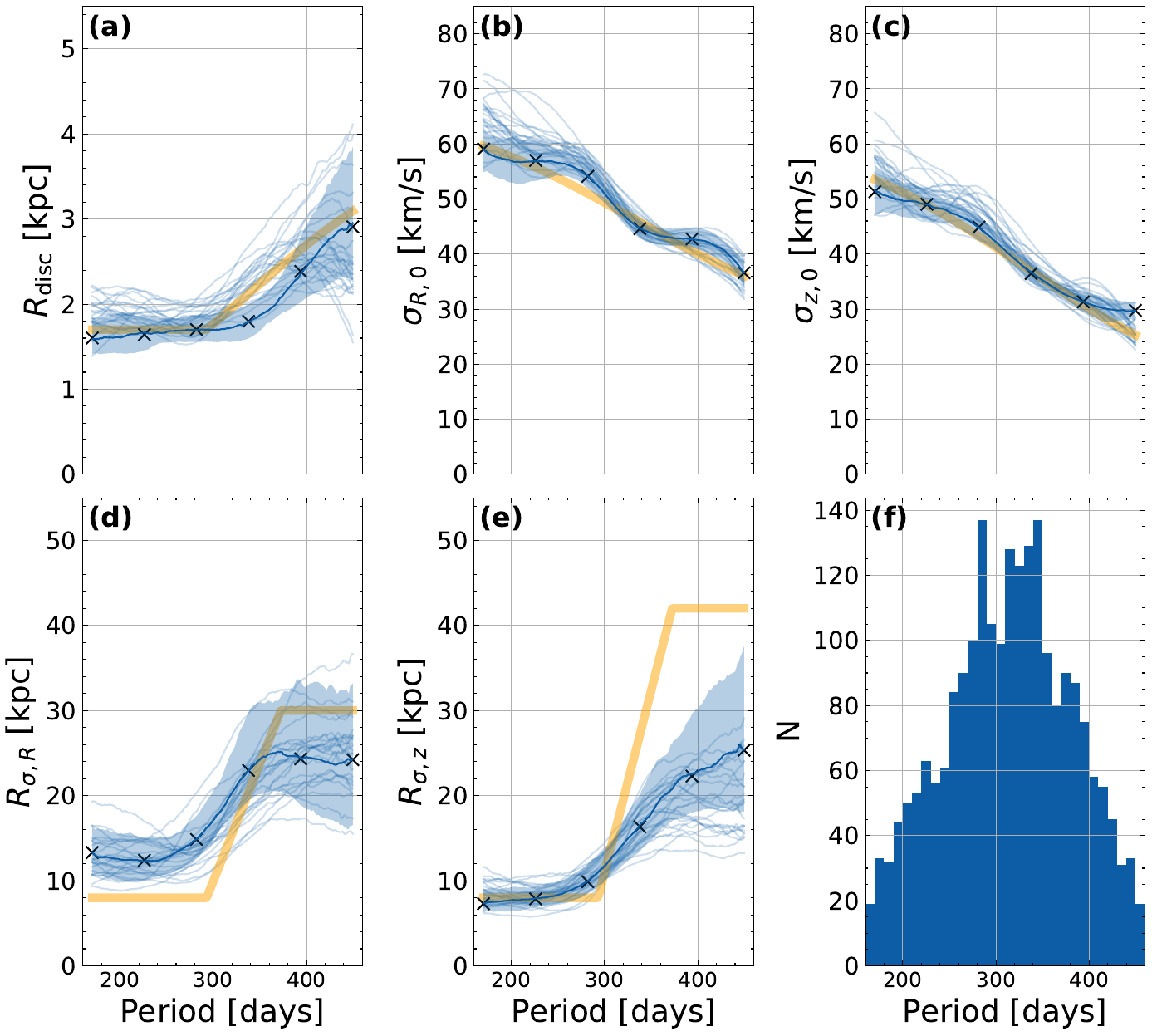}
    \caption{Recovery of the period-dependent distribution-function parameters for one representative mock Mira catalogue. Panels (a), (b), (c), (d) and (e) show the period dependences of $R_{\mathrm{disc}}$, $\sigma_{R,0}$, $\sigma_{z,0}$, $R_{\sigma,R}$ and $R_{\sigma,z}$, respectively. The blue solid curves indicate the posterior medians from the distribution-function fit, while the blue shaded regions represent the corresponding $16$th-$84$th percentile ranges. The thin blue solid curves show the posterior medians obtained from the other 24 mock catalogues but without the indication for the percentiles. The thick orange curves show the input parameter functions used to generate the mock catalogue. Crosses denote the knot positions of the fitted spline functions. Panel (f) shows the stellar period distribution.}
    \label{fig:vali:spline_mock}
\end{figure*}

Panels~(a)-(e) in Fig.~\ref{fig:vali:spline_mock} show the recovery of the period-dependent distribution-function parameters for one representative mock catalogue. The thick orange lines show the period dependence of the input parameters used to generate the mock data, while the solid blue curves and shaded regions show the posterior median and the $16$th-$84$th percentile range from the distribution-function fit to the mock catalogue. The knots used to draw spline curves are shown by crosses. Panel~(f) shows the period distribution of the mock data, which was set to be identical to the observational data.
The constraints are tighter for the velocity-dispersion parameters, $\sigma_{r,0}$ and $\sigma_{z,0}$ than for parameters describing the disc scale length of the stellar density profile, $R_{\rm{disc}}$ and radial scale length of velocity dispersions, $R_{\sigma,R}$ and $R_{\sigma,z}$.

\begin{figure*}	\includegraphics[width=\textwidth]{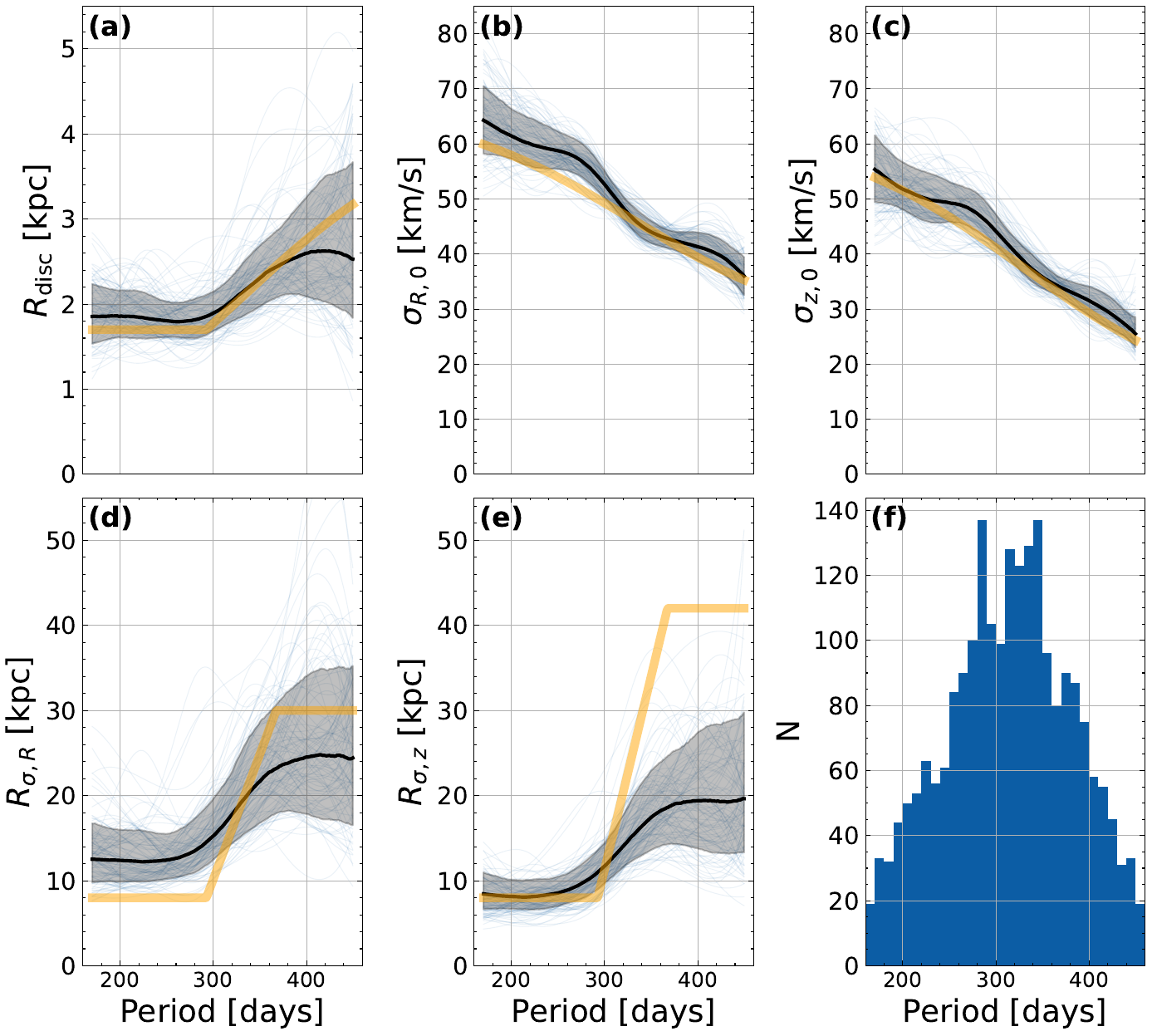}
    \caption{Recovery of the period-dependent distribution-function parameters from the mock Mira catalogues. For each of the 25 independent mock catalogues, 100 posterior spline realisations were randomly selected from the corresponding distribution-function fit to construct the summary. The black solid curves show the median of all these realisations, and the black shaded regions indicate the $16$th-$84$th percentile ranges of the selected realisations over 25 sets of the posterior spline. The thin blue curves show 100 randomly selected spline realisations from the sets of the posterior splines. The thick orange curves show the input parameter functions used to generate the mock catalogues. The panel layout is the same as in Fig.~\ref{fig:vali:spline_mock}: panels (a)--(e) show $R_{\mathrm{disc}}$, $\sigma_{R,0}$, $\sigma_{z,0}$, $R_{\sigma,R}$ and $R_{\sigma,z}$, respectively, and panel (f) shows the period distribution.}
    \label{fig:vali:spline_mock_25}
\end{figure*}

Fig.~\ref{fig:vali:spline_mock_25} summarises the recovery over the 25 independent mock realisations. For this figure, we combined posterior spline realisations drawn from the distribution-function fits to all 25 mock catalogues. The solid black curves show the median of this combined set, and the shaded regions indicate the corresponding $16$th-$84$th percentile ranges. These ranges include both the posterior uncertainty of the individual mock fits and the scatter between independent mock catalogues.

Both $\sigma_{R,0}(P)$ and $\sigma_{z,0}(P)$ are recovered
well in the mock tests. The recovered relations from 25
independent mock realisations (thin solid blue curves) shown in Fig.~\ref{fig:vali:spline_mock} generally reproduce the input trends for both parameters. Nevertheless,
$\sigma_{R,0}(P)$ exhibits slightly larger deviations from
the input relation than $\sigma_{z,0}(P)$. This behaviour
is also reflected in Fig.~\ref{fig:vali:spline_mock_25}.
The median of $\sigma_{R,0}(P)$ for 25 independent mock realisations shows larger offset from the input relation than that of $\sigma_{z,0}(P)$, particularly around
$P\sim260\,\mathrm{days}$. While the difference is small, it
may suggest that $\sigma_{z,0}(P)$ is recovered somewhat more
robustly.

Interestingly, $R_\mathrm{disc}$ is recovered reasonably well for the particular mock realisation highlighted in Fig.~\ref{fig:vali:spline_mock}. The recovered relation reproduces the main features of the input relation, with $R_\mathrm{disc}$ remaining nearly constant at short periods and increasing towards longer periods.
However, the thin blue curves reveal substantial variation among individual mock realisations. Among the many realisations, several realisations produce nearly flat $R_\mathrm{disc}(P)$ relations, while some show a decline towards longer periods, similar to that inferred from the observational data in Fig.~\ref{fig:result:spline}. This behaviour is reflected in the large uncertainty of $R_\mathrm{disc}$ seen in Fig.~\ref{fig:vali:spline_mock_25}, indicating that its recovery is strongly dependent on the particular mock realisation.

The mock validation suggests that the recovery of parameters tied to radial structure, such as $R_{\mathrm{disc}}$, $R_{\sigma,R}$ and $R_{\sigma,z}$, remains substantially more uncertain than that of the local velocity dispersions, $\sigma_{R,0}(P)$ and $\sigma_{z,0}(P)$. This indicates that the main difficulty lies in recovering quantities that require radial coverage across the Mira sample. The fitting of these parameters depends on how the kinematic profiles vary with Galactocentric radius and therefore requires sufficient radial coverage at each period. In the present Mira sample, the number of stars is limited and the radial coverage varies with period, making these parameters difficult to constrain. The absence of line-of-sight velocities further weakens the separation of the radial and azimuthal kinematic components. These effects help explain why the period dependences of $R_{\mathrm{disc}}$, $R_{\sigma,R}$ and $R_{\sigma,z}$ inferred from the real Mira data in  Fig.~\ref{fig:result:spline} do not agree
well with the age-dependent APOGEE red giant trends from
\citet{2025MNRAS.543.2275F}.

It should be noted that we generate the mock catalogues using the quasi-isothermal distribution-function form and analyse them with the same model. Therefore, this test does not assess possible systematic biases that may arise if the true stellar kinematics deviate from the assumed distribution-function form. Moreover, given the current Mira sample size (2\,394 stars) and the absence of line-of-sight velocities, it is not a surprise that we cannot reconstruct the period dependence of the Galactic disc parameters at the same level of detail as \citet{2025MNRAS.543.2275F}.

Nevertheless, the mock tests demonstrate that we can recover $\sigma_{z,0}$, which is the key kinematic parameter used to calibrate the period-age relation of Mira variables, with sufficient accuracy. This result shows that the current data set and modelling framework are well suited for the main goal of this study, calibrating the period-age relation.

\subsection{Recovery of the period--age relation}
We next tested the second step of the analysis pipeline: the calibration of the period--age relation from the recovered Mira period--velocity-dispersion relation. We used the 25 sets of posterior constraints on $\sigma_{z,0}(P)$ obtained from the mock distribution-function fits described in Appendix~\ref{sec:vali:df}. The reference age-velocity dispersion-relation, $\sigma_{z,0}(\tau)$, was constructed from a mock APOGEE red giant catalogue generated to follow the same input age-dependent distribution function as those used in the \rev{Miras'} mock construction, shown in equation~\eqref{eq:f_tau}. 
We adopted the mock red giant results obtained by \citet{2025MNRAS.543.2275F} using the same distribution-function fitting procedure to match the spatial and age distributions of the observed red giant sample.
The resulting fit provides an age-dependent vertical velocity-dispersion relation, $\sigma_{z,0}^{\rm RG}(\tau)$, which we adopted as the reference age--velocity-dispersion relation for the mock period--age calibration. Unlike the Mira mock tests, we did not generate multiple independent mock realisations for the red giant sample. The red giant sample contains substantially more stars, and the stochastic variations associated with individual mock realisations are therefore expected to be much smaller. A single mock-derived age--velocity-dispersion relation is thus sufficient for constructing the reference period--age calibration.

We applied the same period--age fitting procedure as in Section~\ref{sec:result:PAR} to each of the 25 mock Miras' period--velocity-dispersion relations with the reference mock red giant age--velocity-dispersion relations. This test was designed to assess how the full two-step procedure, from the period-dependent distribution-function fit to the period-age calibration, propagates the uncertainties in the recovered $\sigma_{z,0}(P)$ relation into the inferred period-age relation and whether this process causes any biases. It therefore provides a validation of the accuracy and reliability of the calibration.

\begin{figure}	\includegraphics[width=0.48\textwidth]{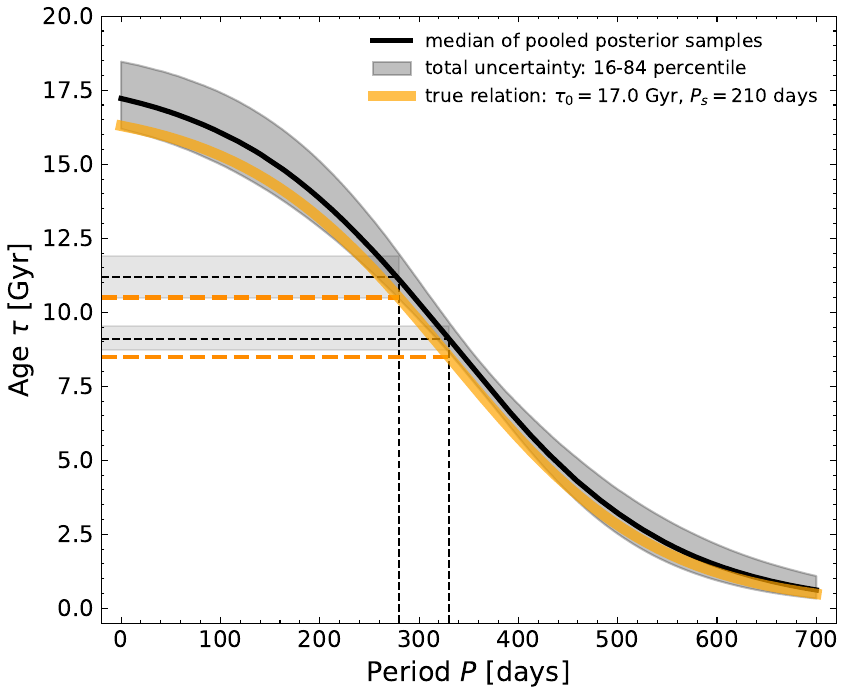}
    \caption{Period--age relation inferred from the combined mock-validation fits. 
Posterior samples of the calibration parameters, $\tau_0$ and $P_s$, were collected from 25 independent mock Mira realisations. 
The black solid line shows the median relation computed from these samples, while the shaded region indicates the 16th--84th percentile range, representing the uncertainty in the inferred period--age relation propagated from the posterior distributions of $\tau_0$ and $P_s$ over the 25 sets of the independent mock realisations. The orange solid lines show the period--age relation drawn from the input parameters. The vertical lines show the bar formation periods $P=280$ and 330 days, and horizontal dashed lines show corresponding ages converted with the recovered mean period--age relation, $\tau=11.2$ and 9.1 Gyr (thin black) and ages converted with the true period--age relation, $\tau=10.5$ and 8.5 Gyr (thick orange).} 
    \label{fig:vali:p-a}
\end{figure}

Fig.~\ref{fig:vali:p-a} summarises the recovery of the period--age relation over the 25 independent mock realisations. 
The black curve shows the median of these recovered period--age relations, and the shaded region indicates their $16$th-$84$th percentile range across the results from the 25 mock realisations. Thus the shaded region in this figure represents the total uncertainty including mock-to-mock scatter and the posterior uncertainty of an individual fit.
The dashed vertical black lines indicate the periods $P=280$ and $330~\mathrm{days}$, which we use as the proxy for the bar-formation epoch.
The dashed horizontal lines indicate the ages corresponding to bar formation periods, when interpreted through the recovered period-age posterior. The horizontal shaded regions show the uncertainty in these inferred ages in the recovered period-age relation over the same periods. For comparison, the orange dotted lines show the ages corresponding to $P=280-330~\mathrm{days}$ under the true input period-age relation (orange thick curves). This comparison suggests that the recovered period-age relation is mildly biased towards older ages by about 0.5~Gyr.

Overall, the mock tests indicate that the analysis pipeline provides a reliable recovery of the imposed period-dependent vertical velocity dispersion, $\sigma_{z,0}(P)$. The subsequent period-age calibration also recovers the true relation well. Although the recovered relation shows a mild systematic shift towards older ages ($\sim$0.5~Gyr), this offset should be regarded as a small additional uncertainty in the inferred Mira ages. The uncertainty in the bar formation epoch is also affected by interpretation of the NSD Mira star formation history and by the uncertainties in the inferred GSE merger epochs. Therefore, this level of systematic offset is unlikely to dominate the comparison between the bar formation epoch and the GSE merger epoch, which is discussed in Section~\ref{sec:discussion}.

\section{Verification of our best-fit model}\label{sec:ap:best}
\begin{figure*}	\includegraphics[width=\textwidth]{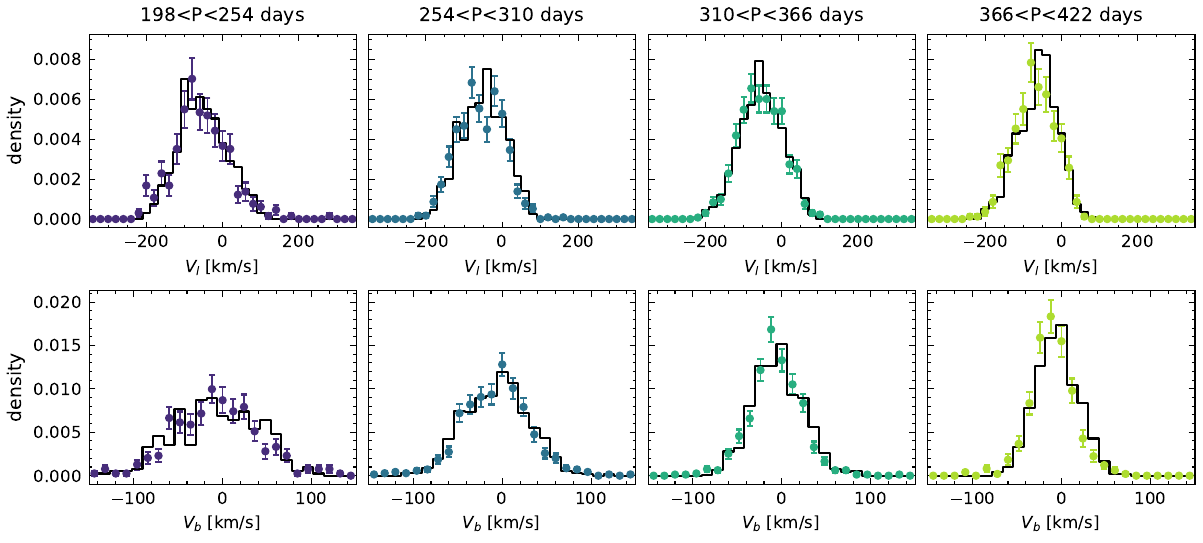}
    \caption{Comparison 
    between the observed tangential velocity distributions (the points with error bars) and those from the best-fit distribution function of the Miras data shown in Table~\ref{tab:mira_df_parameters} (black-solid histograms), computed by taking into account the positions of the stars and the uncertainties in their distance moduli and proper  motions. The top row shows the distributions of $v_\ell$, and the bottom row shows the distributions of $v_b$. The columns correspond to the period bins $198<P<254$, $254<P<310$, $310<P<366$ and $366<P<422$~$\mathrm{days}$. 
}
    \label{fig:result:vlvb_P}
\end{figure*}

\begin{figure*}	\includegraphics[width=\textwidth]{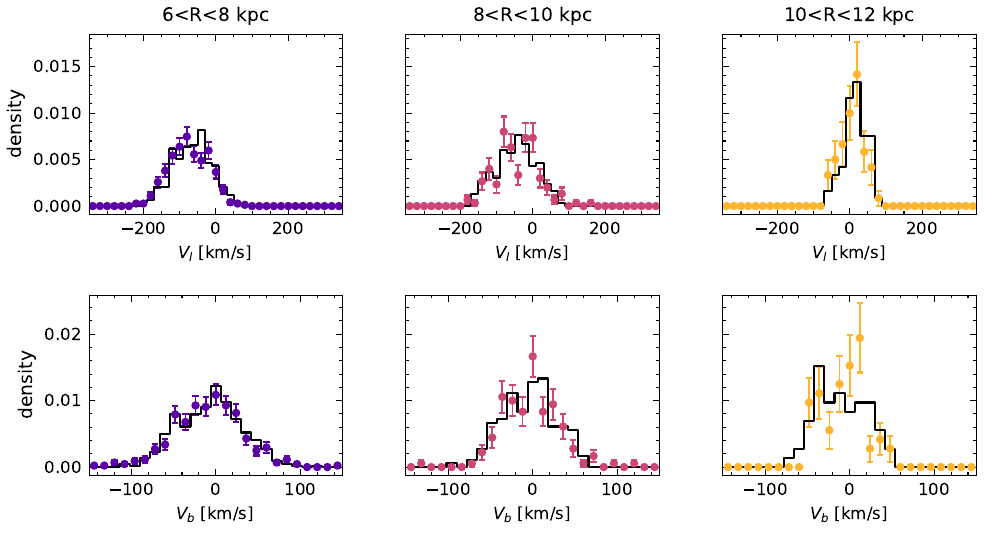}
    \caption{Comparison between the observed and model tangential velocity distributions of Miras in different Galactocentric radial bins. Only stars with $254<P<310~\mathrm{days}$ are shown. The top row shows the distributions of $v_\ell$, and the bottom row shows the distributions of $v_b$. The columns correspond to the radial bins $6<R<8$, $8<R<10$ and $10<R<12~\mathrm{kpc}$. The points with error bars show the observed data, and the black histograms show the best-fit distribution function model at the knot of $P=282$ days.}
    \label{fig:result:vlvb_R1}
\end{figure*}

\begin{figure*}	\includegraphics[width=\textwidth]{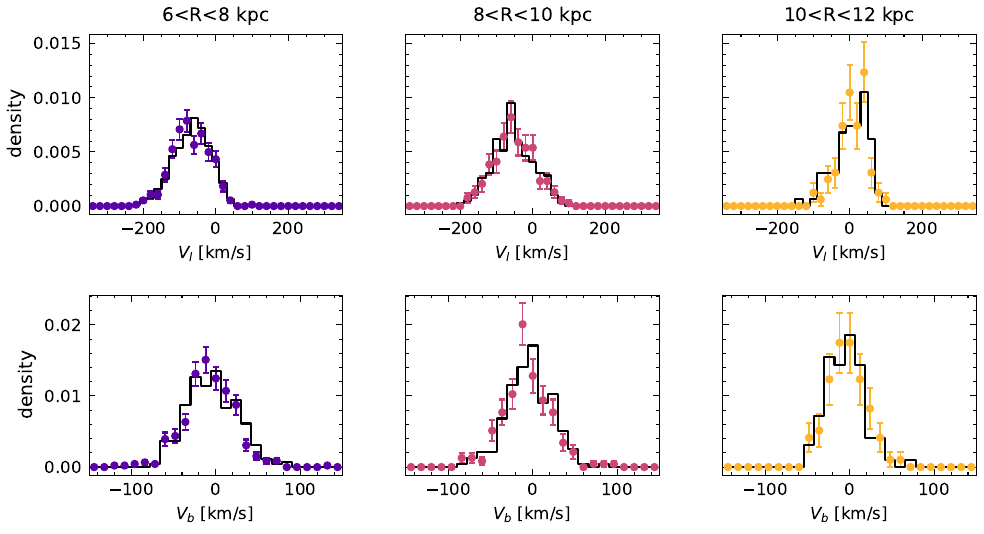}
    \caption{Same as Fig.~\ref{fig:result:vlvb_R1} but for stars with $310<P<366~\mathrm{days}$ compared with the best-fit model at the knot $P=338$ days.}
    \label{fig:result:vlvb_R2}
\end{figure*}

Figs.~\ref{fig:result:vlvb_P}, \ref{fig:result:vlvb_R1} and  \ref{fig:result:vlvb_R2}
compare the observed tangential velocity distributions with those predicted by the best-fit distribution function model. Since we did not use observed line-of-sight velocities for the fitting of Mira sample and instead marginalised the line-of-sight velocity components as explained in Section~\ref{sec:DF}, the model was constrained mainly by the proper-motion information. The distributions of the longitudinal and latitudinal velocities, defined as $v_\ell=4.74047\mu_\ell d$ and $v_b=4.74047\mu_b d$, respectively, where $d$ is the distance, therefore provide a direct diagnostic of the quality of the fit.

Fig.~\ref{fig:result:vlvb_P} compares the observed and modelled velocity distributions in four period bins, $198<P<254, 254<P<310, 310<P<366$ and $366<P<422$ days. The bins are centred on the four interior knots of the period spline shown in Fig.~\ref{fig:result:spline} and Table~\ref{tab:mira_df_parameters}, $P=226,~282,~338$ and $394$~days. For simplicity, the model velocity distributions were computed using the best-fit distribution-function parameters at the corresponding knot periods rather than at the individual periods of the stars. For each observed star, we generated 100 mock realisations by three-dimensional velocities from the best-fit distribution function at the star’s position, including the uncertainties in distance modulus and proper motions to generate mock observations as described in Appendix~\ref{sec:validation} with equations~\eqref{eq:vali:dmerror} and \eqref{eq:vali:pmerror}.
The top row shows the distributions of $v_\ell$, and the bottom row shows those of $v_b$. 
In each panel, the points show the observed Mira data, while the black histograms show the corresponding model predictions. 
The model reproduces the observed distributions reasonably well across the period range. This figure shows the changes in the widths of the $v_\ell$ and $v_b$ distributions with period, which is the main kinematic information used to constrain the period dependence of the velocity-dispersion parameters.

Figs.~\ref{fig:result:vlvb_R1} and \ref{fig:result:vlvb_R2} show the same comparison after dividing the Mira sample into three Galactocentric radial bins for two representative period ranges. Fig.~\ref{fig:result:vlvb_R1} shows the results for $254<P<310~\mathrm{days}$ Miras, and Fig.~\ref{fig:result:vlvb_R2} for $310<P<366~\mathrm{days}$. In each case, the observed distributions of $v_\ell$ and $v_b$ are compared with those from the best-fit model in each radius. The model $v_\ell$ and $v_b$ are generated in the same way as described above.
This tests whether the best-fit model reproduces the spatial variation of the tangential velocity distributions. The model broadly matches the observed $v_\ell$ and $v_b$ distributions in all the radial bins.


\bsp	
\label{lastpage}
\end{document}